\pdfoutput=1 
\documentclass[a4paper, 11pt]{article}
\usepackage[dvipsnames]{xcolor}

\usepackage{jheppub} 

\makeatletter  
\gdef\@fpheader{}  
\makeatother 

\usepackage{tangocolors}
\usepackage{tikz}
\usepackage{tikz-3dplot}
\usetikzlibrary{arrows, decorations,backgrounds, patterns}
\usetikzlibrary{decorations.pathreplacing ,decorations.markings}
\usepackage{amssymb}
\usepackage{amsmath}

\usetikzlibrary{decorations.pathmorphing,backgrounds,shapes,arrows,shadows}
\usetikzlibrary{patterns.meta}
\usetikzlibrary{patterns,decorations.pathmorphing}
\tikzset{
    snake it/.style={decorate, decoration=snake}
}
\usepackage{pgfplots}
\pgfplotsset{compat=1.11}
\usepgfplotslibrary{fillbetween}
\usetikzlibrary{intersections}
\pgfdeclarelayer{bg}
\pgfsetlayers{bg,main}
\tikzset{zigzag/.style={decorate,decoration=zigzag}}
\tikzset{snake it/.style={decorate, decoration=snake}}
\makeatletter
\def\@hex@@Hex#1%
 {\if a#1A\else \if b#1B\else \if c#1C\else \if d#1D\else
  \if e#1E\else \if f#1F\else #1\fi\fi\fi\fi\fi\fi \@hex@Hex}
\makeatother

\usepackage[all]{xy}
\usepackage[percent]{overpic}
\usepackage{slashed}
\usepackage{wrapfig}
\usepackage{tabu}
\usepackage{diagbox}
\usepackage{mathrsfs,amsmath,amssymb,amsthm,amsfonts,tikz,graphicx,accents,hyperref, color}
\usepackage{dsfont,epiolmec, latexsym, stmaryrd, comment}
\usepackage{slashed,ccaption}
\usepackage{mathrsfs, calligra, bbm}
\usepackage{leftidx}
\usepackage{import}
\usepackage{multirow}
\usepackage{amsfonts}
\usepackage{pifont}
\usepackage{tabularx}
\usepackage{cancel}
\usepackage[normalem]{ulem}
\usepackage[utf8]{inputenc}
\usetikzlibrary{intersections,calc}
\usepackage{ifthen}
\usepackage{amsmath}
\usepackage{amsthm}

\usepackage{physics}

\usepackage{tabularray}
\UseTblrLibrary{booktabs}
\usepackage{nicematrix}

\usepackage[T1]{fontenc}
\usepackage{lmodern}

\usepackage{cancel}
\usepackage{caption} 
\usepackage{subcaption}

\usetikzlibrary{patterns.meta}
\usepackage{array}
\usepackage{bm}
\usepackage{stmaryrd}

\hypersetup{ linktoc=all,
    colorlinks, linkcolor={palatinateblue},
    citecolor={brightpink}, urlcolor={amaranth}
}

\graphicspath{{Images/}}

\renewcommand{\d}[1]{\ensuremath{\operatorname{d}\!{#1}}}

\def\sideremark#1{\ifvmode\leavevmode\fi\vadjust{\vbox to0pt{\vss
 \hbox to 0pt{\hskip\hsize\hskip1em
 \vbox{\hsize2cm\tiny\raggedright\pretolerance10000
 \noindent #1\hfill}\hss}\vbox to8pt{\vfil}\vss}}}%
\DeclareSymbolFont{extraup}{U}{zavm}{m}{n}
\DeclareMathSymbol{\varheart}{\mathalpha}{extraup}{86}
\DeclareMathSymbol{\vardiamond}{\mathalpha}{extraup}{87}
\makeatletter
\renewcommand*{\@fnsymbol}[1]{\ensuremath{\ifcase#1\or \clubsuit \or \vardiamond \or \varheart\or
    \spadesuit\or \mathparagraph\or \|\or **\or \dagger\dagger
    \or \ddagger\ddagger \else\@ctrerr\fi}}
\makeatother

\definecolor{rosy}{RGB}{230,235,252}
\definecolor{myframetitle}{RGB}{90,89,170}
\definecolor{myblocktitle}{RGB}{140,185,249}
\definecolor{mytitle}{RGB}{10,80,26}

\definecolor{darkgreen}{RGB}{27,130,45}
\definecolor{darkblue}{rgb}{0,0,0.3}
\definecolor{darkred}{rgb}{0.7,0,0}

\definecolor{light gray}{RGB}{220,220,220}
\definecolor{dark purple}{RGB}{108,0,217}
\definecolor{pink}{RGB}{190,20,100}
\definecolor{orang}{RGB}{193,63,0}
\definecolor{green}{RGB}{11,98,17}
\definecolor{darkpink}{RGB}{153,0,76}
\definecolor{bluegreen}{RGB}{0,102,102}
\definecolor{greenlagan}{RGB}{0,102,0}
\definecolor{redgreen}{RGB}{102,102,0}
\definecolor{Redgreen}{RGB}{153,76,0}
\definecolor{vividviolet}{rgb}{0.62, 0.0, 1.0}
\definecolor{amaranth}{rgb}{0.9, 0.17, 0.31}
\definecolor{palatinateblue}{rgb}{0.15, 0.23, 0.89}
\definecolor{brightpink}{rgb}{1.0, 0.0, 0.5}
\definecolor{cornflowerblue}{rgb}{0.39, 0.58, 0.93}
\definecolor{deepcarminepink}{rgb}{0.94, 0.19, 0.22}
\definecolor{radicalred}{rgb}{1.0, 0.21, 0.37}
\definecolor{darkmagenta}{rgb}{0.67, 0, 0.67}

\newcommand{\tA}{\text{A}}
\newcommand{\tB}{\text{B}}
\newcommand{\tC}{\text{C}}
\newcommand{\tD}{\text{D}}

\newcommand{\hll}{\hat{[}}
\newcommand{\hrr}{\hat{]}}

\newcommand{\fspartial}{%
  \text{
    \ooalign{%
      $\boldsymbol{\Game}$\cr
      \hidewidth\raisebox{0.60ex}{\rule{0.40em}{0.16ex}}\hidewidth\cr
    }%
  }%
}
\makeatletter
\newif\ifkp@upRm
\makeatother
\usepackage{mathptmx}

\usepackage[most]{tcolorbox}

\tcbset{highlight math style={left=02mm,right=02mm,top=02mm,bottom=02mm}} 
\usepackage{empheq}

\newcommand\inbox[1]{\tcbset{fonttitle=\scriptsize} \tcboxmath[colback=white,colframe=black!70]{#1}}

\DeclareFontFamily{OT1}{rsfs}{}

\DeclareFontShape{OT1}{rsfs}{m}{n}{ <-7> rsfs5 <7-10> rsfs7 <10->rsfs10}{} 

\DeclareMathAlphabet{\mycal}{OT1}{rsfs}{m}{n}
\def\mathbi#1{\textbf{\em #1}}

\newcommand{\B}{\mathcal{B}}

\newcommand{\be}{\begin{equation}}
\newcommand{\ee}{\end{equation}}
\newcommand{\bea}{\begin{eqnarray}}
\newcommand{\eea}{\end{eqnarray}}
\makeatletter \@addtoreset{equation}{section}

\begin{document}


\newcommand{\mytitle}{\begin{center}{\textbf{\LARGE{Geometric Aspects of Covariant Phase Space Formalism}: \\ 
\Large{Solution Space Slicings and Surface Charge Integrability }}}
\end{center}}

\title{{\mytitle}}
\author[a]{ M.~Golshani}
\author[b]{, M.M.~Sheikh-Jabbari}
\author[a,b]{, V.~Taghiloo}
\author[a, b]{, M.H.~Vahidinia}

\affiliation{$^a$ Department of Physics, Institute for Advanced Studies in Basic Sciences (IASBS),\\ 
P.O. Box 45137-66731, Zanjan, Iran}
\affiliation{$^b$ School of Physics, Institute for Research in Fundamental
Sciences (IPM),\\ P.O.Box 19395-5531, Tehran, Iran}
\emailAdd{
mahdig@iasbs.ac.ir, 
jabbari@theory.ipm.ac.ir \& shahin.s.jabbari@gmail.com, v.taghiloo@iasbs.ac.ir, vahidinia@iasbs.ac.ir
}
\abstract{
{The Covariant Phase Space Formalism (CPSF) provides a robust framework for deriving symplectic structures and surface charges in diffeomorphism-invariant theories. By construction, the CPSF operates on two distinct manifolds: the spacetime and the Solution Phase Space (SPS). In this paper, we advance the formalism by establishing a strictly parallel geometric formulation for both manifolds. Within this framework, we systematically analyze diffeomorphisms and frame changes on both spaces. While spacetime diffeomorphisms have been extensively studied in the literature, transformations on the SPS have been largely overlooked; we rigorously define and investigate these as \emph{changes of slicing} on SPS. We demonstrate that the standard Wald-Zoupas criterion for the integrability of surface charge variations is inherently slicing-dependent. To resolve this issue, we develop the Frobenius theorem on the SPS and use it to extend the Wald-Zoupas condition into an inherently slicing-independent criterion for integrability. The Frobenius theorem on the SPS also yields a rigorous and natural definition of fundamental geometric quantities on the solution space, specifically the SPS connection, torsion, and curvature. Furthermore, this geometric machinery naturally distinguishes between fundamentally different surface fluxes: “fake” fluxes are identified mathematically as pure gauge artifacts of the SPS connection, while “genuine” fluxes manifest as non-vanishing SPS torsion, which directly relates to the physical gravitational News tensor. Finally, we present a geometric formulation of the Liouville theorem on the SPS, offering a unified classification scheme for theories with and without propagating bulk degrees of freedom.}}

\maketitle

\section{Introduction}\label{sec:Intro}

Noether's theorem stands as one of the most beautiful and profound results in theoretical physics, formally establishing that symmetries and conserved quantities are two sides of the same coin~\cite{Noether:1918zz}. From this theorem, we understand that the conservation laws of energy, momentum, and electric charge are intrinsically rooted in underlying symmetry principles. More precisely, Noether's first theorem dictates that \emph{global} continuous symmetries yield conserved currents. However, while this applies elegantly to global symmetries, the story becomes significantly more complex when considering local (gauge) symmetries. Noether's second theorem states that local symmetries yield Noether currents that identically vanish on-shell--amounting to identities among the equations of motion. Consequently, it was long held that local symmetries are merely redundancies in the physical description of a system. 

The crucial insight developed over the last half-century is that {this paradigm shifts significantly} in the presence of boundaries (e.g., asymptotic boundary of AdS space, spatial infinity, null infinity, black hole horizons, or general boundaries at finite distance). At a boundary, a measure-zero subset of local gauge symmetries yields physical, asymptotic, or boundary symmetries, leading to non-vanishing, physically meaningful charges. A notable example is the Brown-Henneaux charges for asymptotically AdS spacetimes in three dimensions~\cite{Brown:1986nw}, which yield two copies of the Virasoro algebra--a result that is usually viewed as a precursor of the AdS/CFT correspondence~\cite{Maldacena:1997re, Gubser:1998bc, Witten:1998qj}. Another prominent example involves the BMS (Bondi--van der Burg--Metzner--Sachs) charges at null infinity in asymptotically flat spacetimes~\cite{Bondi:1962px, Sachs:1962wk}. Today, the BMS group and its extensions serve as the principal guiding framework for constructing flat space holography, driving vibrant contemporary research programs in both celestial holography~\cite{Strominger:2017zoo, Pasterski:2021rjz, Raclariu:2021zjz} and Carrollian holography~\cite{Bagchi:2012cy, Donnay:2022aba, Bagchi:2023cen, Ruzziconi:2026bix, Bagchi:2025vri}.

In contrast to Noether's first theorem--which serves as a universal tool for associating conserved currents to global symmetries--associating charges with local symmetries in the presence of boundaries has spawned various methodologies. Notable examples include the ADM~\cite{Arnowitt:1962hi} and ADT~\cite{Abbott:1981ff, Deser:2002rt} Hamiltonian approaches, the Brown-York Hamilton-Jacobi method~\cite{Brown:1992br}, Komar integrals in gravity~\cite{Komar:1958wp}, and the Anderson homotopy operator approach~\cite{Anderson:1992kla, Barnich:2001jy}. Among these, the Covariant Phase Space Formalism (CPSF)~\cite{Crnkovic:1987tz, Lee:1990nz, Iyer:1994ys} (for a modern pedagogical treatment, see e.g.,~\cite{Grumiller:2022qhx}) stands out for its { manifestly covariant nature}. Unlike Hamiltonian approaches that rely heavily on foliation-dependent splittings of spacetime into space and time, the CPSF circumvents the need for a background time direction by relying entirely on the symplectic structure and the flow of vector fields directly on the space of solutions; CPSF is hence the right framework to study charges in generally covariant theories.

A foundational, yet less appreciated, aspect of the CPSF is that it inherently operates on \emph{two distinct manifolds}: the physical spacetime $\mathcal{M}$ and the Solution Phase Space (SPS) $\mathcal{P}$. In this work, alongside reviewing the standard formalism, we systematically reformulate it to treat both manifolds in parallel. Consequently, the framework is governed by two independent sets of transformations: spacetime diffeomorphisms (coordinate transformations on $\mathcal{M}$) and SPS gauge symmetries (coordinate transformations on $\mathcal{P}$). While spacetime gauge symmetries have been {extensively} studied, we primarily focus on the (gauge) symmetries of the SPS. We argue that understanding these phase-space coordinate transformations--which we refer to as changes of \emph{slicing}~\cite{ Adami:2020ugu, Adami:2021nnf, Adami:2022ktn, Taghiloo:2024ewx, Ruzziconi:2020wrb, Geiller:2021vpg, Geiller:2024amx, Ciambelli:2023ott, Ciambelli:2024vhy}--is crucial to understand and resolving the physics of surface charges and their integrability.

CPSF is a mathematically (geometrically) robust formulation that can be used to compute the \emph{variation} of a charge (a 1-form on the SPS $\mathcal{P}$) associated with symmetries, in contrast to Noether's theorem that yields the charge itself. Consequently, before one can assign a definitive physical charge to a state, one must first confront the problem of \emph{integrability}.  The most direct and, in a sense, natural integrability criterion is the Wald-Zoupas (WZ) criterion~\cite{Wald:1999wa}:\footnote{The idea of integrability was originally introduced by Lee and Wald \cite{Lee:1990nz} for field-independent symmetry generators, which was more explicitly spelled out and discussed in the Wald-Zoupas paper \cite{Wald:1999wa}. Its extension to cases with field-dependent symmetry generators was later developed in \cite{Compere:2015knw}.\label{footnote-integrability}} charge is integrable if the charge variation is an exact 1-form on the SPS. It is important to recognize that non-integrability is not  a ``flaw'' of the formalism; rather, it is a feature with physical significance, it indicates presence of flux in the system, and the notions of charge (non)integrability and charge (non)conservation are closely related to each other, see e.g. \cite{Adami:2020ugu, Adami:2021nnf, Golshani:2024fry, Adami:2024gdx} for further discussions. The CPSF also sheds light on the fact that why charges in the general relativity are either not covariant on spacetime or not background independent (non-covariant on the field space) \cite{Golshani:2024fry}. 

The WZ criterion may indicate a non-integrable charge where one physically expects charges to be integrable due to the absence of physical flux of bulk propagating degrees of freedom through the boundary~\cite{Barnich:2010eb, Compere:2015knw}. There are several examples of  such cases that, despite being non-integrable by the WZ criterion, the charges can be rendered integrable upon appropriate redefinition of  fields and symmetry generators, a change-of-slicing~\cite{Adami:2020ugu, Adami:2021nnf, Adami:2022ktn, Taghiloo:2024ewx, Ruzziconi:2020wrb, Geiller:2021vpg, Geiller:2024amx, Ciambelli:2023ott, Ciambelli:2024vhy}. One of the objectives of this work is to provide an integrability criterion that is covariant and slicing-independent on the SPS. We do so by elevating the CPSF into a rigorously geometric, coordinate-free framework on the SPS and working out an SPS Frobenius theorem, generalizing the standard  Frobenius theorem for usual spacetime (and vector fields) \cite{Wald:1984rg, Lee2013}. Note that, as stressed in \cite{Golshani:2024fry}, invariance under field redefinitions on SPS and change-of-slicing is a manifestation of ``background independence,'' which is one of the salient features of general relativity. To achieve this, we recognize that the collection of boundary charge variation 1-forms, which we denote as $\boldsymbol{\beta}_a$, acts mathematically as a {boundary} co-frame (or vielbein) on the SPS manifold $\mathcal{P}$. A choice of ``slicing'' is simply a gauge choice of these local frames. 

By applying Cartan's geometric machinery to the phase space, the variation of the charge aspects naturally obeys the first Cartan structure equation:
\begin{equation*}
    \inbox{\delta \boldsymbol{\beta}_a = {\mathscr{A}}_a{}^b \curlywedge \boldsymbol{\beta}_b + {\mathscr{F}}_a\, .}
\end{equation*}
This single geometric equation beautifully separates the coordinate kinematics of the phase space from the physical dynamics of the system:
\begin{itemize}
    \item \textbf{SPS connection ${\mathscr{A}}_a{}^b$ and Fake Flux:} ${\mathscr{A}}_a{}^b$ is a 1-form on the SPS and dictates how the charge aspects rotate into one another under an SPS coordinate transformation (change-of-slicing). Non-zero terms arising from this connection represent \emph{fake flux}--apparent non-integrability that is purely an SPS gauge artifact. Fake flux can always be removed by an appropriate frame transformation (a flat SPS connection with vanishing curvature).
    \item \textbf{Torsion ${\mathscr{F}}_a$ and Genuine Flux:} The torsion 2-form ${\mathscr{F}}_a$ represents the  gauge-invariant obstruction to integrability. When ${\mathscr{F}}_a \neq 0$, no SPS gauge transformation can render the charges integrable. This represents a \emph{genuine flux}: the true, physical radiation of bulk propagating modes passing through the boundary.
\end{itemize}

We identify the source of violation of the Frobenius integrability condition as the presence of SPS torsion, and use this {to provide a strictly covariant integrability criterion}: a system is physically integrable if and only if the genuine flux (torsion) strictly vanishes. When our generalized criterion indicates non-integrability, it guarantees that this is entirely due to the presence of physical bulk radiation, extending the WZ integrability condition that may have slicing-dependent artifacts. 

{
\paragraph{Summary of the framework and main new results.} 
{In this work, we review, refine and reformulate some previously known notions and statements in the CPSF framework build new and novel results upon them.} }

\noindent {{Review, refinement and reformulation:}
\begin{itemize}
    \item A rigorous geometric formulation of the Solution Phase Space (SPS) constructed in parallel to the spacetime geometry, highlighting the {double}-manifold nature of the CPSF.
    \item {A precise definition of \emph{covariance} for generic quantities taking into account the``double-manifold'' nature of the CPSF.}
\end{itemize}}

\noindent {{Main Novel Results:}
\begin{itemize}
    \item A robust generalization of the Wald-Zoupas integrability criterion via the Frobenius theorem, rendering the condition strictly independent of the SPS coordinate choice (slicing).
    \item A mathematically rigorous formulation of the ``change-of-slicing'' technique as a local frame (gauge) transformation on the SPS.
    \item An unambiguous geometric distinction between \emph{fake flux} (SPS connection/gauge artifacts) and \emph{genuine flux} (SPS torsion/physical radiation) utilizing the Frobenius theorem.
    \item A novel geometric formulation of the Liouville Theorem for the boundary sector of the SPS, carefully specifying regimes with and without bulk propagating degrees of freedom.
\end{itemize}
We study the implications of this general analysis and these results by applying them to some physically interesting examples.}


{\paragraph{Comparison with the existing literature.}
{Several ingredients of the present work build upon well-established developments in the covariant phase space literature. We review notions like covariant phase space formalism, symplectic currents and surface charges, the geometric description of field space, field-space covariance, Cartan calculus,
as well as bulk/boundary decompositions and boundary degrees of freedom, to establish notation and provide a unified geometric framework (see, e.g. \cite{Lee:1990nz, Iyer:1994ys, Wald:1999wa, Anderson:1992kla, 
Barnich:2001jy, Donnelly:2016auv, Compere:2018aar, Ciambelli:2022vot, Grumiller:2022qhx, Speziale:2025lkm} and references therein).  
We also make use of concepts related to field-space connections and curvature that have 
appeared in previous studies that we cited above, see also \cite{Gomes:2016mwl, Hopfmuller:2018fni, Riello:2019tad, Chandrasekaran:2020wwn}. It is important to explicitly distinguish the conceptual positioning of our framework within the wider field: while much of the recent literature 
relies heavily on the corner/bulk (codimension-2 vs.\ codimension-0) perspective to restore gauge covariance and address charge integrability via edge modes 
\cite{Donnelly:2016auv, Speranza:2017gxd, Geiller:2021vpg, Freidel:2021dxw, Ciambelli:2023ott, Ciambelli:2024vhy}, our present formulation builds squarely on the complementary boundary/bulk (codimension-1 vs.\ codimension-0) viewpoint explored extensively in our 
 previous papers \cite{Adami:2020ugu, Adami:2021nnf, Adami:2022ktn, Adami:2023wbe, Taghiloo:2024ewx, Golshani:2024fry, Adami:2024gdx}.} Within this setting, 
 we systematically formulate solution phase space (SPS) slicings and changes of slicing as geometric frame transformations and analyze the dependence of 
 the Wald–Zoupas integrability condition on such slicings. The main novel elements of this work are the Frobenius/Cartan formulation based on the charge aspect 
 co-frame $\boldsymbol{\beta}_a$, the decomposition $\delta \boldsymbol{\beta}_a = \mathscr{A}_{a}{}^{b} \curlywedge \boldsymbol{\beta}_b + \mathscr{F}_{a}$ 
 and its interpretation as a rigorous mathematical separation between slicing-removable and genuine non-integrability, the resulting geometric characterization of fake and genuine fluxes, 
 and the formulation of a Liouville theorem on the solution phase space, together with its implications for classifying theories with and without propagating bulk degrees of freedom.}

\paragraph{Outline of the paper.}
This paper begins with a review and geometric reformulation of the Covariant Phase Space Formalism (CPSF), developed in parallel with spacetime and field space geometries in Sections~\ref{sec: CPSF-kinematics}. Using the structures introduced there, we mathematically define the notion of covariance for generic quantities on the field space. In Section \ref{sec: CPSF-dynamics}, we apply the mathematical structure of the previous section to physical theories defined by an action. We discuss the solution space as a subspace of the field space and the solution phase space (SPS) as a subspace of the solution space. The latter, as the name indicates, is a phase space equipped with a symplectic form. We also discuss surface charges and their algebra. Section~\ref{sec: slicing} introduces the concept of slicing in the SPS, characterizing it as an SPS gauge choice, and analyzes its critical role in the definition of surface charges. In Section~\ref{sec: integrability}, we study the integrability of surface charge variations within the CPSF. By exploiting the Frobenius theorem, we derive generalized, slicing‑independent integrability criteria,  extending the Wald–Zoupas condition and leading to a geometric characterization of charge non‑integrability and flux via phase-space torsion. Section~\ref{sec:examples} applies this framework to a range of explicit examples, clearly illustrating the distinction between fake and genuine flux. We conclude with a discussion and outlook in Section~\ref{sec: discussion}. Technical details and supplementary derivations are collected in the appendices.

\section{Covariant phase space formalism: Geometric structure and mathematical tools}\label{sec: CPSF-kinematics}

In the covariant phase space formalism (CPSF), we deal with two distinct manifolds: spacetime \(\mathcal{M}\) and field space \(\Gamma\). In this section, we introduce the geometric structures associated with the field space, drawing a parallel with those of the spacetime manifold. Throughout, we assume that the spacetime dimension is \(d\). {Before proceeding, we emphasize that the geometric structures on the field space—such as the variational bicomplex and jet bundle calculus—have been extensively developed and discussed in the literature \cite{vinogradov1977algebro, vinogradov1978spectral,vinogradov1984cspectral1, vinogradov1984cspectral2, Anderson:1992kla, Anderson:1996sc, Khudaverdian:2001qe, Barnich:2000zw, Barnich:2001jy, Donnelly:2016auv} (see also \cite{Compere:2018aar, Ciambelli:2022vot, Ciambelli:2023bmn, Speziale:2025lkm} for comprehensive reviews). We present this overview primarily to unify these concepts under our specific conventions and to establish the precise notation required for our subsequent analysis.}
\subsection{Geometric tools of CPSF: Variational bi-complex}\label{sec:geometric-tools-CPSF}
In this {sub}section, we introduce the geometric tools of both spacetime and field space in parallel. 
\paragraph{Coordinates.} 
{To perform calculus on manifolds, we need to introduce charts or coordinate systems. In the spacetime, we use coordinates denoted by \( x^\mu \), where \( \mu = 0,1,\dots,d-1 \). Similarly, in the field space, the coordinates are given by the fundamental or dynamical fields of a given theory. We denote these dynamical fields as \( \phi^{I}(x) \), {where $I=1,\cdots,D$ labels the $D$ discrete field components}. Note that $\phi^I(x)$ may be scalars, vectors, forms, or a generic tensor on spacetime. 
{In our notation,  \( \phi^I \) denotes a coordinate on the field space, whereas \( \phi^I(x) \) denotes the corresponding spacetime field  evaluated at the spacetime point \(x\). 
Nonetheless, after \eqref{One-form-field-space} and since there will be no confusion, we will ease the notation and simply use $\phi^I$, dropping an explicit mention of the $(x)$-dependence.}
Just as spacetime is described by a finite set of coordinates, field space is a manifold whose coordinates form an infinite-dimensional set. The collection of all dynamical fields will be denoted by \( \Phi = \{\phi^I\} \). 

\paragraph{Tangent space.}
At each point on a manifold, there exists a vector space of the same dimension as the manifold, known as the tangent space. In the case of spacetime, the tangent space is spanned by the basis vectors \(\partial_\mu\), and any vector field is written as a linear combination of these basis vectors $V(x)=V^{\mu}(x)\,\partial_{\mu}$. In parallel with spacetime, we introduce a basis on the tangent space of field space as follows {$\fspartial_I :=\frac{\fspartial}{\fspartial \phi^{I}(x)}$}. Then, any vector field on the field space is written as follows
 \begin{equation}
     \hat V[\Phi] :=\hat{V}^{I}\fspartial_I = \sum_{I}\int \d{}^d x\, \hat V^{\phi^{I}(x)} \frac{\fspartial}{\fspartial\phi^{I}(x)} \, .
 \end{equation}
As the notation implies, we use Einstein's summation convention also in the field space. This notation will be used consistently throughout the paper.
\paragraph{Cotangent space.}
The dual space of the tangent space is known as the cotangent space. In spacetime, the basis dual to the tangent space basis vectors is given by \( \mathrm{d}x^\mu \), satisfying the property $\partial_{\mu}(\mathrm{d}x^{\nu}) = \delta^{\nu}_{\mu}$. Any arbitrary 1-form in the spacetime can be written as $\omega(x)=\omega_{\mu}(x)\, \d{}x^{\mu}$. One can similarly introduce the cotangent space of the field space. We denote the corresponding basis on the cotangent space with $\delta \phi^I$ with the following property 
\begin{equation}
    \frac{\fspartial}{\fspartial \phi^{I}(x)}\left(\delta \phi^{J}(y)\right) = \delta^{J}_{I}\, \delta^{(d)} (x-y)\,  \qquad \Longrightarrow \qquad \fspartial_{I}(\delta \phi^{J})=\delta_{I}^{J}\, .
\end{equation}
Any arbitrary 1-form in phase space can hence be expanded as follows
 \begin{equation}\label{One-form-field-space}
\hat\omega[\Phi]=\sum_{I}\int \d{}^d x\, \hat{\omega}_{\phi^{I}(x)}\,\delta{\phi^{I}(x)} \equiv \hat{\omega}_{_{I}}\, \delta \phi^{I}\, . 
 \end{equation}

\paragraph{Tensors.}
A tensor field of rank \((m,n)\) in spacetime is defined as follows
\begin{equation}
    T (x)= T^{\mu_1 \cdots \mu_m}{}_{\nu_1 \cdots \nu_n}(x) \, \partial_{\mu_1} \otimes \cdots \otimes \partial_{\mu_m} \otimes \d{x}^{\nu_1} \otimes \cdots \otimes \d{x}^{\nu_n} \, ,
\end{equation}
where \(\otimes\) is the tensor product in spacetime. Similarly, a tensor of rank \((M,N)\) in the field space is defined as
\begin{equation}
    \hat{T}[\Phi] = \hat{T}^{I_1 \cdots I_M}{}_{J_1 \cdots J_N}[\Phi] \, \fspartial_{I_1} \circledast \cdots \circledast \fspartial_{I_M} \circledast \delta \phi^{J_1} \circledast \cdots \circledast \delta \phi^{J_N} \, ,
\end{equation}
where \(\circledast\) denotes the tensor product in the field space.
\paragraph{Forms.}
In spacetime, a \( p \)-form field is a \((0,p)\) tensor that is completely antisymmetric
\begin{equation}
    \omega(x) = \frac{1}{p!} \omega_{\mu_1 \cdots \mu_p}(x) \d{}x^{\mu_1} \wedge  \cdots \wedge \d{}x^{\mu_p}\, ,
\end{equation}  
where \(\wedge\) denotes the wedge product in spacetime.  

Similarly, we define \((0,P)\) tensors in field space that are totally antisymmetric 
\begin{equation}
    \hat{\omega}[\Phi] = \frac{1}{P!} \hat{\omega}_{I_1 \cdots I_P}[\Phi]  \delta \phi^{I_1} \curlywedge  \cdots \curlywedge \delta \phi^{I_P}\, ,
\end{equation}  
where \(\curlywedge\) represents the wedge product in the field space.  
\paragraph{$(p,Q)$-form notation.}
In the previous part, we introduced tensor fields in both spacetime and field space. In the context of the CPSF, we naturally encounter tensors of mixed rank, in particular $(p,Q)$-forms, $p$-forms in spacetime, and $Q$-forms in field space. A $(p,Q)$-form can be expanded in the field space basis as follows
\begin{equation}
    \hat{X} =\frac{1}{Q!} \hat{X}_{I_1 \cdots I_Q} \, \delta \phi^{I_1} \curlywedge \cdots \curlywedge \delta \phi^{I_Q}\, ,
\end{equation}
where \( \hat{X}_{I_1 \cdots I_Q} \) are $p$-forms in spacetime. Similarly, we can expand \( \hat{X} \) in the spacetime basis
\begin{equation}
    \hat{X} =\frac{1}{p!} \hat{X}_{\mu_1 \cdots \mu_p} \, \d{}x^{\mu_1} \wedge \cdots \wedge \d{}x^{\mu_p}\, ,
\end{equation}
where \( \hat{X}_{\mu_1 \cdots \mu_p} \) are $Q$-forms in the field space. Finally, we can expand \( \hat{X} \) in both field space and spacetime as \footnote{{We use independent spacetime and field-space exterior algebras, so $\d{}x^\mu\delta\phi^I=\delta\phi^I \d{}x^\mu$.}}
\begin{equation}
    \hat{X} =\frac{1}{p! Q!} \hat{X}_{I_1 \cdots I_Q; \mu_1 \cdots \mu_p} \, (\delta \phi^{I_1} \curlywedge \cdots \curlywedge \delta \phi^{I_Q}) \, (\d{}x^{\mu_1} \wedge \cdots \wedge \d{}x^{\mu_p})\, .
\end{equation}

{We adopt the following notation. If a quantity $\textbf{X}$ is a $(p,Q)$-form, then $X$ denotes its Hodge dual, which is a $(d-p)$-vector object on spacetime (i.e., it carries $(d-p)$ antisymmetric upper indices) and simultaneously a $Q$-form on field space. For example, if $\boldsymbol{\Theta}$ is a $(d-1,1)$-form, then $\Theta^\mu$ denotes the corresponding 1-form density on field space. We may switch between the two notations, depending on which notation we find more convenient.}
\paragraph{Derivative operators on spacetime.}
In our analysis, we encounter the following  derivative operators on spacetime 
\begin{itemize}
    \item[(1)] {Exterior derivative \(\mathrm{d}\):} This operator increases the rank of differential forms by one.  
\item[(2)]  {Interior derivative \(i_{\xi}\):} This operator decreases the rank of differential forms by one and is associated with contraction along a vector field \(\xi\).  
\item[(3)] { Lie derivative \(\mathcal{L}_\xi\):} This operator preserves the rank of differential forms and describes the infinitesimal change of a form along the flow generated by a vector field \(\xi\). 
\end{itemize}
These operators satisfy the Cartan magic formula, an identity relating the Lie derivative to the exterior and interior derivatives
\begin{equation}\label{cartan-id-spacetime}
    \mathcal{L}_{\xi} = i_{\xi}\, \dd + \dd\, i_{\xi}\, .
\end{equation}
We note that the above is written for forms over spacetime and that, according to our conventions,
\begin{equation}\label{x-mu-i-d}
    i_\xi x^\mu=0\, , \qquad i_{\xi} \d{}x^{\mu} = \xi^{\mu}\, ,
\end{equation}
where $x^\mu$ denote the spacetime coordinates, and $\xi$ (or $\xi^\mu$) represents a diffeomorphism generator.

\paragraph{Derivative operators on field space.}
Similar derivative operators can be constructed on the field space. 
\begin{itemize}
    \item[(1)] {Exterior derivative \(\delta\),} which 
    increases the rank of differential forms on the field space by one.  
\item[(2)] {Interior derivative \(\mathrm{I}_{\hat{\xi}}\):} Consider the vector $\hat{\xi}[\Phi]$ on the field space,
\begin{equation}
    \hat{\xi}[\Phi]= 
    \hat{\xi}^{I} \fspartial_{I}\, . 
\end{equation}
Then, for a 1-form $\hat\omega[\Phi]$
\begin{equation}
     \mathrm{I}_{\hat{\xi}} \,\hat\omega[\Phi]:= 
     \hat{\xi}^{I} \hat{\omega}_{I}\, ,
\end{equation}
yields a scalar and for a generic $(Q+1)$-form over the field space $\hat X[\Phi]$ with components $\hat X_{IJ_1 J_2\cdots J_Q}[\Phi]$, 
\begin{equation}
     \begin{split}
     \mathrm{I}_{\hat{\xi}} \hat X[\Phi]
     &=\frac{1}{Q!} \hat{\xi}^{I} \hat{X}_{I J_{1} \cdots J_{Q}} \delta \phi^{J_1} \curlywedge \cdots \curlywedge \delta \phi^{J_Q}\, ,
     \end{split}
\end{equation}
yields a $Q$-form. 
\item[(3)]  {Lie derivative \(\mathrm{L}_{\hat{\xi}}\),} which preserves the rank of differential forms and describes the infinitesimal change of a form along the flow generated by a vector field on solution space \(\hat{\xi}\). 
\end{itemize}
These operators satisfy an important and widely used identity known as the Cartan magic formula, which relates the Lie derivative to the exterior and interior derivatives
\begin{equation}\label{Cartan-id-field-space}
    \mathrm{L}_{\hat\xi} = \mathrm{I}_{\hat\xi}\, \delta  + \delta\, \mathrm{I}_{\hat\xi}\, .
\end{equation}
We note that the above is written for forms over the field space. 
\subsection{Diffeomorphisms on spacetime and field space}
{Consider a finite coordinate transformation in the spacetime, $x^{\mu}=x^{\mu}(y)$, which is generated by successive infinitesimal spacetime diffeomorphisms, $x^\mu\to x^\mu+\xi^\mu$.} Under this transformations, a covariant rank $(m,n)$ tensor transforms as follows
\begin{equation}
     \tilde{T}^{\mu_1 \cdots \mu_m}{}_{\nu_1 \cdots \nu_n} = \frac{\partial y^{\mu_1}}{\partial x^{\alpha_1}} \cdots \frac{\partial y^{\mu_m}}{\partial x^{\alpha_m}}\, \frac{\partial x^{\beta_1}}{\partial y^{\nu_1}} \cdots \frac{\partial x^{\beta_n}}{\partial y^{\nu_n}}\, {T}^{\alpha_1 \cdots \alpha_m}{}_{\beta_1 \cdots \beta_n}\, ,
\end{equation}
Similarly, one may consider coordinate transformations in the field space, which are field redefinitions
\begin{equation}
    \phi^{I}=\phi^{I}[\psi^{I},\partial\psi^I,\cdots] \, .
\end{equation}
A covariant rank $(P,Q)$ tensor transforms as follows
\begin{equation}
    \tilde{T}^{I_1 \cdots I_P}{}_{J_1 \cdots J_Q} =  \frac{\fspartial \psi^{I_1}}{\fspartial \phi^{N_1}}\cdots \frac{\fspartial \psi^{I_P}}{\fspartial \phi^{N_P}}\, \frac{\fspartial \phi^{M_1}}{\fspartial \psi^{J_1}} \cdots \frac{\fspartial \phi^{M_Q}}{\fspartial \psi^{J_Q}}\, {T}^{N_1 \cdots N_P}{}_{M_1 \cdots M_Q}\, .
\end{equation}
These field redefinitions will play a central role in the remainder of the paper, where we discuss solution phase space slicings, analyze the integrability of surface charges, and the change-of-slicing technique.
\subsection{Covariance as the fundamental link between spacetime and field space} 
So far, we have treated spacetime and field space as independent manifolds. Now, we establish a fundamental connection between these two structures. We assume our \textit{fundamental} fields \(\phi^{I}\) are covariant in the following sense
\begin{equation}\label{covariance-fund-field}
     {
     \delta_{{\xi}} \phi^{I}:= \mathrm{L}_{\hat{\xi}} \phi^{I}\, . 
     }
\end{equation}  
The left-hand side is defined on the spacetime, while the right-hand side acts on the field space.\footnote{{For instance, when $\xi$ represents a diffeomorphism, $\delta_{\xi} \equiv \mathcal{L}_\xi$, whereas for internal gauge symmetries, taking $\xi \equiv \lambda$, $\delta_\lambda$ 
corresponds to the infinitesimal gauge transformation \cite{Grumiller:2022qhx}. Furthermore, it is important to situate this within the broader context of covariant phase space literature. The difference between the field-space Lie derivative $\mathrm{L}_{\hat{\xi}}$ and the spacetime Lie derivative $\delta_{\xi}$ evaluated on a given quantity characterizes the phase-space anomaly of that quantity \cite{Hopfmuller:2018fni, Chandrasekaran:2020wwn}. In this language, imposing \eqref{covariance-fund-field} is equivalent to demanding that the fundamental fields $\phi^{I}$ are non-anomalous, thereby directly tying spacetime covariance to field-space covariance.}} We take this equation as the defining relation for \(\hat{\xi}\).  In other words, there is an analogy between the diffeomorphisms on the spacetime (denoted by $\xi^\mu$) and diffeomorphisms on the field space (denoted by {$\hat\xi^I$}):
\begin{equation}
\begin{split}
       x^\mu\to x^\mu+\xi^\mu \qquad &\equiv \qquad x^\mu\to x^\mu+{\cal L}_\xi x^\mu\, ,\\ 
       \phi^I\to \phi^I+\delta_{\xi}\phi^I \qquad &\equiv \qquad \phi^I\to \phi^I+\mathrm{L}_{\hat{\xi}}\phi^I\, .
\end{split}
  \end{equation}  
Next, using the Cartan formula in the field space, we obtain  
\begin{equation}
    \mathrm{L}_{\hat{\xi}} \phi^{I} = (\delta\, \mathrm{I}_{\hat{\xi}} + \mathrm{I}_{\hat{\xi}}\, \delta) \phi^{I} = \mathrm{I}_{\hat{\xi}} \, \delta \phi^{I}\, ,
\end{equation}  
where we used the identity $\mathrm{I}_{\hat{\xi}}\phi^I=0$. Since \(\delta \phi^{I}\) forms a basis for the cotangent space of the field space, the last equality can be compared to the similar spacetime relation \eqref{x-mu-i-d}.
From this analogy, and recalling \eqref{covariance-fund-field}, we arrive at the key relation \footnote{{This identification of the field-space vector components via the spacetime transformation of the fields has been utilized in the other references (see, e.g., \cite{Gomes:2016mwl, Freidel:2021dxw}).}}
\begin{equation}\label{fundmental-rel}
    \inbox{ \delta_{{\xi}} \phi^{I}=\mathrm{L}_{\hat{\xi}} \phi^{I}=
    \hat{\xi}^{I} \, .}
\end{equation}  
This establishes \(\hat{\xi}\) as the generator of covariance, bridging the structures of spacetime and field space. In particular,  if $\xi$ is a spacetime diffeomorphism, 
$\hat{\xi}^{I} =\delta_{{\xi}} \phi^{I}={\cal L}_\xi\phi^I$. 

Moreover, we have the following useful identities, 
\begin{equation}
\begin{split}
 \delta \delta_{{\xi}}&=\delta_{\xi}\delta + \delta_{\delta{\xi}},\\       \delta \mathrm{L}_{\hat\xi}=\delta(\delta\, \mathrm{I}_{\hat{\xi}} + \mathrm{I}_{\hat{\xi}}\, \delta)= &\delta\mathrm{I}_{\hat{\xi}}\, \delta=(\delta\, \mathrm{I}_{\hat{\xi}} + \mathrm{I}_{\hat{\xi}}\, \delta)\delta
        =\mathrm{L}_{\hat\xi}\delta\, ,    
\end{split}
\end{equation} 
where in the first equality/identity it is assumed that $\delta \xi:=\frac{\fspartial \xi}{\fspartial \Phi} \delta \Phi$ is a 1-form in the field space and the second holds when acting on forms on the field space (see Appendix \ref{appen:covariance} for a proof).  
We also emphasize that, although $\hat{\xi}$ is a vector on field space, as \eqref{fundmental-rel} makes manifest, its components transform under spacetime in the same representation as $\phi^I$. That is, if $\phi^I$ includes scalars, vectors, tensors, or forms, then $\hat{\xi}^I$ are, respectively, scalars, vectors, tensors, and forms in spacetime.

\subsection{Adjusted Lie bracket on the spacetime and Lie bracket on the field space}
The Lie bracket for two vector fields $\eta^{\mu}$ and $\zeta^{\mu}$ in spacetime is defined as follows
\begin{equation}\label{Lie-bracket-spacetime}
    [\eta,\zeta]=[\eta,\zeta]^{\mu}\partial_{\mu}\, , \qquad  [\eta,\zeta]^{\mu}:={\cal L}_\eta \zeta^{\mu}=- {\cal L}_\zeta \eta^{\mu}= \eta^{\nu}\partial_{\nu}\zeta^{\mu}-\zeta^{\nu}\partial_{\nu}\eta^{\mu}\, .
\end{equation}
If $\eta, \zeta$ are field dependent (i.e., have non-zero derivatives on the field space), one should consider the adjusted Lie bracket \cite{Barnich:2010eb, Barnich:2011mi, Compere:2015knw}:
\begin{equation}\label{adjusted-bracket-spacetime}
    [\eta, \zeta]_{\text{\tiny{adj.}}}:= \llbracket \eta, \zeta\rrbracket :=  [\eta, \zeta]  - \mathrm{L}_{\hat{\eta}} \zeta + \mathrm{L}_{\hat{\zeta}} \eta\, .
\end{equation}
Here, to compute the last terms, we treat $\eta[\Phi]$ and $\zeta[\Phi]$ as scalars on the field space, i.e. 
\begin{equation}
    \mathrm{L}_{\hat{\eta}} \zeta = \fspartial_{I}\zeta \delta_\eta\phi^I=\frac{\fspartial \zeta}{\fspartial \Phi} \delta_\eta\Phi.
\end{equation} 
It can be shown that the adjusted bracket, analogously to the standard Lie bracket, satisfies the Jacobi identity \cite{Barnich:2010eb, Barnich:2011mi, Compere:2015knw}.

Given the notations, conventions, and discussions above, one can define a bracket on the field space that parallels the {Lie} bracket on the spacetime:
\begin{equation}\label{Lie-deriv-FS}
    \hll \hat{\eta}, \hat{\zeta} \hrr 
    =\hll \hat{\eta}, \hat{\zeta} \hrr{}^{I} \fspartial_{I}\, , \qquad  \hll \hat{\eta}, \hat{\zeta} \hrr{}^{I}:=
    \hat{\eta}^{J}\fspartial_{J} \hat{\zeta}^{I}-\hat{\zeta}^{J}\fspartial_{J} \hat{\eta}^{I}= \mathrm{L}_{\hat{\eta}} \hat{\zeta} =- \mathrm{L}_{\hat{\zeta}} \hat{\eta}\, ,
\end{equation}
which exactly parallels the similar equation in spacetime \eqref{Lie-bracket-spacetime}.

\underline{\textbf{Lemma}.} 
The Lie bracket in field space and the adjusted bracket are related by
\begin{equation}\label{Lie-brackets}
    \inbox{\hll \hat{\eta}, \hat{\zeta} \hrr{}^{I}
    =-\delta_{\llbracket \eta, \zeta\rrbracket} \phi^I=-\widehat{\llbracket\eta, \zeta\rrbracket}^I\, .}
\end{equation}
See Appendix \ref{adj-bracket-proof} for the proof. It is evident from the definition above that the field space bracket $\hll\ \cdot,\ \cdot\ \hrr$ satisfies the Jacobi identity, as the adjusted bracket satisfies Jacobi. {As explicitly shown in \eqref{Lie-brackets}, the Lie bracket of vector fields on the field space and the adjusted bracket of their corresponding spacetime generators map to each other with a relative minus sign, a structural feature also emphasized in \cite{Gomes:2016mwl}.}

{We close this part by noting the identities 
\begin{equation}\label{I-L-bracket}
\begin{split}
  i_{[\eta, \zeta]}  = \mathcal{L}_{\eta} i_{\zeta}  - i_{\zeta} \mathcal{L}_{\eta}\, ,\\
    \text{I}_{\hll \hat{\eta}, \hat{\zeta} \hrr} = \text{L}_{\hat{\eta}} \text{I}_{\hat{\zeta}} - \text{I}_{\hat{\zeta}} \text{L}_{\hat{\eta}}\, .
\end{split}\end{equation}}
For later reference, we summarize the corresponding differential‑geometric structures on spacetime and on field space in Table~\ref{tab:spacetime-fieldspace-dictionary} {(A similar comparative summary of these dual geometric structures can be found in \cite{Speziale:2025lkm})}.
\begin{table}[h]
\centering
\renewcommand{\arraystretch}{1.3}

\begin{NiceTabular}{c c c}[
  hlines={2-8},  
  vlines={2,3},  
  rules/color=black, 
  rules/width=1pt
]
\textbf{Manifold} & \textbf{Spacetime} & \textbf{Field space} \\
Symmetry generator & $\xi$ & $\hat{\xi}$ \\
Exterior derivative & $\dd$ & $\delta$ \\
Interior product & $i_\xi$ & $\mathrm{I}_{\hat{\xi}}$ \\
Lie derivative & $\delta_{\xi}\equiv\mathcal{L}_{\xi}$ & $\mathrm{L}_{\hat{\xi}}$ \\
Vector basis & $\partial_\mu \equiv \frac{\partial}{\partial x^\mu}$ & $\fspartial_I \equiv \frac{\fspartial}{\fspartial \phi^I}$ \\
1-form basis & $\dd x^\mu$ & $\delta \phi^I$ \\
(adjusted) Lie brackets & $\llbracket\,\cdot,\cdot\,\rrbracket$ & $\hll\,\cdot,\cdot\,\hrr$ \\
\CodeAfter
  \tikz \draw[line width=1pt, rounded corners=4pt] (1-|1) rectangle (9-|4);
\end{NiceTabular}
\caption{{Dictionary between differential-geometric structures on spacetime and on field space.}}
\label{tab:spacetime-fieldspace-dictionary}
\end{table}

Bianchi identities in spacetime and field space are given by \( \mathrm{d}^2 = 0 \) and \( \delta^2 = 0 \), respectively.
{Note that we use separate spacetime and field-space gradings, so $\d{}x^\mu$ commutes with $\delta\phi^I$ and $\d{}\delta=\delta \d{}$, in contrast with the usual graded convention where $\d{}\delta+\delta \d{}=0$.}

\subsection{Definition of covariance}
As discussed in the introduction, the notion of covariance over spacetime was the guiding principle in the formulation of general relativity and the equivalence principle. Similarly, the same notion over the field space may be attributed to the background independence. Of course, covariance on the field space has not been as extensively studied as the same notion on spacetime. We have established the covariance of the fundamental field $\Phi^{I}$ in \eqref{covariance-fund-field}. In what follows, we generalize it for more general objects. Indeed, if diffeomorphisms are allowed to have field-dependence, one should revisit the notion of covariance. Here, we study the covariance properties of a broad class of natural quantities defined on the field space. 

The most general object we will be dealing with in our analysis takes the form
\begin{equation}\label{def-X}
    X = X(\delta_1\Phi,\delta_2\Phi,\xi_1,\xi_2;\Phi)\, ,
\end{equation}
where $X$ is linear in its first four arguments and may depend nonlinearly on the background fields $\Phi$. We say that $X$ is a covariant quantity if its transformation under a field space vector $\hat{\zeta}$ is given by (see Appendix \ref{appen:covariance} for a proof)
\begin{equation}\label{def-covariance}
    \inbox{\hspace*{2mm}
    \begin{split}
   \mathrm{L}_{\hat{\zeta}}X(\delta_1\Phi,\delta_2\Phi,\xi_1,\xi_2;\Phi)
    & = \mathcal{L}_{\zeta} X(\delta_1\Phi,\delta_2\Phi,\xi_1,\xi_2;\Phi) + \mathrm{I}_{\widehat{\delta \zeta}}\,X(\delta_1\Phi, \delta_2 \Phi,\xi_1,\xi_2;\Phi) \\
    & \hspace{-1 cm} + X(\delta_1\Phi, \delta_2\Phi,\textrm{[}\xi_1,\zeta\textrm{]}+\mathrm{L}_{\hat\zeta}\xi_1, \xi_2 ; \Phi) + X(\delta_1\Phi, \delta_2\Phi, \xi_1, \textrm{[}\xi_2,\zeta\textrm{]}+\mathrm{L}_{\hat\zeta}\xi_2 ; \Phi)\, ,
    \end{split}}
\end{equation}
where $\widehat{\delta\zeta}$ is the vector corresponding to $\delta\zeta$ on the field space, which is defined as follows
\begin{equation}\label{hat-delta-zeta}
    \widehat{\delta\zeta} = \left(\delta\, \delta_\zeta -\delta_\zeta\, \delta\right) \Phi= \delta \hat{\zeta}-\delta_\zeta (\delta \Phi)\,.
\end{equation}
As we see $ \widehat{\delta\zeta}\neq \delta \hat{\zeta}$ and for field independent transformations $\delta \hat{\zeta}=\delta_\zeta (\delta \Phi)$.  

For field independent diffeomorphisms, $\delta \xi =0 = \delta\zeta$, equation \eqref{def-covariance} reduces to
\begin{equation}
    \mathrm{L}_{\hat{\zeta}}X=\mathcal{L}_{\zeta} X(\delta_1\Phi,\delta_2\Phi,\xi_1,\xi_2;\Phi)+ X(\delta_1\Phi, \delta_2\Phi,\textrm{[}\xi_1,\zeta\textrm{]}, \xi_2 ; \Phi) + X(\delta_1\Phi, \delta_2\Phi, \xi_1, \textrm{[}\xi_2,\zeta\textrm{]} ; \Phi)\, .
\end{equation}
See Appendix \ref{appen:covariance} for more detailed discussions. 

Equation \eqref{def-covariance} is one of the main equations of this paper and shows that, under a transformation, the quantity $X$ defined in \eqref{def-X} receives contributions from several distinct sources:
\begin{enumerate}
    \item $X$ transforms as a spacetime tensor. This contribution is captured by the spacetime Lie derivative $\mathcal{L}_{\zeta}$ and reflects the transformation of the spacetime indices carried by $X$.
    \item $X$ depends explicitly on the diffeomorphism generators $\xi$, which themselves transform under the action of $\zeta$. This induces additional contributions governed by the commutator $[\xi,\zeta]$.
    \item Further contributions arise from the field dependence of the generators $\xi[\Phi]$ and $\zeta[\Phi]$, leading to additional variations associated with their implicit dependence on the dynamical fields.
\end{enumerate}
Note also that to calculate the Lie derivative along $ \widehat{\delta\zeta}$, we need to use the generalization of the Lie derivative for vector-valued forms, the Nijenhuis-Lie derivative \cite{
FrolicherNijenhuis1956} (see \emph{e.g.} \cite{KolarMichorSlovak1993} for review), which is defined in the spacetime for any vector‑valued forms $X^\mu_{[\nu_1\nu_2\cdots\nu_k]}$ as follows 
\begin{equation}
    \mathcal{L}_{X} = i_{X}\, \dd +(-1)^{k} \dd\, i_{X}\, ,
\end{equation}
and in the field space for any vector‑valued forms $\hat X^I_{[J_1 J_2\cdots J_k]}$ 
\begin{equation}\label{Cartan-soln-space}
    \mathrm{L}_{\hat X} = \mathrm{I}_{\hat X}\, \delta +(-1)^{k} \delta\, \mathrm{I}_{\hat X}\, .
\end{equation}
\paragraph{Some consistency checks.}
To understand better \eqref{def-covariance}, we perform some consistency/sanity checks for some specific $X$ choices. For $X \equiv \xi^{\mu}$, we have
\begin{equation}
   \mathcal{L}_{{\zeta}}\xi^{\mu} = \textrm{[}\zeta, \xi\textrm{]}^\mu\, ,  \qquad  \mathrm{I}_{\widehat{\delta \zeta}}\, \xi^{\mu} =0\, .
\end{equation}
Therefore,
\begin{equation}
    \mathrm{L}_{\hat{\zeta}} \xi^{\mu} = \textrm{[}\zeta, \xi\textrm{]}^\mu + \textrm{[}\xi,\zeta\textrm{]}^\mu+\mathrm{L}_{\hat\zeta}\xi^\mu = \mathrm{L}_{\hat\zeta}\xi^\mu\, .
\end{equation}
As another check,  consider $X \equiv \textrm{[}\xi_1,\xi_2\textrm{]}^\mu$. In this case, we have
\begin{equation}
    \mathcal{L}_{{\zeta}}\textrm{[}\xi_1,\xi_2\textrm{]}^\mu = \textrm{[}\zeta, \textrm{[}\xi_1,\xi_2\textrm{]} \textrm{]}^\mu\, ,  \qquad  \mathrm{I}_{\widehat{\delta \zeta}}\, \textrm{[}\xi_1,\xi_2\textrm{]}^\mu = 0\, .
\end{equation}
Then,
\begin{equation}
    \begin{split}
         \mathrm{L}_{\hat{\zeta}} \textrm{[}\xi_1,\xi_2\textrm{]}^\mu & = \textrm{[}\zeta, \textrm{[}\xi_1,\xi_2\textrm{]} \textrm{]}^\mu + \textrm{[}\textrm{[}\xi_1,\zeta\textrm{]}+\mathrm{L}_{\hat\zeta}\xi_1, \xi_2\textrm{]}^\mu + \textrm{[}\xi_1,  \textrm{[}\xi_2,\zeta\textrm{]}+\mathrm{L}_{\hat\zeta}\xi_2 \textrm{]}^{\mu} \\
         & = \textrm{[}\mathrm{L}_{\hat\zeta}\xi_1, \xi_2\textrm{]}^\mu + \textrm{[}\xi_1,  \mathrm{L}_{\hat\zeta}\xi_2 \textrm{]}^{\mu} 
         \, .
    \end{split}
\end{equation}
See Appendix \ref{appen:covariance} for more examples and sanity checks.

\section{Solution phase space, symplectic form and surface charges}\label{sec: CPSF-dynamics}
Having established a parallel geometric framework for both the field space and the spacetime manifold, we now introduce three crucial structures defined on the field space: (1) the \emph{solution phase space}, as a subsector in the field space which is a symplectic manifold; (2) the \emph{symplectic form}, an invertible $(0,2)$-form over the solution phase space; and (3) the \emph{surface charge variation}, a $(0,1)$-form over the solution phase space. It is important to note that we will be working with covariant quantities, such as $(p, Q)$-forms, which remain well-defined on the solution phase space. We assume that the solution phase space has a trivial topology. Before proceeding to these concepts, we briefly review the fundamental definitions of the symplectic form and the Poisson bracket. {This section primarily serves as a review of well-established results to provide a self-contained foundation for the subsequent analysis. For the original developments of these concepts, we refer the reader to the foundational works \cite{Lee:1990nz, Wald:1993nt, Iyer:1994ys, Wald:1999wa}, and to \cite{Compere:2018aar, Grumiller:2022qhx} for comprehensive modern reviews.}

\paragraph{Symplectic structure and Poisson bracket.}
As a warm-up, we recall that manifolds may be supplemented with additional structures, e.g., we can have a metric on the manifolds (metric spaces) and/or a symplectic form, symplectic manifolds (phase space). A symplectic manifold, or a phase space in physics terminology, is a manifold equipped with a non-degenerate, invertible, and conserved symplectic 2-form
\begin{equation}
    \Omega[\Phi]
    = \frac{1}{2} 
 \Omega_{IJ}\,\delta\phi^I\curlywedge\delta\phi^J\, ,
\end{equation}
where $\d{}\Omega[\Phi]=0$ and $\delta \Omega[\Phi]=0$, and there exists an antisymmetric bi-vector
\begin{equation}
    \begin{split}
    &\Omega^{-1}[\Phi]
    =\Omega^{IJ}\fspartial_I \circledast\fspartial_J\, ,
    \end{split}
\end{equation}
such that 
\begin{equation}
    \Omega_{IK}\Omega^{KJ}
    =\delta_I^J\, .
\end{equation}

Using the symplectic structure, we can define the Poisson brackets. Let \( F[\Phi] \) and \( G[\Phi] \) be two arbitrary functions. Their Poisson bracket is defined as follows
\begin{equation}\label{poisson-bracket-def}
    \{F,G\}_{\text{\tiny{P.B.}}}
    = \Omega^{{{IJ}}}
    \fspartial_I F \fspartial_J G\,.
\end{equation}
\subsection{Explicit construction of symplectic form over solution phase space}
As mentioned, field space is not (necessarily) a symplectic manifold; however, there exists a sector of it, the solution phase space, that is  equipped with a symplectic form. In the following, we review the systematic method to  construct the solution phase space and compute the symplectic form for a given theory starting from its Lagrangian.

The action for a physical system on a manifold $\mathcal{M}$ is given by
\begin{equation}
    S = \int_{\mathcal{M}} \mathrm{d}^{d}x \, L[\Phi] \,,
\end{equation}
where the Lagrangian $L[\Phi]$ is a scalar density over $\mathcal{M}$. One may either work directly with $L[\Phi]$ or adopt the language of differential forms \cite{Lee:1990nz, Iyer:1994ys} by utilizing its Hodge dual, $\textbf{L}[\Phi]$, which is a top-form over $\mathcal{M}$.

Variation of the Lagrangian $\delta {L}$ upon a generic variation of fields $\delta\Phi$, is given by
\begin{equation}
    \delta {L} = {E}_{\Phi}[\Phi]\, \delta \Phi+\partial_{\mu}\Theta^{\mu}[\delta \Phi ; \Phi]\, ,
\end{equation}
where ${E}_\Phi[\Phi]=0$ gives equations of motion and $ \Theta^{\mu}$ is the Hodg-dual of the pre-symplectic potential $\boldsymbol{\Theta}$ which is a $(d-1,1)$-form. 
We can rewrite the variation of the action as
\begin{equation}\label{variation-action}
            \delta S=\int_{\mathcal{M}} \d{}^{d}x \, {E}_{\Phi}\,\delta \Phi+\int_{\B}\d{}x_{\mu}\, \Theta^{\mu}\, ,
\end{equation}
where $\B$ denotes the boundary of $\mathcal{M}$. Therefore, if we restrict our field configurations to those satisfying field equations,  $\delta S$ reduces to a boundary integral, $\delta S=\int_{\B}\d{}x_{\mu}\, \Theta^{\mu}$. Next, we need to define the cotangent space for the manifold consisting of the set of all solutions to ${E}_\Phi[\Phi]=0$. We do so, by taking $\Phi+\delta \Phi$ to be also satisfying ${E}_\Phi[\Phi+\delta\Phi]=0$. That is, $\delta\Phi$ should satisfy linearized field equations around the point $\Phi$ with $E(\Phi)=0$ in the field space. Explicitly, we define the solution space through,  
\begin{equation}\label{def-on-shell}
    {E}_{\Phi}=0\, , \quad \text{and} \quad \frac{\delta {E}_\Phi}{\delta\Phi}\delta\Phi =0\, .
\end{equation}

According to the CPSF \cite{Lee:1990nz}, the solution space is equipped with a $(0,2)$-form that is defined as the variation of the integral of the total  pre-symplectic potential over a codimension-1 hypersurface
\begin{equation} 
\Omega[\delta \Phi\, , \delta \Phi; \Phi]=\delta\int_{\B}\d{}x_{\mu}\, \Theta^{\mu} \, .
\end{equation}
$\Omega$, while by definition a closed 2-form over the solution space, is not necessarily invertible on the solution space defined by \eqref{def-on-shell}. As discussed in \cite{Lee:1990nz, Wald:1999wa, Grumiller:2022qhx}, in systems with gauge/diffeomorphism symmetries, the equations defining the solution space \eqref{def-on-shell} are gauge covariant and hence the solution space involves gauge-equivalent configurations over which $\Omega$ is not invertible. However, one can make $\Omega$ invertible by modding out the solution space by the gauge-equivalent classes \cite{Lee:1990nz}. In what follows,  we use \textit{solution phase space (SPS)} to denote the subset of solution space over which $\Omega$ has been made invertible. That is, the solution phase space is a symplectic manifold.

In standard Hamiltonian mechanics, the symplectic structure--such as the symplectic form and the Poisson bracket--is defined kinematically, conceptually independent of the specific dynamics or Hamiltonian of the system. It is therefore desirable to formulate a definition of the symplectic potential that is similarly independent of the underlying Lagrangian. Equation \eqref{variation-action} implies that $\mathrm{d} \delta \boldsymbol{\Theta} = 0$ on the solution phase space. Consequently, one is led to the following general definition: A symplectic potential $\boldsymbol{\Theta}$ is a $(d-1,1)$-form satisfying
\begin{equation}\label{Theta-def}
    \inbox{\delta \mathrm{d} \boldsymbol{\Theta} = \mathrm{d}\delta \boldsymbol{\Theta} = 0 \qquad \text{(on the solution phase space)} \, .}
\end{equation}

The most general solution to \eqref{Theta-def} is
\begin{equation}\label{Theta-soln-general}
   \boldsymbol{\Theta} = \boldsymbol{\Theta}_0+ \delta \boldsymbol{W} + \d{} \boldsymbol{Y}\, ,
\end{equation}
where $\boldsymbol{\Theta}_0$ is a specific solution to \eqref{Theta-def} and $\boldsymbol{W}, \boldsymbol{Y}$, respectively a $(d-1,0)$ and $(d-2,1)$ are two integration constants. If we require these freedoms to be covariant, then $\boldsymbol{W}$ and $\boldsymbol{Y}$ freedoms will have the following covariance properties \eqref{def-covariance},
\begin{equation}   
\mathrm{L}_{\hat\xi}\,W^\mu=\mathcal{L}_{\xi}\,W^\mu
\,,\qquad \mathrm{L}_{\hat\xi}\,Y^{\mu\nu}=\mathcal{L}_{\xi}\,Y^{\mu\nu}+\mathrm{I}_{\widehat{\delta\xi}}\,Y^{\mu\nu}\,,
\end{equation}
where $W^\mu, Y^{\mu\nu}$ are spacetime Hodg duals of the corresponding forms $\boldsymbol{W}, \boldsymbol{Y}$, and $\widehat{\delta\xi}$ is a $(1,1)-$form on the field space \eqref{hat-delta-zeta}. $\boldsymbol{\Theta}_0$ may be chosen to be the (canonical) Lee-Wald symplectic potential $\boldsymbol{\Theta}_{\text{\tiny{LW}}}$ \cite{Lee:1990nz} and $\boldsymbol{W}$ may be viewed as a boundary Lagrangian, while $\boldsymbol{Y}$ has no appearance in the original Lagrangian of the theory. 

The $(0,2)$ symplectic form may then be defined as
\begin{equation}
    \Omega[\delta \Phi, \delta \Phi; \Phi] = \delta\int_{\B} \, \boldsymbol{\Theta} = \delta \int_{\B} (\boldsymbol{\Theta}_0+ \delta \boldsymbol{W}+ \d{} \boldsymbol{Y}) \,.
\end{equation}
{A subtle point arises regarding whether the variation operator $\delta$ commutes with the boundary integral. As shown in \cite{Adami:2024gdx} (see also \cite{Ciambelli:2021nmv, Freidel:2021dxw, Carrozza:2022xut, Liu:2026bao}), one cannot simply pass $\delta$ through the integral if the boundary fluctuates.} In other words, {$\mathcal{B}$} may be a codimension-1 surface which depends on the field configurations and its $\delta$-variation may not be zero. In this work, we consider \textit{rigid} boundaries and do not take their fluctuations into account. Therefore,
\begin{equation}\label{Omega-freedom}
    \Omega[\delta \Phi, \delta \Phi; \Phi] = \int_{\B} (\delta \boldsymbol{\Theta}_0 + \d{} \delta \boldsymbol{Y}) \,.
\end{equation}
As we see, although the $W$-freedom contributes to the symplectic potential, it drops out in the symplectic form; only the $Y$-freedom gives a nontrivial contribution to the symplectic form.

\subsection{Physical interpretation of \texorpdfstring{$W$}{W} and \texorpdfstring{$Y$}{Y} freedoms}\label{sec:W-Y-freedoms}
In this subsection, we briefly discuss the physical interpretations of the $W$ and $Y$ freedoms in the CPSF. For extensive discussion, refer \cite{Parvizi:2025shq} and references therein.

The $W$-freedom plays a crucial role in the variational principle, as it generates changes in the boundary conditions imposed on the hypersurface $\mathcal{B}$. Consequently, the choice of $W$ is partially constrained by the requirement of a well-posed action principle \cite{skenderis2002lecture, deHaro:2000vlm, Henningson:1998gx, Mann:2005yr, Papadimitriou:2005ii, Emparan:1999pm}. In the context of holography, this freedom acts as a boundary Lagrangian and corresponds to multitrace deformations of the dual boundary theory \cite{Witten:2001ua, Parvizi:2025shq}.

By contrast, the $Y$-freedom does not affect the bulk action or its boundary conditions. Instead, it modifies the symplectic current/form itself and can be interpreted as the symplectic structure intrinsic to the boundary theory \cite{Parvizi:2025shq}. Specifically, it governs the canonical pairing of sources and responses at the corner (the codimension-2 entangling surface). {As \eqref{Omega-freedom} shows, $\Omega$ involves two kinds of terms: codimension-2 integrals over the boundary of $\mathcal{B}$ (the “corner terms”), and codimension-1 integrals over $\mathcal{B}$ that cannot be reduced to corner terms \cite{Donnelly:2016auv, Speranza:2017gxd, Harlow:2019yfa, Adami:2021kvx, Carrozza:2022xut, Adami:2023wbe, Ball:2024hqe}.} The $Y$-terms are corner terms, while the $\boldsymbol{\Theta}_0$ term may also involve corner terms. In general, one can choose the $Y$-term so that all corner terms are in the $Y$-term, and the $\boldsymbol{\Theta}_0$ term only involves codimension-1 terms. Using this, we can decompose the SPS coordinates (fields) $\phi^I$ into ``boundary modes'' $\varphi^a$ {(with $a=1,\cdots, P$)} that span the codimension-2 part of $\Omega$, and the rest of them, the ``bulk modes'' $\varphi^A$ {(with $A=1,\cdots, D-P$)} that span the codimension-1 terms. Hence, in general $\Omega$ involves three kinds of components, $\Omega_{ab}, \Omega_{aA}, \Omega_{AB}$. The components $\Omega_{ab}$ correspond to corner terms, while $\Omega_{aA}$ and $\Omega_{AB}$ generally arise from codimension‑1 integrals.

We note that there is a redundancy or overlap between the $W$ and $Y$ freedoms: a $W$-term that is exact in spacetime, $\boldsymbol{W}=\mathrm{d} \boldsymbol{Z}$, can be equivalently represented as a shift in the $Y$-freedom via $\boldsymbol{Y}=\delta \boldsymbol{Z}$, where $\boldsymbol{Z}$ is a $(d-2)$-form. Physically, just as $\boldsymbol{W}$ can be viewed as a boundary Lagrangian, $\boldsymbol{Z}$ corresponds to a corner Lagrangian term.

Let us summarize what we discussed in this section. The presymplectic form  $\delta \Omega = \delta^2 \int_{\B} \, \boldsymbol{\Theta}  = 0$ is by definition a closed (0,2)-form in field space. Furthermore,  $\mathrm{d}\boldsymbol{\Theta} \approx 0$ (modulo boundary terms) on the solution space, guaranteeing the conservation of the symplectic form, $\mathrm{d}\Omega \approx 0$. However, to establish $\Omega$ as a proper symplectic form on the physical phase space, it must be non-degenerate (invertible) in addition to being closed. In theories with gauge or diffeomorphism invariance, this may be achieved by modding out the solution space by these gauge directions, yielding \textit{solution phase space}, which is equipped  with a well-defined symplectic 2-form \cite{Lee:1990nz, Wald:1999wa, Grumiller:2022qhx}. 

\subsection{Surface charges} 
Given the symplectic 2-form, one can associate a surface charge variation to a symmetry generator $\xi$. The surface charge variation, which is a 1-form on the SPS, is defined as
\begin{equation}\label{charge-variation-def}
    \slashed{\delta} Q(\xi)
    := - \mathrm{I}_{\hat{\xi}}\, \Omega[\Phi]
    = \int_{\mathcal{B}} \left[ \delta(\mathrm{I}_{\hat{\xi}} \boldsymbol{\Theta})
    - \mathrm{L}_{\hat{\xi}} \boldsymbol{\Theta} \right] \, ,
\end{equation}
where in the second equality we have used Cartan’s identity \eqref{Cartan-id-field-space}. We emphasize that the above expression must be evaluated using the SPS, as defined in the previous subsection.

If $\xi$ belongs to the set of local (gauge) symmetries of the theory, it is said to generate a non-trivial (physical) gauge transformation whenever $\slashed{\delta} Q(\xi)$ is finite and non-vanishing. {These are (measure zero) subsets of gauge transformations/diffeomorphisms which, unlike ``proper gauge transformations/diffeomorphisms'' do not render symplectic form non-invertible; explicitly, they parameterize a part of SPS. We will discuss this point further below.} Although $\Omega$ is defined as an integral over a codimension-1 boundary $\mathcal{B}$, when $\xi$ generates a gauge symmetry the resulting surface charge variation reduces to a local codimension-2 integral, i.e., an integral over a surface $\mathcal{S}$. This feature justifies the terminology \emph{surface charge}, in contrast with generic Noether charges, which are volume integrals over codimension-1 constant-time hypersurfaces. This localization property is a fundamental and distinguishing characteristic of local symmetries.

The charge variation \eqref{charge-variation-def} may be expressed in terms of its components as
\begin{equation}\label{charge-variation}
    \slashed{\delta} Q(\xi)
    = - \mathrm{I}_{\hat{\xi}} \bigl( \Omega_0[\Phi] + \delta Y[\delta \Phi] \bigr) \, ,
    \qquad
    Y[\delta \Phi]
    = \int_{\mathcal{S}} \d{}x_{\mu\nu}\, Y^{\mu\nu}[\delta \Phi]
    := Y_I \, \delta \phi^I \, .
\end{equation}
Equivalently, in component form, one finds
\begin{equation}
    \begin{aligned}
    \slashed{\delta} Q(\xi)
    &= - \Omega^0_{IJ}\, \mathrm{L}_{\hat{\xi}} \phi^I \, \delta \phi^J
       - \mathrm{L}_{\hat{\xi}} Y_I \, \delta \phi^I
       + \mathrm{L}_{\hat{\xi}} \phi^I \, \delta Y_I \\
    &= - \bigl( \Omega^0_{IJ} \, \hat{\xi}^I + \mathrm{L}_{\hat{\xi}} Y_I \bigr) \delta \phi^I
       + \hat{\xi}^I \, \delta Y_I \, .
    \end{aligned}
\end{equation}
The right-hand side of \eqref{charge-variation-def} consists of two terms: the first is an exact 1-form on the SPS, while the second term, $\mathrm{L}_{\hat{\xi}} \boldsymbol{\Theta}$, is not manifestly exact. Consequently, the surface charge variation $\slashed{\delta} Q(\xi)$ is not in general integrable over SPS. Addressing this integrability issue is therefore necessary to define finite charges. We will study integrability conditions in detail in the following sections.

\paragraph{Noether charge.} 
It is instructive to compare the surface charge variation $\slashed{\delta} Q(\xi)$ with the Noether charge $Q_{\text{\tiny N}}(\xi)$:
\begin{itemize}
    \item The Noether charge is, in general, a volume integral defined over a constant-time hypersurface. It is computed directly from the symplectic potential $\boldsymbol{\Theta}$, rather than from the symplectic form $\Omega$. The associated Noether current is
    \begin{equation}
        J^\mu_{\xi}
        = \mathrm{I}_{\hat{\xi}} \, \Theta^{\mu}[\delta \Phi; \Phi]
        - \xi^{\mu} \, L \, ,
    \end{equation}
    where $J^\mu_{\xi}$ is the Noether current associated with the diffeomorphism generated by $\xi$. For local (gauge) symmetries, the Noether current, when evaluated on the SPS, can be expressed as a total divergence,
    \begin{equation}\label{Noether-surface-charge}
        J^\mu_{\xi}
        = \partial_{\nu} Q_{\text{\tiny N}}^{\mu\nu}(\xi) \, ,
        \qquad
        Q_{\text{\tiny N}}(\xi)
        := \int_{\mathcal{S}} \d{}x_{\mu\nu}\, Q_{\text{\tiny N}}^{\mu\nu}(\xi) \, .
    \end{equation}
    Well-known examples include Gauss’s law in electromagnetism and Komar charges in general relativity.

    \item Integrability is not an issue in the Noether charge construction, since one computes the charge itself rather than its variation.

    \item The Noether charge is not necessarily a covariant quantity over the SPS \cite{Adami:2024gdx, Golshani:2024fry}, whereas the charge variation $\slashed{\delta} Q(\xi)$ is a $(0,1)$-form.

    \item Since it is constructed from $\Theta^\mu$, the Noether charge density $Q^{\mu\nu}_{\text{\tiny N}}(\xi)$ generally depends on both the $W$ and $Y$ freedoms. In contrast, the surface charge variation $\slashed{\delta} Q(\xi)$, being defined from $\Omega$, is independent of $W$ but retains a dependence on $Y$.

    \item In general,  $\delta Q_{\text{\tiny N}}(\xi) \neq \slashed{\delta} Q(\xi)$, even when the surface charge is integrable and/or $\xi$ is field independent.
\end{itemize}

\subsection{Barnich--Troessaert bracket} 
As discussed above, the surface charge variation \eqref{charge-variation-def} is not, in general, an exact 1-form on the SPS. It can always be decomposed into an integrable part and a non-integrable (flux) term \cite{Barnich:2010eb, Barnich:2011mi},
\begin{equation}\label{charge-decomposition}
    \slashed{\delta} Q(\xi)
    = \delta Q^{\text{\tiny I}}(\xi) + F(\xi;\delta\Phi)\, ,
\end{equation}
where $Q^{\text{\tiny I}}(\xi)$ denotes the integrable part of the charge, which we assume to be a scalar over spacetime and  SPS.\footnote{\label{fn:Q0-noncov} Although we have assumed $Q^{\text{\tiny I}}(\xi)$ to be covariant (a scalar over SPS), it may  contain a non-covariant contribution, particularly when a background subtraction is performed to renormalize the surface charge, or a reference background charge is chosen \cite{Golshani:2024fry}. In such cases one may write $Q^{\text{\tiny I}}(\xi)=\bar Q^{\text{\tiny I}}(\xi)+Q_0(\xi),$
where $Q_0(\xi)=Q^{\text{\tiny I}}(\xi,\Phi_0)$ denotes the background-dependent (and hence non-covariant) part, the reference point charge, while $\bar Q^{\text{\tiny I}}(\xi)$ is the covariant contribution. The non-covariance of $Q_0$ can be seen from
\begin{equation}\label{Qbar-Q0}
    \mathcal{L}_\xi Q_0(\eta)
=Q_0([\xi,\eta])
+\frac{\fspartial Q^{\text{\tiny I}}(\xi)}{\fspartial \Phi}\bigg|_{\Phi=\Phi_0}\,
\mathcal{L}_\xi \Phi_0\, .
\end{equation}
The last term arises from the explicit dependence on the background configuration $\Phi_0$ and is the source of non-covariance: although $\delta\Phi_0=0$, in general $\mathcal{L}_\xi\Phi_0\neq0$. {See \cite{Golshani:2024fry} for further discussion and its implications for the charge algebra; see also the notion of anomaly for non-covariant quantities on the SPS \cite{Chandrasekaran:2020wwn}.}}

We emphasize three important properties of this decomposition:
\begin{enumerate}
    \item Both $\delta Q^{\text{\tiny I}}(\xi)$ and $F(\xi;\delta\Phi)$ are linear in the symmetry generator $\xi$.
    \item Both are covariant 1-forms on the SPS.
    \item Both are expressed as codimension-2 integrals.
\end{enumerate}
The third feature mentioned above is not usually explicitly stressed in the literature, e.g., in \cite{Barnich:2010eb, Barnich:2011mi}. However, in our analysis, we assume it. See appendix~\ref{appen: CEVD} for an alternative choice which allows $ Q^{\text{\tiny I}}(\xi), F(\xi;\delta\Phi)$ to have codimension-1 parts, while $ \slashed{\delta} Q(\xi)$ is still given by a codimension-2 integral.

The Barnich--Troessaert bracket \cite{Barnich:2010eb, Barnich:2011mi} between the integrable charges in our notions and conventions is written as, 
\begin{equation}\label{MOSPB}
\begin{split}
\left\{ Q^{\text{\tiny I}}(\xi_1), Q^{\text{\tiny I}}(\xi_2) \right\}_{\text{\tiny BT}}
:=\;&
\mathrm{L}_{\hat{\xi}_2} Q^{\text{\tiny I}}(\xi_1)
+ \mathrm{I}_{\hat{\xi}_1} F(\xi_2)
\\
= &
\mathrm{I}_{\hat{\xi}_1} \mathrm{I}_{\hat{\xi}_2} \Omega
- \mathrm{I}_{\hat{\xi}_2} F(\xi_1)
+ \mathrm{I}_{\hat{\xi}_1} F(\xi_2)\, .
\end{split}
\end{equation}
The antisymmetry of the bracket under $\xi_1 \leftrightarrow \xi_2$ is manifest from the second line. As is obvious, the BT-bracket of two charges is covariant (scalar) on SPS.

Consistency of the BT-bracket, that it satisfies the Jacobi identity, requires it to take the general form \cite{Barnich:2010eb, Barnich:2011mi}
\begin{equation}\label{QI-BT-bracket}
\inbox{
\begin{split}
\left\{ Q^{\text{\tiny I}}(\xi_1), Q^{\text{\tiny I}}(\xi_2) \right\}_{\text{\tiny BT}}
&=  Q^{\text{\tiny I}}(\llbracket \xi_1, \xi_2 \rrbracket)
+ K_{\xi_1,\xi_2}\, ,
\\
\left\{ Q^{\text{\tiny I}}(\xi_3), K_{\xi_1,\xi_2} \right\}_{\text{\tiny BT}}
&:= - \mathrm{L}_{\hat{\xi}_3} K_{\xi_1,\xi_2}\, ,
\end{split}
}
\end{equation}
where the antisymmetric {(possibly field-dependent) two-cocycle}
$K_{\xi_1,\xi_2} = - K_{\xi_2,\xi_1}$
satisfies the two-cocycle condition
\begin{equation}\label{2-cocycle-K}
    K_{\llbracket \xi_1, \xi_2 \rrbracket,\xi_3}
    - \mathrm{L}_{\hat{\xi}_3} K_{\xi_1,\xi_2}
    + \text{cyclic permutations}
    = 0\, .
\end{equation}

The decomposition \eqref{charge-decomposition} is not unique
\cite{Barnich:2010eb, Barnich:2011mi, Adami:2020amw, Adami:2024gdx}:
\begin{equation}\label{A-freedom-BT}
    Q^{\text{\tiny I}}(\xi) \rightarrow Q^{\text{\tiny I}}(\xi) + A(\Phi;\xi)\, ,
    \qquad
    F(\xi;\delta\Phi) \rightarrow F(\xi;\delta\Phi) - \delta A(\Phi;\xi)\, ,
\end{equation}
where $A(\Phi;\xi)$ is an arbitrary {scalar} on SPS and spacetime, linear in $\xi$. To preserve the codimension-2 nature of both terms, $A(\Phi;\xi)$ is also required to be expressible as a codimension-2 integral over spacetime. 
{One can readily show that the {two-cocycle} transforms as follows under the $A$ transformation \eqref{A-freedom-BT}}
\begin{equation}
   \inbox{ K_{\xi_1,\xi_2}
    \rightarrow
    K_{\xi_1,\xi_2}
    + \mathrm{L}_{\hat{\xi}_2} A(\Phi;\xi_1)
    - \mathrm{L}_{\hat{\xi}_1} A(\Phi;\xi_2)
    - A\!\left(\Phi;\llbracket \xi_1, \xi_2 \rrbracket\right)\, .}
\end{equation}
{By covariance of $A$, one can readily observe that the induced shift in $K_{\xi_1,\xi_2}$ automatically satisfies the cocycle condition \eqref{2-cocycle-K}.}

\paragraph{Noetherian decomposition.}
A particularly natural choice of the decomposition \eqref{charge-decomposition} is the Noetherian one:
\begin{equation}
    Q^{\text{\tiny I}}(\xi) = Q^{\text{\tiny N}}(\xi)\, ,
    \qquad
    F(\xi;\delta\Phi)
    = \slashed{\delta} Q(\xi) - \delta Q^{\text{\tiny N}}(\xi)\, ,
\end{equation}
{where $Q^{\text{\tiny N}}(\xi)$ denotes  the Noether surface charge \eqref{Noether-surface-charge}.}
This choice satisfies all three requirements stated at the beginning of this subsection: $Q^{\text{\tiny N}}(\xi)$ and $ F(\xi;\delta\Phi)$  are linear in $\xi$, the former is a scalar and the latter a 1-form  on SPS, and both can be expressed as codimension-2 integrals. {One may compute the BT-bracket of the above and obtain (see appendix \ref{appen: charge-algebra} for more details of the computation)
\begin{equation}
 \inbox{ \left\{ Q^{\text{\tiny N}}(\xi_1), Q^{\text{\tiny N}}(\xi_2) \right\}_{\text{\tiny BT}}
=  Q^{\text{\tiny N}}(\llbracket \xi_1, \xi_2 \rrbracket)
+ K_{\xi_1,\xi_2}\, , \qquad  K_{\xi_1,\xi_2} = -2\int_{\cal S}\ \d{}x_{\mu\nu}\ \xi_1^\mu \xi_2^\nu L\, .}
\end{equation}
See also \cite{Freidel:2021cbc, Freidel:2021dxw} for a similar result.}

\paragraph{$W$-freedom vs. $A$-freedom.} 
The $W$-freedom shifts the Noether charge as
$Q^{\text{\tiny N}}(\xi) \rightarrow Q^{\text{\tiny N}}(\xi) + \int_{\mathcal{S}} i_\xi \boldsymbol{W}$,
which corresponds to an $A$-freedom with
$A(\xi) = \int_{\mathcal{S}} i_\xi \boldsymbol{W}$.
Of particular interest is the intersection of the $W$- and $Y$-freedoms, the so-called $Z$-freedom, defined by $\boldsymbol{W} = \d{} \boldsymbol{Z}$, where $\boldsymbol{Z}[\Phi]$ is a $(d-2,0)$-form. In this case,
\begin{equation}\label{A-W-xi}
     A(\xi)
    = \int_{\mathcal{S}} i_\xi \d{} \boldsymbol{Z}
    = \int_{\mathcal{S}} \mathcal{L}_\xi \boldsymbol{Z}
    = \int_{\mathcal{S}} \mathrm{L}_{\hat{\xi}} \boldsymbol{Z}\, ,
\end{equation}
where we have used the covariance of $\boldsymbol{Z}$. The contribution of this $Z$-freedom to the {two-cocycle} is then
\begin{equation}\label{Z-central-term}
\left.
    \mathrm{L}_{\hat{\xi}_2} A(\Phi;\xi_1)
    - \mathrm{L}_{\hat{\xi}_1} A(\Phi;\xi_2)
    -  A\!\left(\Phi;\llbracket \xi_1, \xi_2 \rrbracket\right)
\right|_{ A(\xi) = \int_{\mathcal{S}} \mathrm{L}_{\hat{\xi}} \boldsymbol{Z}}
= 0\, .
\end{equation}
This follows from
$[\mathrm{L}_{\hat{\xi}_2}, \mathrm{L}_{\hat{\xi}_1}]
= \mathrm{L}_{\llbracket \hat{\xi}_2, \hat{\xi}_1 \rrbracket}$
and
$\llbracket \hat{\eta}, \hat{\zeta} \rrbracket
= - \widehat{\llbracket \eta, \zeta \rrbracket}$.
{Remarkably, while a generic $W$ contributes to the {two-cocycle}, the covariant $Z$-freedom does not contribute to the {two-cocycle}.}

We close this section with the remark that in appendix~\ref{appen: CEVD} we construct a codimension-1 decomposition for which the {two-cocycle} vanishes. In appendix~\ref{appen: charge-algebra}, we derive the BT algebra of surface charges for the Noetherian decomposition and determine the associated {two-cocycle}.

\section{Slicing of solution phase space and change-of-slicing}\label{sec: slicing}

As pointed out, SPS is an infinite-dimensional symplectic manifold spanned by solutions to \eqref{def-on-shell} modded out by the orbits of proper gauge transformations. The cotangent space of SPS may be spanned by $\delta\Phi$. Moreover, we are dealing with geometric quantities over the SPS, like $(p,Q)$-forms, that are independent of the choice of coordinates/slicing on the field space. In other words, mathematical or physical observables should be invariant under field redefinitions over SPS. On the other hand, as discussed in section \ref{sec:W-Y-freedoms}, SPS directions may be further decomposed into bulk and boundary modes, respectively spanned by $\varphi^A, \varphi^a$ {(with $a=1,\dots, P$ and $A=1,\dots, D-P$)}.  By the very definition, that bulk and boundary  modes, respectively contribute to the codimension-1 and codimension-2 terms in the symplectic form $\Omega$, the change of coordinates/slicings over the SPS should not mix $\varphi^a$ with $\varphi^A$; i.e. the field redefinitions that appear in changes of slicings rotate $\varphi^a$ among themselves, while $\varphi^A$ can be mixed with $\varphi^A$ and $\varphi^a$. 

{In the SPS analysis, besides the bulk and boundary modes (as the basis) and relevant forms, like $\Omega, \boldsymbol{\Theta}$,  we also deal with symmetry generators $\xi$ and the associated surface charges. On very general grounds and recalling  the construction of SPS, the number of boundary modes $\varphi^a$ should be equal to the number of independent physical gauge transformations, symmetry parameters $\mu^a$ (see Eq.~\eqref{Xi-def} below). Besides the field redefinitions discussed above, we have the freedom to choose {$\mu^a$} to be field dependent or not. So, there are two freedoms/choices: Choice of fields that span the SPS and field-dependence of {$\mu^a$}. The combination of these two is what constitute the \textit{change-of-slicing} over SPS \cite{Adami:2020ugu, Adami:2021nnf, Adami:2022ktn, Taghiloo:2024ewx, Ruzziconi:2020wrb, Geiller:2021vpg, Geiller:2024amx, Ciambelli:2023ott, Ciambelli:2024vhy}.}

We use the change-of-slicing freedoms in two steps: 
\begin{itemize}
    \item Field redefinitions over the bulk modes $\varphi^A\to \tilde{\varphi}^A = \tilde{\varphi}^A (\varphi^a, \varphi^A)$ may be used to bring them to ``homogeneous'' frame for bulk modes, such that, $\delta_\xi\varphi^A$ vanishes when $\varphi^A=0$. Note that this is independent of the choice for field-dependence of {$\mu^a$}.
    \item Field redefinitions over the boundary modes $\varphi^a\to \tilde{\varphi}^a=\tilde{\varphi}^a (\varphi^b)$,  and the choice of field-dependence of {$\mu^a$} may be used to conveniently adjust the choice of $Q^{\text{\tiny I}}(\xi), F(\xi;\delta\Phi)$ in \eqref{charge-decomposition}.
\end{itemize}
In what follows, we assume that we are working in a homogeneous frame for the bulk modes that we parametrize by $\varphi^A$, and hence focus on the \emph{boundary} change-of-slicings. {Despite this point, our analysis, discussions, and results below that are concerning ``boundary SPS'' are independent of this choice, which is essentially concerning the ``bulk SPS'' sector.}

Let us decompose the generic surface charge variation into a basis of symmetry parameters and their conjugate charge aspects:\footnote{{In this paper, we use $a, A$ type indices in two different ways: when they are on $\varphi^a, \varphi^A$ or are coming from these quantities, they respectively account for components of boundary and bulk directions of the SPS. However, when sitting on \textbf{boldface} objects, they are  counting labels (e.g., counting a collection of differential forms or vector fields). Thus, $\boldsymbol{\beta}_a$ represents the $a$-th coordinate-free 1-form, rather than the $a$-th component of a single 1-form $\beta$. $\boldsymbol{\beta}_a$ may be viewed as a basis for 1-forms over the boundary part of the SPS.}\label{footnote-aA}}
\begin{equation}\label{surface-charge-1}
    \slashed{\delta} Q_\xi = \oint_{\mathcal{S}} \d{}^{d-2}x\, \mu^{a} \boldsymbol{\beta}_a\, ,
\end{equation}
where $\mu^a$ ($a=1,\dots,P$) represent $P$ independent symmetry parameters (which are allowed to be field-dependent), and $\boldsymbol{\beta}_a$ are their corresponding SPS 1-form charge aspects. More precisely, the parameters $\mu^{a}$ act as free functions that govern the asymptotic/boundary diffeomorphism $\xi$:
\begin{equation}\label{Xi-def}
     \xi^{\mu} \partial_\mu = \big(\Xi^{\mu}_{a}[\Phi,\partial]\mu^{a}\big) \partial_{\mu}\, ,
\end{equation}
where $\Xi^{\mu}_{a}[\Phi,\partial]$ is an operator (potentially containing derivatives) acting on $\mu^{a}$.  The parameters $\mu^{a}$ serve as arbitrary symmetry generators that dictate directions within the SPS. We can choose their field-dependence as a part of the  \textit{change-of-slicing} procedure outlined above. We will explore concrete examples of this decomposition in Section \ref{sec:examples}.

Mathematically, a change-of-slicing is implemented by redefining the symmetry generators using an invertible, field-dependent transition matrix $\Lambda^{a}{}_{b}$:
\begin{equation}\label{change-slicing-1}
   \tilde{\mu}^{a} = \Lambda^{a}{}_{b}\, \mu^{b}\, , \qquad  \tilde{\boldsymbol{\beta}}_{a} = (\Lambda^{-1})^{b}{}_{a}\, \boldsymbol{\beta}_b\, .
\end{equation}
By construction, the total surface charge variation is a geometric invariant and remains unchanged under this transformation, thus, 
\begin{equation}
    \tilde{\mu}^a \tilde{\boldsymbol{\beta}}_{a} = \big(\Lambda^{a}{}_{b}\mu^{b}\big) \big((\Lambda^{-1})^{c}{}_{a}{\beta}_{c}\big) = \delta^{c}_{b} \mu^{b} \boldsymbol{\beta}_c = \mu^{a} \boldsymbol{\beta}_a \, .
\end{equation}

The field-dependence of the symmetry generators is entirely a matter of choice; we can hence, for instance, define a specific slicing where the new generators are explicitly field-independent, i.e., $\delta \tilde{\mu}^{a}=0$. In this newly chosen frame, the original generators $\mu^{a}$ are forced to be field-dependent, with their variations strictly constrained by the transformation matrix via the condition $\delta(\Lambda^{a}{}_{b}\, \mu^{b})=0$.

As a final remark, we note that performing a change-of-slicing generally modifies the Lie bracket of the symmetry generators and, consequently, alters the algebra of the associated surface charges \cite{Adami:2020ugu, Adami:2021nnf, Adami:2022ktn, Taghiloo:2024ewx, Ruzziconi:2020wrb, Geiller:2021vpg, Geiller:2024amx, Ciambelli:2023ott, Ciambelli:2024vhy}. In the following section, we will explicitly demonstrate the profound mathematical relationship between charge integrability and this choice of phase space slicing.

\section{Charge Integrability and Slicings}\label{sec: integrability}

In this section, we study the integrability of surface charges. As discussed earlier, the infinitesimal charge variation $\slashed{\delta} Q(\xi)$ is not, in general, an exact 1-form on the SPS; that is, it cannot always be written as $\slashed{\delta} Q(\xi)=\delta Q(\xi)$. Using the Frobenius theorem and the geometry of the covariant phase space, we will derive the necessary and sufficient conditions for the integrability of surface charge variations and establish a rigorous framework to distinguish coordinate artifacts from genuine physical radiation.

\subsection{Criteria for Charge Integrability}

\emph{What is the precise criterion for charge integrability?} This question has a classic answer within the covariant phase space formalism \cite{Wald:1999wa}. From a geometric viewpoint, the surface charge variation $\slashed{\delta} Q_{\xi}$ defines a 1-form on the SPS. Integrability--namely, the existence of a state function $Q(\xi)$ such that $\slashed{\delta} Q_{\xi}=\delta Q(\xi)$--is therefore equivalent to the condition that this 1-form be closed:
\begin{equation}\label{WZ-IC}
    \delta \slashed{\delta} Q_{\xi} = 0\, ,
    \qquad
    \text{(Wald--Zoupas integrability criterion)}\, .
\end{equation}
We refer to \eqref{WZ-IC} as the strict Wald--Zoupas (WZ) integrability criterion. {(See footnote \ref{footnote-integrability} for more comments on the naming).}

Applying this criterion to our generalized charge expression \eqref{surface-charge-1}, one finds the integrability conditions:
\begin{equation}\label{integrability-in-slice-1}
    \delta \mu^{a} = 0\, , \qquad
    \delta \boldsymbol{\beta}_a = 0
    \qquad \Longrightarrow \qquad
    \boldsymbol{\beta}_a = \delta \mathrm{Q}_a \, .
\end{equation}
The first condition reflects the usual specific choice of field dependence for the symmetry generators (here taken to be strictly field-independent), while the second condition enforces the integrability of the charge aspect 1-forms.

Even though it is widely used--particularly for exact symmetries and field-independent diffeomorphisms--the strict WZ criterion \eqref{WZ-IC} is not universally satisfactory and may lead to physically counterintuitive  conclusions. Two notable situations illustrate its critical limitations:
\begin{itemize}
    \item \textbf{Theories without bulk flux:}  
    The non-integrability of surface charges is commonly attributed to a flux of propagating bulk degrees of freedom (e.g., gravitational radiation) escaping through the boundary. Nevertheless, in certain two- and three-dimensional gravitational theories that do not have propagating bulk modes (gravitational waves), the surface charges are not integrable by the WZ criterion \cite{Adami:2020ugu, Adami:2021nnf, Adami:2022ktn, Taghiloo:2024ewx, Ruzziconi:2020wrb, Geiller:2021vpg, Geiller:2024amx, Ciambelli:2023ott, Ciambelli:2024vhy}. The WZ integrability condition seems to be unphysically tight. 
    \item \textbf{Dependence on the choice of slicing:}  
    As illustrated by \eqref{integrability-in-slice-1}, the WZ criterion is sensitive to the SPS slicing. There are cases in which \eqref{WZ-IC} indicates non-integrability, yet a change-of-slicing \eqref{change-slicing-1} renders the same surface charge integrable \cite{Adami:2020ugu, Adami:2021nnf, Adami:2022ktn, Taghiloo:2024ewx, Ruzziconi:2020wrb, Geiller:2021vpg, Geiller:2024amx, Ciambelli:2023ott, Ciambelli:2024vhy}. This demonstrates that the standard WZ criterion can be a coordinate/slicing-dependent artifact rather than an invariant physical statement.
\end{itemize}

In the subsequent subsections, we resolve these issues by reformulating charge integrability geometrically. We  first generalize the WZ condition \eqref{WZ-IC} using the Frobenius theorem over the SPS to obtain a slicing-independent integrability criterion. We then construct a complete Cartan geometric framework on the covariant phase space. This framework allows us to rigorously define the geometric {connection, curvature, and torsion} of the solution space and definitively separate ``fake flux'' (a pure slicing artifact) from ``genuine flux'' (true physical radiation).

\subsection{Charge integrability and Frobenius theorems}

Let us begin with a slicing \eqref{integrability-in-slice-1} in which the surface charge is integrable in the strict Wald--Zoupas sense, meaning $\delta \boldsymbol{\beta}_a = 0$. We now perform a field-dependent change-of-slicing on the SPS, parametrized by an invertible matrix $\Lambda^{a}{}_{b}$ \eqref{change-slicing-1}, such that the old charge aspects are related to the new ones as $\boldsymbol{\beta}_b = \Lambda^{c}{}_{b}  \tilde{\boldsymbol{\beta}}_c$, or equivalently $ \tilde{\boldsymbol{\beta}}_a = (\Lambda^{-1})^{b}{}_{a} \boldsymbol{\beta}_b$. 

Using the integrability condition of the original slicing ($\delta \boldsymbol{\beta}_a = 0$), we compute the phase-space exterior derivative of the charge aspect in the new slicing:
\begin{equation}\label{frob-2-condition}
    \delta \tilde{\boldsymbol{\beta}}_{a}
    = \delta\!\left( (\Lambda^{-1})^{b}{}_{a} \right) \curlywedge \boldsymbol{\beta}_b
    = \delta\!\left( (\Lambda^{-1})^{b}{}_{a} \right)
      \curlywedge \left( \Lambda^{c}{}_{b}  \tilde{\boldsymbol{\beta}}_c \right)\,.
\end{equation}
Using the identity $\delta(\Lambda^{-1} \Lambda) = 0$, one can define the matrix-valued phase-space connection 1-form $\mathcal{D}_{a}{}^{c}$, \begin{equation}\label{D-eq}
    \mathcal{D}_{a}{}^{c}
    := \delta\!\left( (\Lambda^{-1})^{b}{}_{a} \right)\Lambda^{c}{}_{b}
    = - (\Lambda^{-1})^{b}{}_{a}\, \delta \Lambda^{c}{}_{b}\, .
\end{equation}
This result has a profound implication. We conclude that an integrable frame exists \emph{if and only if} there exists a matrix-valued 1-form $\mathcal{D}_{a}{}^{b}$ on the SPS such that the charge aspects satisfy the generalized differential equation:
\begin{equation}\label{frob-general-connection}
    \inbox{
    \delta \tilde{\boldsymbol{\beta}}_a = \mathcal{D}_{a}{}^{b} \curlywedge \tilde{\boldsymbol{\beta}}_b\, .
    }
\end{equation}
Equation \eqref{frob-general-connection} provides a fully slicing-independent, covariant criterion for charge integrability. The Wald--Zoupas criterion \eqref{WZ-IC} is a special case where one starts with the  ``correct'' integrable slicing, corresponding to $\Lambda^{a}{}_{b} = \delta^{a}_{b}$ and consequently $\mathcal{D}_{a}{}^{b} = 0$. Whenever a non-zero connection $\mathcal{D}_{a}{}^{b}$ satisfying \eqref{frob-general-connection} can be found, the apparent non-integrability is an artifact of the slicing. The integrating factor--namely, the change-of-slicing matrix $\Lambda^{a}{}_{b}$ that returns the system to the strict WZ frame--can be reconstructed by solving the differential equation \eqref{D-eq}. 

Additionally, the symmetry generator condition $\delta \mu^{a} = 0$ in the original slicing dictates the required field-dependence in the new slicing. Recalling \eqref{change-slicing-1}, ${\tilde{\mu}^{a} = \Lambda^{a}{}_{b}\, \mu^{b}}$, and \eqref{D-eq} we learn that,
\begin{equation}\label{sym-gen-PT}
    \delta \tilde{\mu}^{a}+ \mathcal{D}_{b}{}^{a}\, \tilde{\mu}^{b}=0\, .
\end{equation}
{In an integrable slicing (if it exists), $\mu^a$ will be called chemical potentials associated with the charges $\mathrm{Q}_a$, $\boldsymbol{\beta}_a=\delta \mathrm{Q}_a$. }

Although conceptually powerful, condition \eqref{frob-general-connection} is not always convenient in practical calculations, as it requires explicitly guessing or constructing the connection $\mathcal{D}_{a}{}^{b}$. This  can be bypassed by projecting $\mathcal{D}_{a}{}^{b}$ out of the equation altogether. Taking the wedge product of \eqref{frob-general-connection} with all available $\boldsymbol{\beta}$ 1-forms trivially annihilates the right-hand side, yielding a purely algebraic, connection-free criterion:
\begin{equation}\label{frob-2-2-condition}
    \inbox{
    \delta \boldsymbol{\beta}_a \curlywedge \mathscr{B} = 0\, ,
    \qquad
    \mathscr{B}
    := \boldsymbol{\beta}_{1} \curlywedge \boldsymbol{\beta}_{2} \curlywedge \cdots \curlywedge \boldsymbol{\beta}_{P}\, .
    }
\end{equation}

\paragraph{Relation to the Dual Frobenius Theorem.}
The criteria derived above are not accidental algebraic constructions; rather, they are the direct application of the Frobenius theorem formulated in terms of differential forms--often referred to as the \emph{dual Frobenius theorem} \cite{Wald:1984rg, Lee2013}.

In differential geometry, a set of $P$ linearly independent 1-forms $\{\boldsymbol{\beta}_1,\dots,\boldsymbol{\beta}_P\}$ generates an algebraic ideal $\mathcal{I}$. The Frobenius theorem states that this ideal is a \emph{differential ideal} (i.e., $\d{}\mathcal{I} \subset \mathcal{I}$) if and only if
\[
    \d{}\boldsymbol{\beta}_a \wedge \boldsymbol{\beta}_1 \wedge \cdots \wedge \boldsymbol{\beta}_P = 0\, .
\]
In the covariant phase space formalism, the exterior derivative is the phase-space variation $\delta$. Consequently, condition \eqref{frob-2-2-condition} is the exact mathematical requirement for the charge aspect 1-forms $\boldsymbol{\beta}_a$ to generate a differential ideal on the SPS.

Physically, this guarantees that the SPS admits a smooth foliation by codimension-$P$ submanifolds. On these submanifolds, the appropriately sliced charge aspects become locally exact, rendering the surface charges integrable. The change-of-slicing procedure $\Lambda^{a}{}_{b}$ is the physical realization of the integrating factors whose existence is guaranteed by the Frobenius theorem. 

The essential takeaway is this: when the standard Wald--Zoupas criterion fails ($\delta \boldsymbol{\beta}_a \neq 0$), physical integrability is not necessarily lost. Instead, one must test the algebraic Frobenius condition \eqref{frob-2-2-condition} to determine whether an integrable foliation of the SPS exists, hidden behind a non-optimal choice of slicing.

\subsection{Change-of-Slicing as SPS Parallel Transport}

The appearance of the connection 1-form $\mathcal{D}_{a}{}^{b}$ allows us to recast the integrability conditions into a profoundly geometric language: that of parallel transport over the SPS. 

In differential geometry, a tensor field is parallel transported along a manifold if its covariant derivative vanishes. If we define the phase-space covariant exterior derivative $D_{\mathcal{D}}$ associated with the connection $\mathcal{D}_{a}{}^{b}$, we can rewrite the generalized Frobenius condition \eqref{frob-general-connection} as:
\begin{equation}\label{parallel-transport-beta}
    D_{\mathcal{D}} \tilde{\boldsymbol{\beta}}_a := \delta \tilde{\boldsymbol{\beta}}_a - \mathcal{D}_{a}{}^{b} \curlywedge \tilde{\boldsymbol{\beta}}_b = 0\, .
\end{equation}
This equation implies that for an integrable system, the charge aspect 1-forms $\boldsymbol{\beta}_a$ are \emph{covariantly constant} over the SPS. As we vary the physical fields (moving from one solution to another via the exterior derivative $\delta$), their variation is completely governed by the connection. The charge aspects are effectively parallel transported across the solution space.

This parallel transport mechanism extends to the other fundamental objects defining our slicing. The chemical potentials $\tilde{\mu}^a$, which parameterize the asymptotic symmetry generators, also obey a strict transport law \eqref{sym-gen-PT}. From the integrability requirement, their variation across SPS is dictated by $\delta \tilde{\mu}^{a} = \mathcal{D}_{b}{}^{a} \tilde{\mu}^{b}$, which can be rewritten as:
\begin{equation}\label{parallel-transport-mu}
    D_{\mathcal{D}} \tilde{\mu}^{a} := \delta \tilde{\mu}^{a} + \mathcal{D}_{b}{}^{a} \tilde{\mu}^{b} = 0\, .
\end{equation}
Equation \eqref{parallel-transport-mu} states that the symmetry generators are covariantly constant with respect to $\mathcal{D}_{a}{}^{b}$. As the physical configuration evolves along the phase space, the generators are parallel-transported by the connection to preserve their alignment with the underlying integrable foliation.

Similarly, the transformation matrix $\Lambda^{a}{}_{b}$ that connects different slicings satisfies a transport equation derived directly from the definition of the connection. By rearranging $\mathcal{D} = -\Lambda^{-1} \delta \Lambda$, we find:
\begin{equation}\label{parallel-transport-lambda}
    \delta \Lambda^{a}{}_{b} + \mathcal{D}_{c}{}^{a} \Lambda^{c}{}_{b} = 0\, .
\end{equation}

Collectively,  \eqref{parallel-transport-beta}, \eqref{parallel-transport-mu}, and \eqref{parallel-transport-lambda} depict a cohesive geometric picture of an integrable system. Integrability implies the existence of a highly structured ``charge bundle'' over the SPS, where the fundamental objects--the charge aspects, the symmetry generators, and the slicing frames--are all parallel transported along the solution space using the same connection $\mathcal{D}_{a}{}^{b}$. 

Recognizing this structure immediately invites a fundamental geometric question: Is this parallel transport \emph{path-independent}? If we vary the fields along two different paths in phase space, do the transported slicing frames match? To answer this, we must evaluate the curvature of the connection $\mathcal{D}_{a}{}^{b}$.

\subsection{The Maurer--Cartan equation and fake flux}

To address the question posed above, we investigate the geometric properties of the SPS connection $\mathcal{D}_{a}{}^{b}$ introduced in \eqref{frob-general-connection}. Hereafter, and for the ease of notation,  we drop the tilde on integrable slicing quantities and hence  the integrability condition reads as $\delta \boldsymbol{\beta}_a = \mathcal{D}_{a}{}^{b} \curlywedge \boldsymbol{\beta}_b$. Let us apply the SPS exterior derivative $\delta$ on this. Since $\delta$ is nilpotent ($\delta^2 \equiv 0$), we obtain:
\begin{align}
    0 = \delta^2 \boldsymbol{\beta}_a 
    &= \delta \mathcal{D}_{a}{}^{b} \curlywedge \boldsymbol{\beta}_b - \mathcal{D}_{a}{}^{c} \curlywedge \delta  \boldsymbol{\beta}_c \nonumber\\
    &= \delta \mathcal{D}_{a}{}^{b} \curlywedge \boldsymbol{\beta}_b - \mathcal{D}_{a}{}^{c} \curlywedge \left(\mathcal{D}_{c}{}^{b} \curlywedge \boldsymbol{\beta}_b\right) \nonumber\\
    &= \left( \delta \mathcal{D}_{a}{}^{b} - \mathcal{D}_{a}{}^{c} \curlywedge \mathcal{D}_{c}{}^{b} \right) \curlywedge \boldsymbol{\beta}_b\, . \label{nilpotency-beta}
\end{align}
The term inside the parentheses naturally defines the curvature 2-form $\mathcal{R}_{a}{}^{b}$ associated with the connection $\mathcal{D}_{a}{}^{b}$:
\begin{equation}\label{curvature-D}
    \mathcal{R}_{a}{}^{b} := \delta \mathcal{D}_{a}{}^{b} - \mathcal{D}_{a}{}^{c} \curlywedge \mathcal{D}_{c}{}^{b}\, .
\end{equation}
Equation \eqref{nilpotency-beta} implies that $\mathcal{R}_{a}{}^{b} \curlywedge \boldsymbol{\beta}_b = 0$. {In other words, $\mathcal{R}_{a}{}^{b} \curlywedge \boldsymbol{\beta}_b = 0$ is the condition that integrability condition \eqref{D-eq} is integrable (has solutions)} and the Frobenius theorem provides a stronger statement. Recall from \eqref{D-eq} that the connection was explicitly constructed from a change-of-slicing matrix:
\begin{equation}
    \mathcal{D} = -\Lambda^{-1} \delta \Lambda\, .
\end{equation}
Substituting this explicit form into the definition of the curvature \eqref{curvature-D}, we find:
\begin{align}
    \mathcal{R} &= \delta(-\Lambda^{-1} \delta \Lambda) - (-\Lambda^{-1} \delta \Lambda) \curlywedge (-\Lambda^{-1} \delta \Lambda) \nonumber\\
    &= -(\delta \Lambda^{-1}) \curlywedge \delta \Lambda - \Lambda^{-1} \delta^2 \Lambda - \Lambda^{-1} \delta \Lambda \curlywedge \Lambda^{-1} \delta \Lambda\, .
\end{align}
Using $\delta^2 \Lambda = 0$ and the identity $\delta \Lambda^{-1} = -\Lambda^{-1} \delta \Lambda \Lambda^{-1}$, this evaluates exactly to zero:
\begin{equation}\label{Maurer-Cartan-eq}
    \inbox{
   \mathcal{R}_{a}{}^{b} =  \delta \mathcal{D}_{a}{}^{b} - \mathcal{D}_{a}{}^{c} \curlywedge \mathcal{D}_{c}{}^{b} = 0\, .
    }
\end{equation}
This is the celebrated \emph{Maurer--Cartan structure equation}. It proves that the SPS curvature $\mathcal{R}_{a}{}^{b}$ associated with a Frobenius-integrable system is identically zero globally ($\mathcal{R}_{a}{}^{b} \equiv 0$). 

\paragraph{Physical interpretation: Fake flux.}
The geometric flatness and the Maurer--Cartan equation \eqref{Maurer-Cartan-eq} perfectly encapsulate the physics of systems without genuine bulk radiation. The connection $\mathcal{D}_{a}{}^{b}$ is a \emph{pure-gauge} configuration. It contains no independent dynamical information; it merely tracks how the observer has chosen to parametrize (slice) the SPS. 

Consequently, if a physical system satisfies the algebraic Frobenius condition \eqref{frob-2-2-condition}, any observed variation in the charge aspects ($\delta \boldsymbol{\beta}_a \neq 0$) is devoid of physical bulk degrees of freedom. {In the terminology of BT decomposition \eqref{charge-decomposition}, the Frobenius condition \eqref{frob-2-2-condition} implies that there are slicings in which the flux term $F$ vanishes. We hence call them \emph{fake flux}. Explicitly, the flux term in \eqref{charge-decomposition} is slicing-dependent, and if \eqref{frob-2-2-condition} holds, it means that there exist slicings in which the flux is zero. In a more geometric language, }
if the associated boundary SPS curvature vanishes, the parallel transport defined by $\mathcal{D}_{a}{}^{b}$ is path-independent. This guarantees that there is no geometric obstruction preventing us from performing a finite gauge transformation (a change-of-slicing $\Lambda^{a}{}_{b}$) to reach a globally defined canonical frame where the variations strictly vanish ($\delta \tilde{\boldsymbol{\beta}}_{a} = 0$), rendering the integrability of the system manifest.

This elegantly resolves the first limitation of the standard Wald--Zoupas criterion: in theories like 3D gravity without bulk propagating modes, the charge aspect 1-forms  always satisfy the generalized Frobenius condition \eqref{frob-2-2-condition}, yielding a flat boundary SPS geometry and confirming the absence of genuine physical flux, regardless of the initial slicing choice.

However, a critical question remains: how do we mathematically describe a system where genuine physical radiation \emph{is} escaping through the boundary? In such scenarios, the Frobenius condition must fail, and the boundary SPS geometry is no longer flat. This requires us to move beyond pure-gauge connections and introduce a more general Cartan geometric structure.

\subsection{Genuine flux and the Cartan geometric framework}

We have established that when a system lacks bulk propagating degrees of freedom, the apparent {non-integrability} of surface charges is merely a slicing artifact (fake flux), characterized by the Frobenius condition $\delta \boldsymbol{\beta}_a = \mathcal{D}_{a}{}^{b} \curlywedge \boldsymbol{\beta}_b$ and a globally flat pure-gauge connection ($\mathcal{R}_{a}{}^{b} = 0$). 

However, in cases where, besides the boundary modes, we also have bulk modes--such as those encountered in higher-dimensional Einstein gravity--genuine physical radiation can cross the boundary. In such cases, the algebraic Frobenius condition \eqref{frob-2-2-condition} is not satisfied. It becomes impossible to find a pure-gauge connection $\mathcal{D}_{a}{}^{b}$ that satisfies the integrability equation; i.e., there is no choice of slicing that can globally eliminate the variations in the charge aspects $\delta \boldsymbol{\beta}_a$. Driven by the dynamics of bulk degrees of freedom, the covariant phase space intrinsically resists foliations by integrable submanifolds.

To mathematically capture this physical obstruction, we must abandon the assumption of a pure-gauge connection and elevate the SPS 
geometry to a complete Cartan geometric framework, allowing for non-vanishing curvature. {This is achieved by considering both boundary and bulk parts of the SPS.} 
We introduce a general \emph{charge bundle connection} 1-form, $\mathscr{A}_{a}{}^{b}$, and generalize the integrability equation \eqref{frob-general-connection} to the 
\emph{first Cartan structure equation}: \footnote{{We note that the decomposition in \eqref{cartan-structure-1} is not uniquely fixed \textit{a priori}; it inherently possesses an 
algebraic ambiguity $\mathscr{A}' = \mathscr{A} + \Delta \mathscr{A}$ and $ \mathscr{F}' =  \mathscr{F} - \Delta \mathscr{A} \curlywedge \boldsymbol{\beta}$. 
Physically, $\Delta \mathscr{A} \curlywedge \boldsymbol{\beta}$ represents a shift in how one identifies the purely kinematic gauge artifact (fake flux). However, 
this ambiguity is canonically fixed in our framework by demanding that the torsion $\mathscr{F}$ transforms strictly homogeneously as a genuine tensorial flux under 
phase-space coordinate transformations (changes of slicing), as we will explicitly demonstrate  in \eqref{gauge-trans-T} .}}
\begin{equation}\label{cartan-structure-1}
    \inbox{
    \delta \boldsymbol{\beta}_a = \mathscr{A}_{a}{}^{b} \curlywedge \boldsymbol{\beta}_b + \mathscr{F}_{a}\, .
    }
\end{equation}
Here, the 2-form $\mathscr{F}_{a}$ is the \emph{torsion} of the SPS.\footnote{{This formulation is the rigorous SPS analogue of the standard spacetime Cartan structure equation, $\d{} e = \omega \wedge e + T$.}} Physically, $\mathscr{F}_{a}$ represents the \emph{genuine flux} of the system. It is the precise mathematical obstruction to Frobenius integrability, measuring the exact amount of physical radiation leaking through the boundary that cannot be transformed away by any change-of-slicing over the SPS.

\paragraph{Gauge covariance and slicing independence.}
A robust physical observable must transform covariantly under a change-of-slicing. Let us perform a change-of-slicing parametrized by $\Lambda^{a}{}_{b}$, such that $\boldsymbol{\beta}_a = \Lambda^{c}{}_{a}  \tilde{\boldsymbol{\beta}}_c$. Substituting this into \eqref{cartan-structure-1} and demanding that the structure equation holds its form in the new slicing ($\delta \tilde{\boldsymbol{\beta}}_{a} = \tilde{\mathscr{A}}_{a}{}^{b} \curlywedge \tilde{\boldsymbol{\beta}}_b + \tilde{\mathscr{F}}_{a}$), we find the transformation rules for the geometric objects:
\begin{align}
    \tilde{\mathscr{A}}_{a}{}^{b} &= (\Lambda^{-1})^{b}{}_{c}\, \mathscr{A}_{d}{}^{c}\, \Lambda^{d}{}_{a} - (\Lambda^{-1})^{b}{}_{c}\, \delta \Lambda^{c}{}_{a}\, , \label{gauge-trans-A}\\
    \tilde{\mathscr{F}}_{a} &= (\Lambda^{-1})^{b}{}_{a}\, \mathscr{F}_{b}\, . \label{gauge-trans-T}
\end{align}
Equation \eqref{gauge-trans-A} shows that the connection $\mathscr{A}_{a}{}^{b}$ transforms inhomogeneously, exactly like a gauge field, absorbing the slicing-dependent artifact $-\Lambda^{-1} \delta \Lambda$. In stark contrast,  \eqref{gauge-trans-T} proves that the genuine flux $\mathscr{F}_{a}$ transforms homogeneously as a genuine SPS tensor. 

This provides a definitive resolution to the second limitation of the Wald--Zoupas criterion: while the strict WZ flux ($\delta \boldsymbol{\beta}_a$) is highly slicing-dependent and mixes pure-gauge artifacts with physical degrees of freedom, the Cartan torsion ($\mathscr{F}_a$) successfully isolates the invariant physical radiation. The statement that genuine flux exists ($\mathscr{F}_{a} \neq 0$) is now an absolute, slicing-independent geometric fact.

\paragraph{The Bianchi identity: Flux as the source of curvature.}
We can now finally address the core geometric nature of genuine flux. If pure-gauge (fake) flux corresponds to a flat phase space, how does genuine physical radiation manifest geometrically? 

To answer this, we enforce the nilpotency of the SPS exterior derivative ($\delta^2 \equiv 0$) on the first Cartan structure equation \eqref{cartan-structure-1}:
\begin{align}
    0 = \delta^2 \boldsymbol{\beta}_a 
    &= \delta \mathscr{A}_{a}{}^{b} \curlywedge \boldsymbol{\beta}_b - \mathscr{A}_{a}{}^{c} \curlywedge \delta  \boldsymbol{\beta}_c + \delta \mathscr{F}_{a} \nonumber\\
    &= \delta \mathscr{A}_{a}{}^{b} \curlywedge \boldsymbol{\beta}_b - \mathscr{A}_{a}{}^{c} \curlywedge \left( \mathscr{A}_{c}{}^{b} \curlywedge \boldsymbol{\beta}_b + \mathscr{F}_{c} \right) + \delta \mathscr{F}_{a} \nonumber\\
    &= \left( \delta \mathscr{A}_{a}{}^{b} - \mathscr{A}_{a}{}^{c} \curlywedge \mathscr{A}_{c}{}^{b} \right) \curlywedge \boldsymbol{\beta}_b + \left( \delta \mathscr{F}_{a} - \mathscr{A}_{a}{}^{c} \curlywedge \mathscr{F}_{c} \right)\, .
\end{align}
By defining the  boundary SPS \emph{curvature 2-form} $\mathscr{R}_{a}{}^{b}$ and the \emph{covariant exterior derivative} of the torsion $\text{D}_{\mathscr{A}} \mathscr{F}_{a}$:
\begin{align}
    \mathscr{R}_{a}{}^{b} &:= \delta \mathscr{A}_{a}{}^{b} - \mathscr{A}_{a}{}^{c} \curlywedge \mathscr{A}_{c}{}^{b}\, , \\
    \text{D}_{\mathscr{A}} \mathscr{F}_{a} &:= \delta \mathscr{F}_{a} - \mathscr{A}_{a}{}^{c} \curlywedge \mathscr{F}_{c}\, ,
\end{align}
we arrive at the \emph{SPS Bianchi identity}:
\begin{equation}\label{bianchi-identity}
    \inbox{
    \mathscr{R}_{a}{}^{b} \curlywedge \boldsymbol{\beta}_b + \text{D}_{\mathscr{A}} \mathscr{F}_{a} = 0\, .
    }
\end{equation}
This equation is {one of the central results} of this section and provides a profound physical interpretation. It reveals that the covariant derivative of the genuine flux $\text{D}_{\mathscr{A}} \mathscr{F}_{a}$ acts as the source for the boundary SPS  curvature $\mathscr{R}_{a}{}^{b}$. 

When genuine flux is present ($\mathscr{F}_{a} \neq 0$), the connection $\mathscr{A}_{a}{}^{b}$ is dynamically forced away from the flat, pure-gauge configuration dictated by the Maurer--Cartan equation. The propagating bulk modes actively curve the boundary part of the SPS, causing the boundary curvature tensor $\mathscr{R}_{a}{}^{b}$ to become non-zero. Consequently, the non-integrability of surface charges is not merely an algebraic failure, but a direct manifestation of the underlying curved geometry of the boundary SPS, sourced by the radiative (bulk) SPS.

{Before proceeding, it is crucial to carefully distinguish our charge bundle connection $\mathscr{A}_{a}{}^{b}$ from the 
field-space connections previously introduced in the literature \cite{Gomes:2016mwl, Riello:2019tad}. 
Earlier work construct connections to trace gauge orbits across the entire field space—typically utilizing principal fiber bundle techniques to define 
a BRST-like ghost that removes redundant gauge directions and ensures the covariance of the symplectic potential, however, our construction is distinct in several fundamental ways. 
First, our connection is defined strictly on the Solution Phase Space, specifically restricting to its boundary sector, rather than the whole SPS. Second, our formulation is derived directly from the dual-Frobenius theorem applied to the charge aspect co-frame $\boldsymbol{\beta}_a$, rather than from the symplectic potential. Consequently, the geometric roles differ significantly: in the principal bundle approach of \cite{Gomes:2016mwl, Riello:2019tad}, physical (gauge-invariant) field-space data is encoded in the curvature of the connection; in contrast, our connection $\mathscr{A}_{a}{}^{b}$ acts exclusively as the generator of slicing-dependent ``fake flux,'' while the genuine, physical radiative flux manifests algebraically as the SPS torsion $\mathscr{F}_a$.}

\subsection{Boundary Liouville theorem}\label{subsec:Liouville}

In this subsection, we establish a Liouville-like theorem governing the evolution of the volume form in the boundary SPS, utilizing the Cartan geometric framework developed in the previous sections. 

Let us define the $P$-form
\begin{equation}
\boldsymbol{\mathcal{B}} := \boldsymbol{\beta}_{1} \curlywedge \boldsymbol{\beta}_{2} \curlywedge \cdots \curlywedge \boldsymbol{\beta}_{P} \, ,
\end{equation}
which represents the natural  volume form for the  $P$-dimensional boundary SPS (the sector spanned by the surface charges). To understand how this volume evolves under arbitrary variations in the solution space, we take the exterior derivative of $\boldsymbol{\mathcal{B}}$:
\begin{equation}
\delta \boldsymbol{\mathcal{B}}
=
\sum_{a=1}^{P}
(-1)^{a-1}
\boldsymbol{\beta}_1 \curlywedge \cdots \curlywedge
\boldsymbol{\beta}_{a-1}
\curlywedge
\delta \boldsymbol{\beta}_a
\curlywedge
\boldsymbol{\beta}_{a+1}
\curlywedge \cdots \curlywedge
\boldsymbol{\beta}_P \, .
\end{equation}
We substitute the variation of the charge aspect using the first Cartan structure equation \eqref{cartan-structure-1}, which explicitly separates the purely kinematic frame transformations (the connection $\boldsymbol{\mathscr{A}}$) from the physical non-integrability (the torsion $\boldsymbol{\mathscr{F}}$). This yields two distinct contributions:
\begin{equation}
\begin{aligned}
\delta \boldsymbol{\mathcal{B}}
&=
\sum_{a,b=1}^{P}
(-1)^{a-1}
\boldsymbol{\beta}_1 \curlywedge \cdots \curlywedge
\boldsymbol{\beta}_{a-1}
\curlywedge
\big(\boldsymbol{\mathscr{A}}_{a}{}^{b} \curlywedge \boldsymbol{\beta}_b\big)
\curlywedge
\boldsymbol{\beta}_{a+1}
\curlywedge \cdots \curlywedge
\boldsymbol{\beta}_P \\
&\quad +
\sum_{a=1}^{P}
(-1)^{a-1}
\boldsymbol{\beta}_1 \curlywedge \cdots \curlywedge
\boldsymbol{\beta}_{a-1}
\curlywedge
\boldsymbol{\mathscr{F}}_a
\curlywedge
\boldsymbol{\beta}_{a+1}
\curlywedge \cdots \curlywedge
\boldsymbol{\beta}_P \, .
\end{aligned}
\end{equation}

Let us evaluate the first summation involving the phase-space connection. Since  $\boldsymbol{\mathscr{A}}_{a}{}^{b}$ is a 1-form on the boundary SPS, commuting it to the front of the wedge product past $(a-1)$ 1-forms introduces a sign factor of $(-1)^{a-1}$, which squares to $+1$. The remaining wedge product of $\boldsymbol{\beta}$ forms will vanish identically unless $b=a$, because any $b \neq a$ implies the form $\boldsymbol{\beta}_b$ appears twice. Thus, only the trace of the connection survives:
\begin{equation}
\delta \boldsymbol{\mathcal{B}}
=
\left(\sum_{a=1}^{P} \boldsymbol{\mathscr{A}}_{a}{}^{a}\right)
\curlywedge
\boldsymbol{\mathcal{B}}
+
\sum_{a=1}^{P}
(-1)^{a-1}
\boldsymbol{\beta}_1 \curlywedge \cdots \curlywedge
\boldsymbol{\mathscr{F}}_a
\curlywedge \cdots \curlywedge
\boldsymbol{\beta}_P \, .
\end{equation}
The term $\sum_{a} \boldsymbol{\mathscr{A}}_{a}{}^{a}$ is simply the trace of the fake flux matrix. Using our explicit definition for the boundary SPS connection derived from the generalized Frobenius condition, $\boldsymbol{\mathscr{A}}_{a}{}^{b} = -(\Lambda^{-1})^{c}{}_{a}\,\delta\Lambda^{b}{}_{c}$, the trace evaluates precisely to the variation of the Jacobian determinant of the slicing transformation:
\begin{equation}
\text{Tr}(\boldsymbol{\mathscr{A}})
=
\sum_{a=1}^{P} \boldsymbol{\mathscr{A}}_{a}{}^{a}
=
-(\Lambda^{-1})^{b}{}_{a}\,\delta\Lambda^{a}{}_{b}
=
-\delta\ln\det(\Lambda) \, .
\end{equation}
Substituting this back into the evolution equation gives:
\begin{equation}
\delta \boldsymbol{\mathcal{B}}
=
-\delta\ln\det(\Lambda) \curlywedge \boldsymbol{\mathcal{B}}
+
\sum_{a=1}^{P}
(-1)^{a-1}
\boldsymbol{\beta}_1 \curlywedge \cdots \curlywedge
\boldsymbol{\mathscr{F}}_a
\curlywedge \cdots \curlywedge
\boldsymbol{\beta}_P \, .
\end{equation}
Multiplying both sides by $\det(\Lambda)$ and rearranging terms, we arrive at the modified boundary Liouville equation:
\begin{equation}\label{bdry-Lio-thrm-modified}
\inbox{
\delta\!\left[\det(\Lambda)\,\boldsymbol{\mathcal{B}}\right]
=
\det(\Lambda)
\sum_{a=1}^{P}
(-1)^{a-1}
\boldsymbol{\beta}_1 \curlywedge \cdots \curlywedge
\boldsymbol{\beta}_{a-1}
\curlywedge
\boldsymbol{\mathscr{F}}_a
\curlywedge
\boldsymbol{\beta}_{a+1}
\curlywedge \cdots \curlywedge
\boldsymbol{\beta}_P \, .
}
\end{equation}

This fundamental relation captures the complete geometric evolution of the boundary SPS (surface charge phase-space) volume. We can draw two vital physical conclusions from this equation:
\begin{enumerate}
    \item \textbf{Boundary Liouville Theorem (integrable case):} If the system exhibits no physical radiation ($\boldsymbol{\mathscr{F}}_a = 0$), the right-hand side identically vanishes, yielding $\delta\!\left[\det(\Lambda)\,\boldsymbol{\mathcal{B}}\right]=0$. This states that under arbitrary variations in the SPS, the boundary volume form is exactly conserved up to the Jacobian $\det(\Lambda)$ associated with the frame transformation. In this regime, the non-integrability is purely an artifact of the slicing (fake flux).
    \item \textbf{SPS Boundary Volume Dissipation (non-integrable case):} When genuine flux is present ($\boldsymbol{\mathscr{F}}_a \neq 0$), the precise failure of the boundary SPS volume conservation is driven entirely by the torsion 2-form, the genuine flux. The right-hand side provides an exact measure of how physical radiation dilates or contracts the boundary SPS volume element.
\end{enumerate}

\subsection{Vector field involution and the standard Frobenius theorem}

In the preceding subsections, we explored integrability predominantly from the perspective of the \textit{dual} Frobenius theorem using differential forms. However, this geometric structure can be equivalently understood via the standard Frobenius theorem, which is formulated in terms of vector fields \cite{Wald:1984rg, Lee2013}.

Consider a $D$-dimensional SPS equipped with a set of $P$ linearly independent charge aspect 1-forms $\boldsymbol{\beta}_a$. One can always define a complementary set of $D-P$ linearly independent vector fields $\hat{\boldsymbol{V}}_A$ that are strictly orthogonal to these 1-forms:
\begin{equation}\label{V-beta-ortho}
   \text{I}_{\hat{\boldsymbol{V}}_{A}} \boldsymbol{\beta}_{a} \equiv \boldsymbol{\beta}_{a}(\hat{\boldsymbol{V}}_{A}) = 0 \qquad \forall\ A=1,2,\dots, D-P,\quad a=1,2,\dots, P\, .
\end{equation}
The vector fields $\hat{\boldsymbol{V}}_{A}$ are $D-P$ vectors spanning tangent to the $(D-P)$-dimensional subspace of the SPS where the boundary charge aspects remain strictly constant; {here we are adopting the same notation discussed in footnote \ref{footnote-aA}: $A$ index on boldface objects is counting the number of such objects.} $\hat{\boldsymbol{V}}_{A}$ represent the directions of the purely bulk (radiative) degrees of freedom. Furthermore, we define $P$ SPS vector fields $\hat{\boldsymbol{\xi}}^a$ associated with the boundary symmetry generators, normalized such that they satisfy the duality relation:
\begin{equation}
   \text{I}_{\hat{\boldsymbol{\xi}}_a} \boldsymbol{\beta}_b \equiv \boldsymbol{\beta}_b(\hat{\boldsymbol{\xi}}^a) = \delta_{b}^{a}\, .
\end{equation}
{\textbf{Note on convention.} When the $a,b$ indices are used as ``counting labels'' (counting independent symmetry generators, like those on $\mu^a, \tilde{\boldsymbol{\beta}}_{a}$), the upper and lower indices may be used interchangeably, and upper or lower indices do not have a specific meaning.}

According to the standard Frobenius theorem, the integrability of the foliation generated by $\hat{\boldsymbol{V}}_A$ amounts to verifying that these vector fields are \textit{involutive} under the SPS Lie bracket. That is, the bracket of any two such vectors must close algebraically onto the same subspace:
\begin{equation}\label{V-brackets}
   \hll \hat{\boldsymbol{V}}_{B}, \hat{\boldsymbol{V}}_{C} \hrr = \mathbi{S}^{A}_{BC}\ \hat{\boldsymbol{V}}_{A}\, ,
\end{equation}
for some SPS functions $\mathbi{S}^{A}_{BC}$. To prove that this involution is exactly equivalent to our dual-Frobenius integrability criterion discussed in previous subsections, we evaluate the SPS exterior derivative $\delta\boldsymbol{\beta}_a$ on two bulk vector fields $\hat{\boldsymbol{V}}_B$ and $\hat{\boldsymbol{V}}_C$. Using the standard identity (Cartan's formula for the exterior derivative), we have:
\begin{equation}
    (\delta\boldsymbol{\beta}_{a})(\hat{\boldsymbol{V}}_{B},\hat{\boldsymbol{V}}_{C})
    = \hat{\boldsymbol{V}}_{B}\!\left(\boldsymbol{\beta}_{a}(\hat{\boldsymbol{V}}_{C})\right)
    - \hat{\boldsymbol{V}}_{C}\!\left(\boldsymbol{\beta}_{a}(\hat{\boldsymbol{V}}_{B})\right)
    - \boldsymbol{\beta}_{a}\big( \hll \hat{\boldsymbol{V}}_{B}, \hat{\boldsymbol{V}}_{C} \hrr \big)\, .
\end{equation}
By virtue of the orthogonality condition \eqref{V-beta-ortho}, the first two terms on the right-hand side identically vanish. Rearranging the equation yields a direct relationship between the exterior derivative of the charge aspect and the Lie bracket of the bulk vectors:
\begin{equation}
    \boldsymbol{\beta}_{a}\big( \hll \hat{\boldsymbol{V}}_{B}, \hat{\boldsymbol{V}}_{C} \hrr \big) = - (\delta\boldsymbol{\beta}_{a})(\hat{\boldsymbol{V}}_{B},\hat{\boldsymbol{V}}_{C})\, .
\end{equation}
Now, assume the dual Frobenius criterion holds--namely, the condition for fake flux without genuine flux \eqref{frob-general-connection}: $\delta\boldsymbol{\beta}_a = \boldsymbol{\mathcal{D}}_a{}^b \curlywedge \boldsymbol{\beta}_b$. When we evaluate the right-hand side, we find:
\begin{equation}
    (\boldsymbol{\mathcal{D}}_a{}^b \curlywedge \boldsymbol{\beta}_b)(\hat{\boldsymbol{V}}_{B}, \hat{\boldsymbol{V}}_{C}) = \boldsymbol{\mathcal{D}}_a{}^b(\hat{\boldsymbol{V}}_B)\boldsymbol{\beta}_b(\hat{\boldsymbol{V}}_C) - \boldsymbol{\mathcal{D}}_a{}^b(\hat{\boldsymbol{V}}_C)\boldsymbol{\beta}_b(\hat{\boldsymbol{V}}_B) = 0\, ,
\end{equation}
because every term in the wedge product contains at least one $\boldsymbol{\beta}$ evaluated on a $\hat{\boldsymbol{V}}$, which is identically zero. Consequently, we obtain $\boldsymbol{\beta}_{a}\big( \hll \hat{\boldsymbol{V}}_{B}, \hat{\boldsymbol{V}}_{C} \hrr \big) = 0$. This implies that the Lie bracket of any two vectors $\hat{\boldsymbol{V}}_{B}$ and $\hat{\boldsymbol{V}}_{C}$ is itself a vector strictly orthogonal to all 1-forms $\boldsymbol{\beta}_{a}$. Therefore, it must be expressible as a linear combination of the basis vectors $\hat{\boldsymbol{V}}_{A}$, confirming that the involution condition \eqref{V-brackets} is satisfied.

We conclude this subsection with three pertinent remarks:
\begin{enumerate}
    \item The standard Frobenius theorem guarantees that the existence of $D-P$ linearly independent vector fields $\hat{\boldsymbol{V}}_{A}$, defined through \eqref{V-beta-ortho} and satisfying the involution condition \eqref{V-brackets}, is both necessary and sufficient for integrability. When satisfied, SPS admits a well-defined foliation, and the vector fields $\hat{\boldsymbol{V}}_{A}$ form a complete basis for the tangent space of the $(D-P)$-dimensional bulk part of SPS, over which the boundary charge aspects $\boldsymbol{\beta}_{a}$ are constant.

    \item Together, the symmetry vector fields $\{ \hat{\boldsymbol{\xi}}_{a} \}$ and the bulk vector fields $\{ \hat{\boldsymbol{V}}_{A} \}$ provide exactly $P + (D-P) = D$ linearly independent vectors. Thus, they constitute a complete, well-defined frame for the tangent space of the entire SPS. This logical completeness relies fundamentally on the non-degeneracy of the symplectic form, the orthogonality condition \eqref{V-beta-ortho}, and the definition of the charge aspects $\boldsymbol{\beta}_{a}$ as the symplectic conjugates to the symmetry generators $\hat{\boldsymbol{\xi}}_{a}$. We will discuss this point further in the next subsection.

    \item As discussed in Section \ref{sec:geometric-tools-CPSF}, the underlying spacetime symmetry generators $\boldsymbol{\xi}_{a}$ satisfy a specific algebra:
    \begin{equation}\label{algebra-sym-gen}
        \llbracket \boldsymbol{\xi}_{a}, \boldsymbol{\xi}_{b} \rrbracket = f_{ab}{}^{c}\ \boldsymbol{\xi}_{c}\, ,
    \end{equation}
    where $\llbracket\ , \ \rrbracket$ denotes {the adjusted bracket \eqref{adjusted-bracket-spacetime}},
    and $f_{ab}{}^{c}$ are the structure constants of the symmetry algebra. Note that we are using the conventions mentioned in footnote \ref{footnote-aA} and the index $a$ on  $\boldsymbol{\xi}_{a}$ counts the number of the symmetry generators. When the system is fully integrable (meaning the genuine flux vanishes, $\boldsymbol{\mathscr{F}}_a = 0$), there exists a well-defined, slicing-independent Hamiltonian charge associated with each symmetry generator. These charges satisfy an isomorphic algebra up to possible central extensions \cite{Brown:1986ed, Brown:1986nw}. Furthermore, through the homeomorphism from spacetime to SPS, the algebra \eqref{algebra-sym-gen} uniquely induces a corresponding algebra over the associated SPS vector fields $\hat{\boldsymbol{\xi}}_{a}$, taking the form (cf. \eqref{Lie-brackets}):
    \begin{equation}\label{algebra-sym-gen-vector}
        \hll \hat{\boldsymbol{\xi}}_{a}, \hat{\boldsymbol{\xi}}_{b} \hrr = - \widehat{\llbracket \boldsymbol{\xi}_{a}, \boldsymbol{\xi}_{b} \rrbracket} \, .
    \end{equation}
\end{enumerate}

\subsection{A theorem on symplectic orthogonality}\label{sec:Omega-orthogonality}

In the previous subsections, we established the necessary and sufficient conditions for the existence of an integrable slicing, culminating in the involution of the bulk vector fields $\hat{\boldsymbol{V}}_{A}$ and the induced algebra of the symmetry generators $\hat{\boldsymbol{\xi}}_{a}$.

Let us now assume that there exists a specific slicing of the SPS where the basis vectors $\{\hat{\boldsymbol{\xi}}_a, \hat{\boldsymbol{V}}_A\}$ render the symplectic form purely block-diagonal. That is, the cross-terms exactly vanish:
\begin{equation}\label{Omega-block-diag}
    \boldsymbol{\Omega}_{ab} := \boldsymbol{\Omega}(\hat{\boldsymbol{\xi}}_{a},\hat{\boldsymbol{\xi}}_{b}), \qquad  
    \boldsymbol{\Omega}_{AB} := \boldsymbol{\Omega}(\hat{\boldsymbol{V}}_{A},\hat{\boldsymbol{V}}_{B}), \qquad
    \boldsymbol{\Omega}_{aA} := - \boldsymbol{\Omega}_{Aa} = \boldsymbol{\Omega}(\hat{\boldsymbol{\xi}}_{a},\hat{\boldsymbol{V}}_{A}) = 0\, .
\end{equation}

We demand that this symplectic block-diagonal structure is strictly preserved when we move along the SPS flows generated by a generic symmetry vector $\hat{\boldsymbol{\xi}}$. Using the SPS Lie derivative $\text{L}_{\hat{\boldsymbol{\xi}}}$, this requirement reads:
\begin{equation}\label{block-diag-pres}
    \text{L}_{\hat{\boldsymbol{\xi}}} \boldsymbol{\Omega}_{aA} = 0\, .
\end{equation}
We aim to find the necessary and sufficient geometric conditions for this requirement to hold.

Since the symplectic form is closed ($\delta \boldsymbol{\Omega} = 0$), applying the Leibniz rule to the Lie derivative yields:
\begin{equation}
\begin{split}
   \text{L}_{\hat{\boldsymbol{\xi}}} \boldsymbol{\Omega}_{aA} 
   &= (\text{L}_{\hat{\boldsymbol{\xi}}} \boldsymbol{\Omega})(\hat{\boldsymbol{\xi}}_{a},\hat{\boldsymbol{V}}_{A}) + \boldsymbol{\Omega}(\text{L}_{\hat{\boldsymbol{\xi}}}\hat{\boldsymbol{\xi}}_{a},\hat{\boldsymbol{V}}_{A}) + \boldsymbol{\Omega}(\hat{\boldsymbol{\xi}}_{a}, \text{L}_{\hat{\boldsymbol{\xi}}}\hat{\boldsymbol{V}}_{A})\\
    &= \boldsymbol{\Omega}( \hll \hat{\boldsymbol{\xi}}, \hat{\boldsymbol{\xi}}_{a} \hrr ,\hat{\boldsymbol{V}}_{A} ) + \boldsymbol{\Omega}( \hat{\boldsymbol{\xi}}_{a}, \hll \hat{\boldsymbol{\xi}}, \hat{\boldsymbol{V}}_{A} \hrr )\, .
\end{split}
\end{equation}
If the symmetry algebra \eqref{algebra-sym-gen-vector} holds, the bracket $\hll \hat{\boldsymbol{\xi}}, \hat{\boldsymbol{\xi}}_{a} \hrr$ is a linear combination of the boundary symmetry generators $\hat{\boldsymbol{\xi}}_c$. Due to the block-diagonal assumption \eqref{Omega-block-diag}, the first term vanishes, since $\boldsymbol{\Omega}(\hat{\boldsymbol{\xi}}_c, \hat{\boldsymbol{V}}_A) = 0$. Thus, condition \eqref{block-diag-pres} is satisfied if and only if the second term vanishes. This requires the SPS Lie bracket $\hll \hat{\boldsymbol{\xi}}, \hat{\boldsymbol{V}}_{A} \hrr$ to lie entirely within the span of the bulk vectors $\hat{\boldsymbol{V}}_B$:
\begin{equation}
    \hll \hat{\boldsymbol{\xi}}, \hat{\boldsymbol{V}}_{A} \hrr = \mathcal{F}_{A}{}^{B} \hat{\boldsymbol{V}}_{B}\, .
\end{equation}
The above can be written in more intuitive and informative forms using the SPS geometric analysis. By definition, the bulk vectors $\hat{\boldsymbol{V}}_A$ are strictly orthogonal to the boundary variations, meaning $\text{I}_{\hat{\boldsymbol{V}}_A} \boldsymbol{\beta}_a = 0$. Therefore, the necessary and sufficient condition to preserve the block-diagonal structure is that this bracket has no component along the charge aspect 1-form:
\begin{equation}\label{bracket-projection-zero}
    \text{I}_{\hll \hat{\boldsymbol{\xi}}, \hat{\boldsymbol{V}}_A \hrr} \boldsymbol{\beta}_a = 0\, .
\end{equation}
We can evaluate this condition covariantly using Cartan's magic formula \eqref{Cartan-id-field-space}. The interior product of a Lie bracket acting on a form satisfies the identity \eqref{I-L-bracket}.
Applying it to our vectors gives:
\begin{equation}
    \text{I}_{\hll \hat{\boldsymbol{\xi}}, \hat{\boldsymbol{V}}_A \hrr} \boldsymbol{\beta}_a = \text{L}_{\hat{\boldsymbol{\xi}}} (\text{I}_{\hat{\boldsymbol{V}}_A} \boldsymbol{\beta}_a) - \text{I}_{\hat{\boldsymbol{V}}_A} (\text{L}_{\hat{\boldsymbol{\xi}}} \boldsymbol{\beta}_a)\, .
\end{equation}
Since $\text{I}_{\hat{\boldsymbol{V}}_A} \boldsymbol{\beta}_a = 0$ identically everywhere on the SPS, the first term vanishes. For the second term, we expand the Lie derivative using Cartan's magic formula \eqref{Cartan-id-field-space} and invoke the first Cartan structure equation \eqref{cartan-structure-1}. Defining the scalar charge variation $\overline{\xi_a} := \text{I}_{\hat{\boldsymbol{\xi}}} \boldsymbol{\beta}_a$, we find:
\begin{equation}
\begin{split}
    \text{L}_{\hat{\boldsymbol{\xi}}} \boldsymbol{\beta}_a 
    &= \text{I}_{\hat{\boldsymbol{\xi}}} \big(\boldsymbol{\mathscr{A}}_a{}^b \curlywedge \boldsymbol{\beta}_b + \boldsymbol{\mathscr{F}}_a \big) + \delta \overline{\xi_a} \\
    &= \boldsymbol{\mathscr{A}}_a{}^b(\hat{\boldsymbol{\xi}}) \boldsymbol{\beta}_b -  \boldsymbol{\mathscr{A}}_a{}^b \overline{\xi_b}+ \text{I}_{\hat{\boldsymbol{\xi}}} \boldsymbol{\mathscr{F}}_a + \delta \overline{\xi_a}\, .
\end{split}
\end{equation}
Applying the interior product $\text{I}_{\hat{\boldsymbol{V}}_A}$ to this result, recalling that $\text{I}_{\hat{\boldsymbol{V}}_A} \boldsymbol{\beta}_b = 0$, $\text{I}_{\hat{\boldsymbol{V}}_A} \delta \overline{\xi_a} = \hat{\boldsymbol{V}}_A(\overline{\xi_a})$, and taking care of the alternating signs in the wedge product, we arrive at the fully covariant condition for symplectic orthogonality $\text{L}_{\hat{\boldsymbol{\xi}}} \boldsymbol{\Omega}_{aA}=0$:
\begin{equation}\label{block-diag-covariant}
  \inbox{  \hat{\boldsymbol{V}}_A(\overline{\xi_a}) - \overline{\xi_b} \boldsymbol{\mathscr{A}}_a{}^b(\hat{\boldsymbol{V}}_A) + \boldsymbol{\mathscr{F}}_a(\hat{\boldsymbol{\xi}}, \hat{\boldsymbol{V}}_A) = 0\, . }
\end{equation}
This geometric expression beautifully separates the slicing-dependent and slicing-independent parts. The off-diagonal terms $\text{L}_{\hat{\boldsymbol{\xi}}} \boldsymbol{\Omega}_{aA}$ can receive contributions from three distinct sources: the evolution of the symmetry parameter along the bulk ($\hat{\boldsymbol{V}}_A(\overline{\xi_a})$), the SPS frame rotations (change-of-slicings) which are captured by the fake flux connection $\boldsymbol{\mathscr{A}}$, and the physical radiation dictated by the genuine flux part $\boldsymbol{\mathscr{F}}$.

To connect this covariant result to explicit SPS coordinates, we evaluate \eqref{block-diag-covariant} in a slicing where the fake flux vanishes ($\boldsymbol{\mathscr{A}}_a{}^b = 0$) and the genuine flux is parameterized by coordinates $\{Q_a, \varphi^A\}$. Note that for bulk modes we use $\varphi^A$ that denotes the homogeneous slicing of the bulk sector, cf. Section \ref{sec: slicing}. In this coordinate basis,  
\begin{equation}\begin{split}
    \boldsymbol{\beta}_a &= \delta Q_a + N_{aA} \delta \varphi^A\, ,\\
    \boldsymbol{\mathscr{F}}_a &= \delta N_{aB} \curlywedge \delta \varphi^B\, ,\\
    \hat{\boldsymbol{\xi}} &= \hat{\xi}^b \fspartial_b\, ,\\ 
    \hat{\boldsymbol{V}}_A &= \fspartial_A - N_{cA} \fspartial_c\, .
\end{split}
\end{equation}
Substituting these coordinate definitions into the non-vanishing terms of \eqref{block-diag-covariant} yields:
\begin{equation}
\begin{split}
    \hat{\boldsymbol{V}}_A(\hat{\xi}^a) &= (\fspartial_A - N_{bA} \fspartial_b) \hat{\xi}^a = \fspartial_A \hat{\xi}^a - N_{bA} \fspartial_b \hat{\xi}^a \\
    \boldsymbol{\mathscr{F}}_a(\hat{\boldsymbol{\xi}}, \hat{\boldsymbol{V}}_A) &= (\delta N_{aB} \curlywedge \delta \varphi^B)(\hat{\xi}^c \fspartial_c, \fspartial_A - N_{dA} \fspartial_d) = \hat{\xi}^b \fspartial_b N_{aA}\, .
\end{split}
\end{equation}
Setting their sum to zero, we exactly recover the explicit coordinate constraint:
\begin{equation}\label{block-diag-condition-final}
    \inbox{ \fspartial_A \hat\xi^a+\hat\xi^b \fspartial_b N_{aA} - N_{bA} \fspartial_b \hat\xi^a   = 0\, . }
\end{equation}
Physically, this represents a constraint on how the genuine flux couples to the symmetry generators. In the absence of genuine flux ($N_{aA} = 0$, implying $\boldsymbol{\mathscr{F}}_a = 0$), and assuming the symmetry vector fields $\hat{\boldsymbol{\xi}}$ are independent of the bulk propagating modes ($\fspartial_A \hat{\xi}^a = 0$), the requirement \eqref{block-diag-condition-final} is trivially satisfied, and the block-diagonal structure of the SPS is automatically preserved.

{\paragraph{Non-uniqueness of integrable slicing.} We close this section by pointing out that the integrable slicing is not unique. Suppose that we have found an integrable slicing where the charge variation takes the exact form $\slashed{\delta} Q_{\xi} = \mu^a \delta Q_a$. Consider a change-of-slicing by redefining the SPS coordinates $Q_a \to \tilde{Q}_a(Q)$, such that the physical charge variation remains invariant \cite{Adami:2020ugu}:
\begin{equation}
  \mu^a \delta Q_a = \tilde{\mu}^a \delta \tilde{Q}_a \qquad \Longrightarrow \qquad   \delta \tilde{Q}_a = \frac{\partial \tilde{Q}_a}{\partial Q_b} \delta Q_b\, , \qquad \mu^b = \tilde{\mu}^a \frac{\partial \tilde{Q}_a}{\partial Q_b}\, .
\end{equation}
The above change-of-slicing obviously takes the original integrable slicing to another integrable slicing. Since this relation holds for any arbitrary, invertible function $\tilde{Q}_a(Q)$, the choice of integrable slicing is manifestly non-unique; once the existence of an integrable slicing is established by the Frobenius theorem, the freedom to perform these SPS gauge transformations (coordinate changes) guarantees that there are infinitely many such integrable slicings.}

\section{Examples}\label{sec:examples} 
In this section, we apply the geometric framework developed above to explicit physical systems. {These systems have been extensively studied in the literature, in particular see \cite{Adami:2021nnf, Adami:2022ktn}. In our analysis here, we use those results and rewrite them in the more geometric language (especially in the SPS sense) we developed in previous sections.} We use these examples to analyze the integrability of charge variations, explicitly construct flat phase-space connections, and distinguish genuine radiative flux from pure gauge artifacts (fake flux).

\subsection{Example 1: 2D Einstein--Dilaton Gravity}\label{sec:3dim-dilaton}

As our first test case, we study two-dimensional Einstein--dilaton gravity. This theory is particularly illuminating because it is known to possess no bulk propagating degrees of freedom; therefore, it cannot support genuine physical radiation. We will demonstrate how our geometric criteria successfully identify its apparent non-integrability as a pure coordinate artifact (fake flux). {We adopt the conventions and results obtained in \cite{Adami:2022ktn}.}

The action for Einstein--dilaton gravity is
\begin{equation}\label{2D-action}
    S = \frac{1}{16\pi G} \int \d{}^2 x \sqrt{-g}  \left[ \Phi \, R - X(\Phi) \right]\, ,
\end{equation}
where $\Phi$ is the dilaton field and $X(\Phi)$ is its potential. The corresponding equations of motion are
\begin{subequations}\label{EOM}
\begin{align}
    \nabla_{\mu}\nabla_{\nu}\Phi - \frac{1}{2}g_{\mu\nu}\Box \Phi &= 0\, , \label{eom-1}\\
    \Box \Phi + X(\Phi) &= 0\, ,\label{eom-2}\\
    R - \frac{\d X}{\d \Phi} &= 0\, . \label{eom-3}
\end{align}
\end{subequations}
The general solution to these equations is given by \cite{Adami:2022ktn}
\begin{equation}\label{2d-solution-GF}
    \d s^2 = -V(v,r)  \d v^2 + 2\eta(v) \d v \d r \,, \qquad  \Phi (v,r)= \Omega(v) + \eta(v) \, \lambda(v) \, r  \, ,
\end{equation}
with the metric function
\begin{equation}
    V = -\frac{1}{\lambda^2} \left( 2\lambda \partial_v \Phi + \mathcal{X} + m \right)\, , \qquad \text{where} \quad X= \frac{\d{} \mathcal{X}}{\d \Phi}\, .
\end{equation}
The SPS is completely labeled by three arbitrary functions of the advanced time, $\Omega(v)$, $\eta(v)$, and $\lambda(v)$, along with the constant parameter $m$.

\subsubsection{Setup and the Frobenius condition}

The vector fields preserving the form of the metric \eqref{2d-solution-GF} are parameterized by three free functions:
\begin{equation}\label{2d-CBS}
    \xi = T(v)\partial_{v} + \left[ Z(v) - \frac{r}{2} W(v) \right]\partial_{r} \, ,
\end{equation}
where $T(v)$, $Z(v)$, and $W(v)$ generate $v$-supertranslations, $r$-supertranslations, and $r$-superscalings, respectively. The surface charge variation associated with these symmetries is \cite{Adami:2022ktn}:
\begin{equation}\label{Charge-Variation-001'}
    \slashed{\delta} Q_\xi
    = \frac{1}{16\pi G} \left[ W\delta\Omega + 2Z\, \delta\bigl(\Omega \,e^{\Pi/2}\bigr) + T\left(\frac{1}{\lambda}\delta m - \partial_{v}\Pi\,\delta\Omega + \partial_{v}\Omega\,\delta\Pi \right) \right]\, ,\qquad \Pi := \ln{\bigl(\frac{\eta \lambda}{\Omega}\bigr)^2\,.}
\end{equation}
Note that we have made a field-redefinition and used $\Pi$ instead of $\eta$.

Let us first examine the integrability of this charge variation using the  Wald--Zoupas criterion \eqref{WZ-IC}. Assuming the symmetry generators are field-independent ($\delta W = \delta Z = \delta T = 0$), we have
\begin{equation}
    16\pi G\,\delta \slashed{\delta} Q_\xi = T \left[ \delta(\lambda^{-1}) \curlywedge \delta m - \delta(\partial_v \Pi) \curlywedge \delta \Omega + \delta(\partial_v \Omega) \curlywedge \delta \Pi \right] \neq 0\, .
\end{equation}
The strict WZ criterion fails. This presents a physical paradox: the charge is non-integrable, implying a flux of energy or information, yet 2D Einstein--dilaton gravity has no bulk degrees of freedom to source such a flux.

To resolve this, we apply our generalized integrability  criterion based on the SPS Frobenius theorem. Comparing \eqref{Charge-Variation-001'} with the generic expansion $16\pi G\,\slashed{\delta} Q_\xi = \mu^a \boldsymbol{\beta}_a$, we identify the symmetry basis $\mu^a = \{W, Z, T\}$ and the corresponding 1-form charge aspects $\boldsymbol{\beta}_a$:
\begin{equation}
    \boldsymbol{\beta}_W = \delta \Omega\, , \qquad 
    \boldsymbol{\beta}_Z = 2 \delta\bigl(\Omega \,e^{\Pi/2}\bigr)\, , \qquad 
    \boldsymbol{\beta}_T = \frac{1}{\lambda}\delta m - \partial_{v}\Pi\,\delta\Omega + \partial_{v}\Omega\,\delta\Pi\, .
\end{equation}
We construct the top-form $\boldsymbol{\mathcal{B}}$ spanning the charge space:
\begin{equation}
    \begin{split}
       \boldsymbol{\mathcal{B}} &= \boldsymbol{\beta}_W \curlywedge \boldsymbol{\beta}_Z \curlywedge \boldsymbol{\beta}_T = 2\, \delta \Omega \curlywedge \delta\bigl(\Omega \,e^{\Pi/2}\bigr) \curlywedge \left(\frac{1}{\lambda}\delta m - \partial_{v}\Pi\,\delta\Omega + \partial_{v}\Omega\,\delta\Pi \right) \\
        &= \lambda^{-1} \Omega\, e^{\Pi/2}\, \delta \Omega \curlywedge \delta \Pi \curlywedge \delta m\, .
    \end{split}
\end{equation}
Next, we evaluate the SPS exterior derivatives (curvatures) of the individual components:
\begin{equation}
    \delta \boldsymbol{\beta}_W = 0\, , \qquad \delta \boldsymbol{\beta}_Z = 0\, , \qquad \delta \boldsymbol{\beta}_T = \delta \left(\frac{1}{\lambda}\delta m - \partial_{v}\Pi\,\delta\Omega + \partial_{v}\Omega\,\delta\Pi \right)\, .
\end{equation}
The Frobenius criterion requires that $\delta \boldsymbol{\beta}_a \curlywedge \boldsymbol{\mathcal{B}} = 0$ for all $a$. The only non-trivial check is for $a=T$:
\begin{equation}\label{Frob-crit-2D}
    \delta \boldsymbol{\beta}_T \curlywedge \boldsymbol{\mathcal{B}} 
    = \left( \lambda^{-1}\Omega\, e^{\Pi/2} \right) \delta \left(\frac{1}{\lambda}\delta m - \partial_{v}\Pi\,\delta\Omega + \partial_{v}\Omega\,\delta\Pi \right) \curlywedge \delta \Omega \curlywedge \delta \Pi \curlywedge \delta m = 0\, .
\end{equation}
This vanishes identically because all 1-forms comprising $\delta \boldsymbol{\beta}_T$ are already present in the measure $\boldsymbol{\mathcal{B}}$. 
Therefore, despite the failure of the WZ condition, the charge variation \emph{is} integrable in the generalized Frobenius sense. This geometrically guarantees the existence of  a foliation of the SPS and that the apparent obstruction is pure fake flux.

\subsubsection{Cartan structure and fake flux}

The Frobenius condition is satisfied; thus, there must exist a change of basis to an integrable slicing where the genuine flux definitively vanishes. We construct the frame matrix $\Lambda^{a}{}_{b}$ that transforms the physical charge aspects $\boldsymbol{\beta}_a$ into an integrable basis $\tilde{\boldsymbol{\beta}}_a$ via $\boldsymbol{\beta}_a = \Lambda^{b}{}_{a} \tilde{\boldsymbol{\beta}}_b$, while the symmetry generators transform inversely as $\tilde{\mu}^a = \Lambda^{a}{}_{b} \mu^b$. 
    
We define the transformed symmetry generators $\tilde{\mu}^a = \{\tilde{W}, \tilde{Z}, \tilde{T}\}$ as:
\begin{equation}
    \tilde{W} = W - T\,\partial_{v}\Pi + 2Z\,e^{\Pi/2}  \, , \qquad 
    \tilde{Z} = Z\,\Omega\, e^{\Pi/2} + T\,\partial_{v}\Omega \, ,  \qquad 
    \tilde{T} = \frac{T}{\lambda} \, .
\end{equation}
This corresponds to the transformation matrix $\Lambda$:
\begin{equation}
  \Lambda^{a}{}_{b} =  \begin{pmatrix}
1 &  2 e^{\Pi/2} & -\partial_v \Pi \\
0 & \Omega e^{\Pi/2} & \partial_{v}\Omega \\
0 & 0 & \lambda^{-1}
\end{pmatrix}\, .
\end{equation}
In this new frame, the charge variation simplifies to the trivially integrable form $16\pi G\, \slashed{\delta} {{Q}}_\xi= \tilde{W} \delta\Omega + \tilde{Z} \delta\Pi + \tilde{T} \delta m$. Thus, the transformed basis 1-forms are simply:
\begin{equation}
    \tilde{\boldsymbol{\beta}}_W = \delta\Omega \, , \qquad 
    \tilde{\boldsymbol{\beta}}_Z = \delta\Pi \, , \qquad 
    \tilde{\boldsymbol{\beta}}_T = \delta m \, .
\end{equation}
Being exact 1-forms on the SPS, their exterior derivatives  identically vanish: $\delta \tilde{\boldsymbol{\beta}}_a = 0$. Using the First Cartan structure equation in the tilde frame:
\begin{equation}
    \delta \tilde{\boldsymbol{\beta}}_a = \tilde{\boldsymbol{\mathscr{A}}}_a{}^b \curlywedge \tilde{\boldsymbol{\beta}}_b + \tilde{\boldsymbol{\mathscr{F}}}_a = 0\, .
\end{equation}
This implies we are in a completely flat frame with $\tilde{\boldsymbol{\mathscr{A}}}_a{}^b = 0$ and $\tilde{\boldsymbol{\mathscr{F}}}_a = 0$. 
    
To isolate the nature of the non-integrability in the original slicing, we apply the inverse transformation. Mapping back to the $\boldsymbol{\beta}_a$ basis yields the structure equation:
\begin{equation}
    \delta \boldsymbol{\beta}_{a} = \boldsymbol{\mathscr{A}}_{a}{}^{b} \curlywedge \boldsymbol{\beta}_{b} + \boldsymbol{\mathscr{F}}_{a} \, .
\end{equation}
$\boldsymbol{\mathscr{F}}$ transforms as a vector-valued 2-form and hence its vanishing  in the tilde frame mathematically proves its vanishing in the physical frame as well: $\boldsymbol{\mathscr{F}}_a = 0$. There is strictly no genuine flux in the system. 
    
The entire WZ non-integrability is thus generated by the connection $\boldsymbol{\mathscr{A}}_a{}^b$, which takes the pure-gauge form:
\begin{equation}
    \boldsymbol{\mathscr{A}}_a{}^b = (\Lambda^{-1})^b{}_c \delta \Lambda^c{}_a \, .
\end{equation}
Explicitly, this fake flux connection evaluates to:
\begin{equation}
    \boldsymbol{\mathscr{A}} =  \begin{pmatrix}
0 &  2 e^{\Pi/2} \frac{\delta\Omega}{\Omega} & \delta (\partial_v \Pi) + 2\frac{\delta (\partial_v\Omega)}{\Omega} + \frac{\delta \lambda}{\lambda}\left( \partial_v\Pi + 2\frac{\partial_v\Omega}{\Omega} \right) \\
0 & -\frac{1}{2}\left( \delta \Pi + \frac{\delta \Omega}{\Omega} \right) & -e^{-\Pi/2} \frac{\delta(\lambda \partial_v\Omega)}{\lambda\Omega} \\
0 & 0 & -\frac{\delta \lambda}{\lambda}
\end{pmatrix}\, .
\end{equation}
Since $\boldsymbol{\mathscr{A}}$ is a pure-gauge connection derived entirely from a change-of-slicing, its SPS curvature identically vanishes ($\boldsymbol{\mathscr{R}} = \delta \boldsymbol{\mathscr{A}} + \boldsymbol{\mathscr{A}} \curlywedge \boldsymbol{\mathscr{A}} = 0$). The apparent non-integrability $\delta \boldsymbol{\beta}_T \neq 0$ is solely given by the term $\boldsymbol{\mathscr{A}}_T{}^b \curlywedge \boldsymbol{\beta}_b$. This rigorously confirms that 2D Einstein--dilaton gravity is completely devoid of physical radiation; the WZ obstruction is entirely an SPS gauge artifact (fake flux).

\subsection{Example 2: 3D Einstein gravity}\label{sec:3dim-Einstein}
As the second example, we consider Einstein gravity in three spacetime dimensions:
\begin{equation}\label{3D-action}
    S= \frac{1}{16\pi G} \int \d{}^3 x \sqrt{-g}  \left( R - 2\Lambda \right).
\end{equation}
Unlike higher-dimensional gravity, 3D gravity lacks propagating bulk degrees of freedom. Therefore, we expect the SPS to admit a foliation into integrable submanifolds, meaning any observed non-integrability is expected to be due to the choice of slicing (fake flux) rather than genuine physical radiation.

We consider a solution space with the following line element \cite{Adami:2022ktn}:
\begin{equation}\label{G-N-T-metric''}
    \d s^2=  -V \d v^2 + 2 \eta \d v \d r + {\cal R}^2 \left( \d \phi + U \d v \right)^2\, ,
\end{equation}
with
\begin{subequations}\label{metric-components''}
    \begin{align}
    {\cal R} &= \Omega + \lambda \, \eta \, r\, , \label{calR-3d}\\
    U &= {\cal U} +  \frac{1}{\lambda \, {\cal R}}\, \frac{\partial_\phi \eta}{ \eta }  + \frac{\Upsilon}{2 \lambda {\cal R}^2}\, , \\
    V &= \frac{1}{\lambda^2}\left( - \Lambda \mathcal{R}^2 -\mathcal{M}  + \frac{\Upsilon^2}{4  {\cal R}^2} - \frac{2\mathcal{R}}{\eta }   \mathcal{D}_v ( \eta \lambda ) +\frac{\Upsilon}{\mathcal{R}}\, \frac{\partial_\phi \eta}{ \eta } \right)\, ,
    \end{align}
\end{subequations}
where $\{\Omega, \lambda, \eta, \Upsilon, \mathcal{U}, \mathcal{M}\}$ are functions of $(v, \phi)$. The metric \eqref{G-N-T-metric''} with \eqref{metric-components''} solves the Einstein field equations if:
\begin{subequations}\label{M-Upsilon-EoM''}
\begin{align}
    &\mathcal{E}_{\hat{\mathcal{M}}}:=\mathcal{D}_{v}\hat{\mathcal{M}}+\Lambda\lambda\partial_{\phi}\left(\frac{\hat{\Upsilon}}{\lambda^2}\right)+2\partial_{\phi}^3\mathcal{U}=0\, , \label{EOM-M}\\
    &\mathcal{E}_{\hat{\Upsilon}}:=\mathcal{D}_{v}\hat{\Upsilon}-\lambda\partial_{\phi}\left(\frac{\hat{\mathcal{M}}}{\lambda^2}\right)+2\partial_{\phi}^3(\lambda^{-1})=0\, , \label{EOM-Upsilon}
\end{align}
\end{subequations}
where the hatted fields are defined as:
\begin{equation}\label{hatM-hatUsp''}
    \begin{split}
       &  \hat{\Upsilon}={\Upsilon+\Omega\partial_{\phi}\Pi}\, , \qquad {\Pi :=\ln \left( \frac{\eta \lambda}{\Omega} \right)^2}\, ,\\
       & {\hat{\mathcal{M}}= \mathcal{M} + {\lambda\Omega\mathcal{D}_{v}\Pi}+ \left(\frac{\partial_{\phi}\eta}{\eta}\right)^2+3\left(\frac{\partial_{\phi}\lambda}{\lambda}\right)^2-2\frac{\partial_{\phi}^2\lambda}{\lambda}}\, .
    \end{split}
\end{equation}
Here, the differential operator $\mathcal{D}_v$ acting on a codimension-1 function $O_w (v,\phi)$ of weight $w$ is defined via $\mathcal{D}_v O_w = \partial_v O_w - \mathcal{L}_\mathcal{U} O_w$, with $\mathcal{L}_{\mathcal{U}} O_w = \mathcal{U} \partial_\phi O_w + w O_w \partial_\phi \mathcal{U}$. The weights for various functions are listed in Table \ref{Table-1}.
\begin{table}[h]
\centering
\begin{tabular}{ |l|l| }
  \hline
  $w= -1$  & $\mathcal{U}$ , $Y$\\
  $w= 0$  & $\eta$ , $T$ , $W$ , $Z$ , $\Pi$ , $\partial_{v}$\\
  $w= 1$  & $\Omega$ , $\lambda$ , $\partial_{\phi}$ \\
  $w= 2$  & $\hat{\mathcal{M}}$ , $\hat{\Upsilon}$ \\
  \hline
\end{tabular}
\caption{Weight $w$ for various quantities defined and used in this section, cf. \cite{Adami:2022ktn}.}\label{Table-1}
\end{table}

\subsubsection{Setup and the Frobenius condition}
The vector fields preserving the line element \eqref{G-N-T-metric''} are generated by \cite{Adami:2022ktn}:
\begin{equation}\label{null-bondary-sym-gen''}
\begin{split}
    \xi &=T\partial_{v}+\left[Z  - \frac{r }{2}\, W -\frac{\Upsilon}{2\eta\lambda^2\mathcal{R}} \, \partial_{\phi}T - \frac{1}{\eta^2\lambda }\partial_{\phi}\left(\frac{\eta\partial_{\phi}T}{\lambda}\right)\right]\partial_{r}+\left(Y+\frac{\partial_{\phi}T}{\lambda\mathcal{R}}\right)\partial_{\phi}\, .
\end{split}
\end{equation}
Evaluating this vector field on the boundary yields the symmetry parameters $\Xi_{a}^{\mu}$ \eqref{Xi-def}:
\begin{equation}
    \begin{split}
       & \Xi_{T}^{\mu} = \delta^{\mu}_{(v)} + \delta^{\mu}_{(r)} \left[ -\frac{\Upsilon}{2\eta\lambda^2\mathcal{R}} \, \partial_{\phi}  - \frac{1}{\eta^2\lambda }\partial_{\phi}\left(\eta \lambda^{-1} \partial_\phi\right) \right] +  \delta^{\mu}_{(\phi)} \left[ \frac{\partial_{\phi}}{\lambda\mathcal{R}}\right]\, , \\
       & \Xi_{Z}^{\mu} = \delta^{\mu}_{(r)}\, , \qquad \Xi_{W}^{\mu} =- \frac{r}{2} \delta^{\mu}_{(r)}\, , \qquad \Xi_{Y}^{\mu} = \delta^{\mu}_{(\phi)}\, . 
    \end{split}
\end{equation}

The associated surface charge variation evaluates to:
\begin{equation}\label{surface-charge-Thermo''}
    \hspace{-0.25 cm}   {\slashed{\delta} Q_\xi = {\frac{1}{16\pi G}}\oint_{\mathcal{S}} \d \phi \left[ 
            W \delta\Omega + 2Z \delta(\Omega \, e^{\Pi/2}) +Y\delta \Upsilon+ T \left( -\mathcal{D}_{v}\Pi \, \delta\Omega+{\mathcal{U}} \delta\Upsilon+\mathcal{D}_{v}\Omega\ \delta\Pi+\lambda^{-1}\delta{\hat{\mathcal{M}}}\right)
            \right].}
\end{equation}
By matching this to our general basis ${16\pi G} \slashed{\delta} Q_\xi = \oint \mu^a \boldsymbol{\beta}_a$, we identify the symmetry generators:
\begin{equation}
    \mu^{1} = W\, ,\qquad \mu^2 = Z\, , \qquad \mu^3 = Y\, , \qquad \mu^4 = T\, ,
\end{equation}
and their corresponding charge aspects:
\begin{equation}
    \boldsymbol{\beta}_{1} = \delta \Omega\, , \qquad \boldsymbol{\beta}_2 = 2 \delta(\Omega \, e^{\Pi/2}) \, , \qquad \boldsymbol{\beta}_3 = \delta \Upsilon\, , \qquad \boldsymbol{\beta}_4 = -\mathcal{D}_{v}\Pi \, \delta\Omega+{\mathcal{U}} \delta\Upsilon+\mathcal{D}_{v}\Omega\ \delta\Pi+\lambda^{-1}\delta{\hat{\mathcal{M}}}\, .
\end{equation}

If we apply the WZ integrability criterion $\delta\slashed{\delta}Q_{\xi} = 0$ \eqref{WZ-IC}, we find:
\begin{equation}
    \delta \slashed{\delta} Q_\xi = T \left( -\delta(\mathcal{D}_{v}\Pi) \curlywedge \delta\Omega + \delta {\mathcal{U}} \curlywedge \delta\Upsilon + \delta(\mathcal{D}_{v}\Omega) \curlywedge \delta\Pi + \delta(\lambda^{-1}) \curlywedge \delta{\hat{\mathcal{M}}}\right) \neq 0,
\end{equation}
where we assumed field-independent parameters. Naively, this non-vanishing result would signal non-integrability. However, we must test this against the generalized Frobenius condition \eqref{frob-2-2-condition}. We construct $\mathcal{B}$:
\begin{equation}
    \begin{split}
       \mathcal{B} &= \boldsymbol{\beta}_1 \curlywedge \boldsymbol{\beta}_2 \curlywedge \boldsymbol{\beta}_3 \curlywedge \boldsymbol{\beta}_4 \\
       &= 2\delta \Omega \curlywedge \delta(\Omega \, e^{\Pi/2}) \curlywedge \delta \Upsilon \curlywedge \left( -\mathcal{D}_{v}\Pi \, \delta\Omega+{\mathcal{U}} \delta\Upsilon+\mathcal{D}_{v}\Omega\ \delta\Pi+\lambda^{-1}\delta{\hat{\mathcal{M}}}\right)\\
        &= \lambda^{-1}\, \Omega\, e^{\Pi/2} \delta \Omega \curlywedge \delta \Pi  \curlywedge \delta \Upsilon \curlywedge \delta{\hat{\mathcal{M}}}\,.
    \end{split}
\end{equation}
Taking the exterior derivative of $\boldsymbol{\beta}_4$, the only non-trivial Frobenius condition yields:
\begin{equation}\label{integrability-con-3D}
    \delta \boldsymbol{\beta}_4 \curlywedge \mathcal{B} = \lambda^{-1}\, \Omega\, e^{\Pi/2} \delta \left( -\mathcal{D}_{v}\Pi \, \delta\Omega+{\mathcal{U}} \delta\Upsilon+\mathcal{D}_{v}\Omega\ \delta\Pi+\lambda^{-1}\delta{\hat{\mathcal{M}}}\right) \curlywedge  \delta \Omega \curlywedge \delta \Pi  \curlywedge \delta \Upsilon \curlywedge \delta{\hat{\mathcal{M}}}=0.
\end{equation}
This condition holds identically; thus, the SPS \textit{can} be foliated into integrable submanifolds. As expected from the absence of bulk propagating degrees of freedom, the apparent non-integrability is an artifact of the chosen slicing.

\subsubsection{Cartan structure and fake flux} 
To study the geometric nature of the slicing freedom manifest, we analyze the system using the Cartan structure equations \eqref{cartan-structure-1}:
\begin{equation}
    \delta \boldsymbol{\beta}_{a} = \boldsymbol{\mathscr{A}}_{a}{}^{b} \curlywedge \boldsymbol{\beta}_{b} + \boldsymbol{\mathscr{F}}_{a}\, .
\end{equation}
As the Frobenius condition is satisfied, we expect the genuine flux  $\boldsymbol{\mathscr{F}}_{a}$ to be identically zero. To prove this and explicitly identify the fake flux generator (the connection $\boldsymbol{\mathscr{A}}_{a}{}^{b}$), we construct the global frame transformation to an integrable slicing \cite{Adami:2022ktn}. 

Consider the transformed basis elements:
\begin{equation}\label{hat-slicing''}
   \tilde{W} =  W -  T \mathcal{D}_v \Pi {+2}e^{\Pi/2} Z\, , \qquad \tilde{Z} = T \, \mathcal{D}_v \Omega + \Omega e^{\Pi/2} Z \, , \qquad   \tilde{Y}= Y+\mathcal{U} T \, , \qquad \tilde{T}= \frac{T}{\lambda} \, .
\end{equation}
This corresponds to a change of frame $\tilde{\boldsymbol{\mu}}^a = (\Lambda^{-1})^a{}_b \boldsymbol{\mu}^b$, mapping to new charge aspects $\tilde{\boldsymbol{\beta}}_a = \Lambda^a{}_b \boldsymbol{\beta}_b$. The transformation matrix $\Lambda$ and its inverse are:
\begin{equation}
  \Lambda^{a}{}_{b} =  \begin{pmatrix}
1 &  2 e^{\Pi/2} & 0 & -\mathcal{D}_v \Pi \\
0 & \Omega e^{\Pi/2} & 0 & \mathcal{D}_{v}\Omega \\
0 & 0 & 1 & \mathcal{U} \\
0 & 0 & 0 & \lambda^{-1}
\end{pmatrix}\, , \qquad
  (\Lambda^{-1})^{a}{}_{b} =  \begin{pmatrix}
1 &  -2\Omega^{-1} & 0 & \lambda(\mathcal{D}_{v}\Pi + 2\Omega^{-1}\mathcal{D}_v \Omega) \\
0 & \Omega^{-1} e^{-\Pi/2} & 0 & -\lambda e^{-\Pi/2}\Omega^{-1}\mathcal{D}_{v}\Omega \\
0 & 0 & 1 & -\lambda \mathcal{U} \\
0 & 0 & 0 & \lambda
\end{pmatrix}\, .
\end{equation}
In the tilde frame, the charge aspects reduce to exact variations of state space functions:
\begin{equation}
    \tilde{\boldsymbol{\beta}}_1 = \delta\Omega\, , \qquad \tilde{\boldsymbol{\beta}}_2 =  \delta\Pi\, , \qquad \tilde{\boldsymbol{\beta}}_3 =\delta {\Upsilon}\, , \qquad \tilde{\boldsymbol{\beta}}_4 =\delta\hat{\mathcal{M}}\, .
\end{equation}
Consequently, their exterior derivatives trivially vanish ($\delta \tilde{\boldsymbol{\beta}}_a = 0$), implying that in the tilde frame,  $\tilde{\boldsymbol{\mathscr{A}}} = 0$ and $\tilde{\boldsymbol{\mathscr{F}}} = 0$.

By applying the gauge transformation rules for the charge bundle, we map these objects back to the original physical frame $\boldsymbol{\beta}_a$. The torsion transforms purely tensorially, so $\boldsymbol{\mathscr{F}}_a = (\Lambda^{-1})^b{}_a \tilde{\boldsymbol{\mathscr{F}}}_b = 0$. As expected, 3D gravity contains absolutely zero genuine flux.

The entire non-integrability is governed by the connection $\boldsymbol{\mathscr{A}}_{a}{}^{b}$, which acts as a pure-gauge configuration defined by $\boldsymbol{\mathscr{A}} = - \Lambda^{-1} \delta \Lambda$. Computing this explicitly yields the fake flux generator matrix:
\begin{equation}
\boldsymbol{\mathscr{A}}^{a}{}_{b} =  \begin{pmatrix}
0 &  2 e^{\Pi/2} \frac{\delta\Omega}{\Omega} & 0 & \delta \mathcal{D}_v \Pi + \frac{2\delta \mathcal{D}_v\Omega}{\Omega} +\frac{\delta \lambda}{\lambda}\left( \mathcal{D}_v\Pi + \frac{2\mathcal{D}_v\Omega}{\Omega} \right) \\
0 & -\frac{1}{2}\left( \delta \Pi + \frac{\delta \Omega}{\Omega} \right) & 0 & -e^{-\Pi/2} \frac{\delta(\lambda \mathcal{D}_v\Omega)}{\lambda\Omega} \\
0 & 0 & 0 & -\left( \delta \mathcal{U} + \mathcal{U}\frac{\delta \lambda}{\lambda} \right) \\
0 & 0 & 0 & \frac{\delta \lambda}{\lambda}
\end{pmatrix}\, .
\end{equation}
This connection precisely isolates the slicing ambiguity for every symmetry generator simultaneously. Being derived from a frame transformation, its Maurer-Cartan form (the SPS curvature) is identically zero:
\begin{equation}
    \boldsymbol{\mathscr{R}} = \delta\boldsymbol{\mathscr{A}} - \boldsymbol{\mathscr{A}} \curlywedge \boldsymbol{\mathscr{A}} \equiv 0\, .
\end{equation}
We conclude that the surface charges in 3D Einstein gravity are inherently integrable; the apparent WZ flux is merely a "fake flux" corresponding to parallel transport across a flat charge bundle.

\subsection{Example 3: Einstein gravity in higher dimensions (\texorpdfstring{$d>3$}{}) with a null boundary}\label{sec:higher-dim-Einstein}

As a final example, we explore Einstein gravity in higher dimensions ($d>3$) in a spacetime with a null boundary \cite{Adami:2021nnf}. Unlike the previous two examples, this theory contains propagating bulk degrees of freedom (gravitational waves). Consequently, alongside any slicing-dependent fake flux, we expect the phase space geometry to inherently capture a non-vanishing genuine flux associated with physical radiation crossing the boundary.

The line element in the Gaussian null-type coordinate system is given by
\begin{equation}\label{G-F-M-01}
    \d s^2 =  -V \d v^2 + 2 \eta \d v \d r + g_{{\tA \tB}} \left( \d x^{\tA} + U^\tA \d v\right) \left( \d x^\tB + U^\tB \d v\right)\, , \qquad \tA=1, \cdots, d-2\, .
\end{equation}
The presence of bulk radiative modes bars us from solving the equations of motion in an exact, general closed form. Instead, we perform a near-boundary expansion around the null hypersurface $r=0$ (the null boundary) \cite{Adami:2021nnf}:
\begin{equation}\label{nearN-expansion}
        V = 2\big(\eta\kappa - \mathcal{D}_v \eta \big)  r + \mathcal{O}(r^2)\,, \quad
        U^\tA = {\cal U}^\tA - \frac{\eta}{\Omega}\Upsilon^\tA r + \mathcal{O}(r^2)\,, \quad
        g_{\tA \tB} = \Omega_{\tA\tB} - 2\eta\lambda_{\tA \tB}\, r + \mathcal{O}(r^2)\, , 
\end{equation}
where all expansion coefficients $\{\eta, \kappa, \Omega, \mathcal{U}^{\tA}, \Upsilon^{\tA}, \Omega_{\tA \tB}, \lambda_{\tA \tB}\}$ are functions of $(v, x^\tA)$ governed by the relations:
\begin{equation}\label{OmegaAB-gammaAB}
    \Omega := \sqrt{\det \Omega_{\tA \tB}}\, ,\qquad \Omega_{\tA \tB} = \Omega^{2/(d-2)}\gamma_{\tA \tB}\, ,\qquad \det \gamma_{\tA \tB} = 1\,,
\end{equation}
and we have defined the derivative $\mathcal{D}_v := \partial_v - \mathcal{L}_{\mathcal{U}}$, with $\mathcal{L}_{\mathcal{U}}$ denoting the Lie derivative along $\mathcal{U}^\tA$. {It is important to note a distinction in notation: the upright capital indices $\tA, \tB$ and the object $\Omega_{\tA\tB}$ appearing in this section are should not be mistaken with the italic index $A$ used to label the bulk modes $\varphi^A$ and the components of the SPS symplectic form discussed in Section \ref{sec:Omega-orthogonality}.}

For later convenience, we introduce the shear tensors:
\begin{equation}\label{N-AB-L-AB-def}    
    N_{\tA\tB} = \frac{1}{2}{\cal D}_v \Omega_{\tA\tB} - \frac{\Theta_l}{d-2} \Omega_{\tA\tB} = \frac{1}{2}\Omega^{\frac{2}{d-2}} \mathcal{D}_v \gamma_{\tA\tB}\, , \qquad L_{\tA\tB} = \lambda_{\tA\tB} - \frac{\Theta_n}{d-2} \Omega_{\tA\tB}\, ,
\end{equation}
with the corresponding expansion scalars:
\begin{equation}\label{Theta-l-Theta-n-def}
    \Theta_l = \frac{{\cal D}_v \Omega}{\Omega} = \frac{\partial_v \Omega}{\Omega} - \bar{\nabla}_\tA \mathcal{U}^{\tA}\, , \qquad\quad \Theta_n = \Omega^{\tA\tB}\lambda_{\tA\tB}\, ,
\end{equation}
where $\bar{\nabla}_\tA$ denotes the covariant derivative compatible with $\Omega_{\tA\tB}$. Note that the shear tensor $N_{\tA\tB}$ directly captures the radiative bulk degrees of freedom (the news).

The equations of motion projected onto the $r=0$ surface dictate the system's evolution and constraints \cite{Adami:2021nnf}:
\begin{subequations}\label{EoM-r0}
\begin{align}
        &{\cal D}_v  \Theta_l - \kappa \Theta_l + \frac{1}{d-2} \Theta_l^2 + N_{\tA\tB}N^{\tA\tB} = 0\, ,\label{EoM-Raychaudhuri}\\
        &{\cal D}_v \Big(\Upsilon_{\tA} + \Omega \frac{\partial_\tA\eta}{\eta}\Big) - 2\Omega\partial_{\tA}\Big( \kappa + \frac{d-3}{d-2}\Theta_l \Big) + 2 \Omega\bar{\nabla}^{\tB}N_{\tA \tB} = 0\, , \label{EoM-Damour}\\
        &{\cal D}_v  \Theta_{n} + \kappa \Theta_{n} + \Theta_l\Theta_{n} - \big(\bar{\nabla}_{\tC}\mathcal{H}^{\tC} + \mathcal{H}^{\tC}\mathcal{H}_{\tC}\big) + \frac{1}{2} \bar{R} - \Lambda = 0\, ,\label{Thetan-EoM}\\
        &2\mathcal{D}_{v}L_{AB} - 4{L_{(\tA}}^{\tC}{N_{\tB)\tC}} + \Theta_{n}N_{\tA\tB} + \Big(2\kappa + \frac{d-6}{d-2}\Theta_l\Big)L_{\tA\tB} + \bar{R}_{\tA\tB} \nonumber\\
        &\quad - 2\mathcal{H}_{\tA}\mathcal{H}_{\tB} - 2\bar{\nabla}_{(\tA}\mathcal{H}_{\tB)} + \Big(2\bar{\nabla}_{\tC}\mathcal{H}^{\tC} + 2\mathcal{H}^{\tC}\mathcal{H}_{\tC} - \bar{R}\Big)\frac{\Omega_{\tA\tB}}{d-2} = 0\, , \label{trace-less-AB}
\end{align}
\end{subequations}
where ${\cal H}_{\tA} := \frac{\Upsilon_\tA}{2\Omega} + \frac{\partial_{\tA}\eta}{2\eta}$. 

\subsubsection{Setup and the Frobenius condition}
The diffeomorphisms that preserve the chosen gauge and keep the null boundary at $r=0$ and form of the line element, i.e. the symmetry generators, are given by the vector fields  \cite{Adami:2021nnf}:
\begin{align}\label{NBS-vector}
    \xi &= T\,\partial_v + \Big[ r(\mathcal{D}_{v}T - W) - r^2 \frac{\eta}{2} \Big(\frac{\Upsilon_\tA}{\Omega} - \frac{\partial_\tA \eta}{\eta}\Big) \partial^{\tA}T + \mathcal{O}(r^3)\Big]\,\partial_r \nonumber \\ 
        &\quad + \Big[Y^\tA - r \eta \partial^\tA T - r^2{\eta^2} \lambda^{\tA\tB}\partial_{\tB}T + \mathcal{O}(r^3)\Big]\,\partial_\tA\, .
\end{align}
Here, $T=T(v,x^\tA)$, $W=W(v,x^\tA)$, and $Y^\tA=Y^\tA(v,x^\tA)$ act as arbitrary symmetry parameters. Evaluated near the $r=0$ boundary, the symmetry vector components $\Xi_{a}^{\mu}$ \eqref{Xi-def} are:
\begin{equation}
     \begin{split}
         \Xi^{\mu}_{T} &= \delta^{\mu}_{(v)} + \delta^{\mu}_{(r)} \left[ r \mathcal{D}_{v} - r^2 \frac{\eta}{2} \Big(\frac{\Upsilon_\tA}{\Omega} - \frac{\partial_\tA \eta}{\eta}\Big) \partial^{\tA} \right] + \delta^{\mu}_{(\tA)} \left[ -r \eta \partial^\tA - r^2{\eta^2} \lambda^{\tA\tB}\partial_{\tB} \right] + \mathcal{O}(r^3) \, , \\
         \Xi^{\mu}_{W} &= -r \delta^{\mu}_{r}\, , \qquad \Xi^{\mu}_{Y^{\tA}} = \delta^{\mu}_{(\tA)}\, . 
     \end{split}
\end{equation}

The surface charge variation associated with these symmetries evaluates to:
\begin{equation}\label{surface-charge-01}
        \slashed{\delta} Q_{\xi} = \frac{1}{16\pi G} \int_{\mathcal{S}} \d{}^{d-2} x \left(  W\delta\Omega + Y^{\tA}\delta\Upsilon_{\tA} + T \slashed{\delta} \mathcal{A}\right)\, , 
\end{equation}
with the phase space 1-form defined by
\begin{equation}
    \begin{split}
    \slashed{\delta} \mathcal{A} &= \Omega \Theta_l \delta \mathcal{P} - \Gamma \delta\Omega + \mathcal{U}^{\tA}\delta \Upsilon_{\tA} - \Omega N^{\tA\tB} \delta\Omega_{\tA\tB}\, ,
    \end{split}
\end{equation}
where we have introduced the field combinations:
\begin{equation}
   \mathcal{P} := \ln \frac{\eta}{\Theta_l^2}\, , \qquad  \Gamma := -2\kappa + \frac{2}{d-2} \Theta{_l} + \frac{\mathcal{D}_v \eta}{\eta}\, .
\end{equation}
{That is, we have used field redefinition freedom to introduce $\mathcal{P}, \Gamma$ instead of $\eta$, $V$ (or $\kappa$)  respectively.}
From the explicit form of the surface charge variation \eqref{surface-charge-01}, we identify the symmetry basis $\mu^a$ and the associated charge aspects $\boldsymbol{\beta}_a$:
\begin{equation}
     \mu^1 = Y^{1} \, , \quad \cdots \quad \mu^{d-2} = Y^{d-2}\, , \qquad \mu^{d-1} = W\, , \qquad \mu^{d} = T\, ,
\end{equation}
\begin{equation}
    \boldsymbol{\beta}_1 = \delta \Upsilon_1 \, , \quad \cdots \quad \boldsymbol{\beta}_{d-2} = \delta \Upsilon_{d-2}\, , \qquad \boldsymbol{\beta}_{d-1} = \delta\Omega\, , \qquad \boldsymbol{\beta}_{d} = \slashed{\delta} \mathcal{A}\, .
\end{equation}

We now examine the SPS  integrability using the generalized Frobenius condition \eqref{frob-2-2-condition}. Assuming field-independent parameters ($\delta T = \delta W = \delta Y^{\tA} = 0$), we first construct the maximal wedge product $\mathcal{B}$:
\begin{equation}
    \begin{split}
       \mathcal{B} &= \boldsymbol{\beta}_{1} \curlywedge \cdots \curlywedge \boldsymbol{\beta}_{d} \\
       &= \delta \Upsilon_1 \curlywedge \cdots \curlywedge \delta \Upsilon_{d-2} \curlywedge \delta \Omega \curlywedge \big(\Omega \Theta_l \delta \mathcal{P} - \Omega N^{\tA\tB} \delta\Omega_{\tA\tB}\big)\, .
    \end{split}
\end{equation}
Notice that terms proportional to $\delta\Omega$ and $\delta\Upsilon_\tA$ drop out of $\mathcal{B}$ due to the antisymmetry of the exterior product. Applying the exterior derivative $\delta$ to $\boldsymbol{\beta}_d$, the only non-vanishing Frobenius condition yields:
\begin{equation}\label{integrability-d-dim}
    \begin{split}
    \delta \boldsymbol{\beta}_{d} \curlywedge \mathcal{B} &= \delta \big( \Omega\Theta_l \delta \mathcal{P} - \Omega N^{\tA\tB} \delta\Omega_{\tA\tB} \big) \\
    &\quad \curlywedge \delta \Upsilon_1 \curlywedge \cdots \curlywedge \delta \Upsilon_{d-2} \curlywedge \delta \Omega \curlywedge \big(\Omega\Theta_l \delta \mathcal{P} - \Omega N^{\tC\tD} \delta\Omega_{\tC\tD}\big) \neq  0\, .
    \end{split}
\end{equation}
This non-vanishing result rigorously proves that the SPS cannot be foliated into integrable submanifolds. Crucially, this failure is due to the presence of propagating bulk modes encoded in the news tensor ($N^{\tA\tB} \neq 0$). Geometrically, this indicates that the SPS possesses a non-trivial torsion 2-form $\boldsymbol{\mathscr{F}}$.

\subsubsection{Cartan Structure and genuine flux} 

To reveal the precise geometric origin of this non-integrability, we explicitly construct the Cartan structure equations for this SPS. We aim to cast the exterior derivative of the charge aspects $\boldsymbol{\beta}_a = (\boldsymbol{\beta}_\tA, \boldsymbol{\beta}_{d-1}, \boldsymbol{\beta}_d)^T$ into the fundamental form:
\begin{equation}\label{Cartan-structure-ex3}
    \delta \boldsymbol{\beta}_{a} = \boldsymbol{\mathscr{A}}_{a}{}^{b} \curlywedge \boldsymbol{\beta}_{b} + \boldsymbol{\mathscr{F}}_{a}\, ,
\end{equation}
where $\boldsymbol{\mathscr{A}}_{a}{}^{b}$ is the charge bundle connection (capturing the fake flux generator) and $\boldsymbol{\mathscr{F}}_{a}$ is the torsion 2-form (the genuine flux).

For the first $d-1$ basis elements, the variations are exact 1-forms ($\boldsymbol{\beta}_\tA = \delta \Upsilon_\tA$, $\boldsymbol{\beta}_{d-1} = \delta \Omega$), naturally implying that $\delta \boldsymbol{\beta}_\tA = 0$ and $\delta \boldsymbol{\beta}_{d-1} = 0$. Therefore, the components $\boldsymbol{\mathscr{A}}_{\tA}{}^{b}$, $\boldsymbol{\mathscr{A}}_{d-1}{}^{b}$, $\boldsymbol{\mathscr{F}}_\tA$, and $\boldsymbol{\mathscr{F}}_{d-1}$ all identically vanish.

The non-trivial dynamics reside entirely in $\boldsymbol{\beta}_d = \slashed{\delta}\mathcal{A}$. Taking its exterior derivative, we find:
\begin{equation}\label{delta-beta-d-initial}
    \delta \boldsymbol{\beta}_{d} = \delta(\Omega \Theta_l) \curlywedge \delta \mathcal{P} - \delta\Gamma \curlywedge \delta\Omega + \delta\mathcal{U}^{\tA} \curlywedge \delta \Upsilon_{\tA} - \delta(\Omega N^{\tA\tB}) \curlywedge \delta\Omega_{\tA\tB}\, .
\end{equation}
To map this into the Cartan framework \eqref{Cartan-structure-ex3}, we must express $\delta \mathcal{P}$ in terms of the basis 1-forms $\boldsymbol{\beta}_a$. From the definition of $\slashed{\delta}\mathcal{A}$, we can isolate $\delta \mathcal{P}$ as:
\begin{equation}\label{delta-P-sub}
    \delta \mathcal{P} = \frac{1}{\Omega \Theta_l} \big( \boldsymbol{\beta}_d + \Gamma \boldsymbol{\beta}_{d-1} - \mathcal{U}^\tA \beta_\tA + \Omega N^{\tA\tB} \delta\Omega_{\tA\tB} \big)\, .
\end{equation}
Substituting \eqref{delta-P-sub} back into \eqref{delta-beta-d-initial} and carefully grouping the terms by the basis elements $\boldsymbol{\beta}_a$, we obtain:
\begin{equation}
\begin{split}
    \delta \boldsymbol{\beta}_d &= \frac{\delta(\Omega \Theta_l)}{\Omega \Theta_l} \curlywedge \boldsymbol{\beta}_d + \left( \Gamma \frac{\delta(\Omega \Theta_l)}{\Omega \Theta_l} - \delta\Gamma \right) \curlywedge \boldsymbol{\beta}_{d-1} + \left( \delta\mathcal{U}^\tA - \mathcal{U}^\tA \frac{\delta(\Omega \Theta_l)}{\Omega \Theta_l} \right) \curlywedge \boldsymbol{\beta}_\tA \\
    &\quad - \left( \delta(\Omega N^{\tA\tB}) - \Omega N^{\tA\tB} \frac{\delta(\Omega \Theta_l)}{\Omega \Theta_l} \right) \curlywedge \delta\Omega_{\tA\tB}\, .
\end{split}
\end{equation}
By matching this expression to the structure equation \eqref{Cartan-structure-ex3}, we can directly extract the non-vanishing components of the connection matrix $\boldsymbol{\mathscr{A}}_{a}{}^{b}$:
\begin{subequations}\label{Cartan-connection-ex3}
\begin{align}
    \boldsymbol{\mathscr{A}}_{d}{}^{\tA} &= \delta\mathcal{U}^\tA - \mathcal{U}^\tA \delta \ln(\Omega \Theta_l)\, , \\
    \boldsymbol{\mathscr{A}}_{d}{}^{d-1} &= \Gamma \delta \ln(\Omega \Theta_l) - \delta\Gamma\, , \\
    \boldsymbol{\mathscr{A}}_{d}{}^{d} &= \delta \ln(\Omega \Theta_l)\, .
\end{align}
\end{subequations}
The remaining piece gives the SPS torsion 2-form $\boldsymbol{\mathscr{F}}_d$, which can be neatly simplified to
\begin{equation}\label{Cartan-torsion-ex3}
    \boldsymbol{\mathscr{F}}_d = - \Omega \Theta_l \delta(\Theta_l^{-1} N^{\tA\tB}) \curlywedge \delta\Omega_{\tA\tB}\, ,
\end{equation}
where we used  the identity $\delta(\Omega N^{\tA\tB}) - \Omega N^{\tA\tB} \delta \ln(\Omega \Theta_l) = \Omega \Theta_l \delta(\Theta_l^{-1} N^{\tA\tB})$. {The fact that $\boldsymbol{\mathscr{F}}_a$ is a 2-form strongly suggests comparing it with the 2-form symplectic flux. In \cite{Adami:2023wbe} (in particular see Section 4 there), the boundary and bulk symplectic fluxes were studied extensively. Specifically, the quantity $(\Theta_l^{-1} N^{\tA\tB}) \curlywedge \delta\Omega_{\tA\tB}$ appeared as the source of the non-closedness of the boundary symplectic 2-form. A more detailed investigation of the relation between the SPS torsion and the (bulk and boundary) SPS symplectic fluxes is instructive and is left for future work.}

This explicit decomposition isolates the true source of non-integrability. The connection $\boldsymbol{\mathscr{A}}_{a}{}^{b}$ acts as the fake flux generator, defining parallel transport across the SPS slices. We can rigorously test if this slicing-dependent connection introduces any intrinsic obstruction by computing the associated boundary SPS curvature, defined via the Maurer-Cartan form $\boldsymbol{\mathscr{R}}_{a}{}^{b} := \delta \boldsymbol{\mathscr{A}}_{a}{}^{b} - \boldsymbol{\mathscr{A}}_{a}{}^{c} \curlywedge \boldsymbol{\mathscr{A}}_{c}{}^{b}$. 

Since only the $d$-th row of $\boldsymbol{\mathscr{A}}$ is non-zero, the wedge product term simplifies to $\boldsymbol{\mathscr{A}}_{a}{}^{c} \curlywedge \boldsymbol{\mathscr{A}}_{c}{}^{b} = \delta_{a}^{d} \boldsymbol{\mathscr{A}}_{d}{}^{d} \curlywedge \boldsymbol{\mathscr{A}}_{d}{}^{b}$. The boundary SPS curvature 2-form components are,
\begin{subequations}
\begin{align}
    \boldsymbol{\mathscr{R}}_{d}{}^{\tA} &= \delta \big( \delta\mathcal{U}^\tA - \mathcal{U}^\tA \delta \ln(\Omega \Theta_l) \big) - \delta \ln(\Omega \Theta_l) \curlywedge \big(\delta\mathcal{U}^\tA - \mathcal{U}^\tA \delta \ln(\Omega \Theta_l)\big) \nonumber\\
    &= - \delta\mathcal{U}^\tA \curlywedge \delta \ln(\Omega \Theta_l) - \delta \ln(\Omega \Theta_l) \curlywedge \delta\mathcal{U}^\tA = 0\, , \\
    \boldsymbol{\mathscr{R}}_{d}{}^{d-1} &= \delta \big( \Gamma \delta \ln(\Omega \Theta_l) - \delta\Gamma \big) - \delta \ln(\Omega \Theta_l) \curlywedge \big(\Gamma \delta \ln(\Omega \Theta_l) - \delta\Gamma\big) \nonumber\\
    &= \delta\Gamma \curlywedge \delta \ln(\Omega \Theta_l) + \delta \ln(\Omega \Theta_l) \curlywedge \delta\Gamma = 0\, , \\
    \boldsymbol{\mathscr{R}}_{d}{}^{d} &= \delta \big( \delta \ln(\Omega \Theta_l) \big) - \delta \ln(\Omega \Theta_l) \curlywedge \delta \ln(\Omega \Theta_l) = 0\, .
\end{align}
\end{subequations}
We find that $\boldsymbol{\mathscr{R}}_{a}{}^{b} \equiv 0$ exactly. The charge bundle connection is Maurer-Cartan flat, confirming that the slicing ambiguity (fake flux) does not intrinsically obstruct integrability. Therefore, the non-integrability is  entirely driven by the torsion $\boldsymbol{\mathscr{F}}_d \neq 0$ given in \eqref{Cartan-torsion-ex3}. From a physical standpoint, this provides a profound geometric conclusion: the existence of radiating bulk gravitational waves (parameterized by the news tensor $N^{\tA\tB}$) manifests purely as a non-vanishing genuine flux (torsion) on the SPS, breaking the Frobenius integrability independent of any gauge or slicing choice.

\section{Discussion}\label{sec: discussion}

A distinctive feature of the Covariant Phase Space Formalism (CPSF) is {its intrinsic geometric nature}, involving two  distinct manifolds: the physical spacetime and the Solution Phase Space (SPS). In this work, we have reformulated CPSF to treat the two manifolds in parallel, demonstrating that the concept of covariance serves as the fundamental link between them. A crucial ingredient that we identified and explored is the nature of coordinate transformations within the SPS. In particular, we established a rigorous mathematical framework for these transformations, the  ``change-of-slicing'' technique as a local frame transformation on the SPS, and obtained several significant results.

We revisited the Wald-Zoupas criterion for integrability of surface charges \cite{Wald:1999wa}. In our geometric framework, we established that the Frobenius theorem (or dual-Frobenius theorem) provides a covariant and SPS slicing-independent integrability condition. This refined condition is robust and physically precise, as it is solely sensitive to the genuine origin of non-integrability: the radiation of bulk propagating degrees of freedom that couple to boundary modes through the flux-balance equations, like Bondi or Damour equations \cite{Bondi:1962px, Sachs:1962wk, PhD:Damour}. {This geometric perspective allows us to distinguish clearly between what we term \emph{fake flux} and \emph{genuine flux}.} We have shown that fake flux arises as a gauge artifact, mathematically represented by the connection on the SPS, which can be altered or eliminated by a suitable choice of slicing; {in the same sense that one can always find a local frame in standard differential geometry where the Christoffel symbols vanish.} In contrast, genuine flux is tied directly to the torsion of the SPS, a geometric invariant representing physical radiation that cannot be removed by a coordinate transformation. Furthermore, this comprehensive geometric framework facilitated a novel formulation of the Boundary Liouville Theorem, enabling us to carefully delineate the different physical regimes that emerge in the presence or absence of bulk propagating modes. The power and utility of this approach were then demonstrated through a detailed analysis of various explicit examples, showing how our methods can be used to systematically remove fake flux and isolate the true physical dynamics at the boundary.

There are several promising directions for further consideration that naturally emerge from the geometric framework established in this work:
    \begin{itemize}
    \item {\textbf{Memory Effects, Holonomies, and Algebras.} One of the main objectives of this study is to develop a rigorous definition of geometric quantities--connection, torsion, and curvature--intrinsic to the boundary sector of the SPS. This formalism provides a natural language to revisit memory effects, which are traditionally categorized into soft and hard parts. A crucial open question is to rigorously distinguish slicing-dependent and slicing-independent parts  of these memory effects. We propose the following geometric identification: within our framework, the soft memory effect manifests as the holonomy associated with the boundary connection on the SPS, which is slicing-dependent, whereas the hard memory corresponds to the SPS torsion, which is slicing-independent. {This perspective warrants further rigorous investigation, especially in comparison with alternative approaches that relate field-space connections directly to soft charges \cite{Riello:2019tad}.}

    Furthermore, this geometric perspective is deeply tied to the algebraic formulation of memory effects. While the standard Barnich-Troessaert (BT) bracket provides a framework for computing the algebra of the integrable parts of surface charges (boundary modes), a complete description of the radiative phase space requires understanding the cross-commutation relations between the boundary and bulk modes. Recent progress has established brackets for the isolated boundary-boundary and bulk-bulk sectors \cite{Adami:2023wbe, Ciambelli:2023mir}, but the boundary-bulk relations remain to be fully explored. Our curvature-based framework cleanly isolates genuine physical flux from the fake flux that is coordinate artifacts, thus, it provides the precise slicing-independent foundation needed to uniquely define these cross-sector brackets. This suggests an algebraic formulation of the memory effect. We postpone the explicit development of the formulation of geometric/algebraic  memory effect to future work.} 

    \item \textbf{Change-of-Slicing and the Angular Momentum Problem.} A long-standing challenge in general relativity is the so-called ``angular momentum problem,'' which refers to the inherent ambiguity in defining a consistent, supertranslation-invariant angular momentum at null infinity. In the standard Bondi-Sachs framework, the definition of angular momentum is sensitive to the choice of origin and is shifted under supertranslations, leading to a variety of competing definitions \cite{Penrose:1982wp}. Recent progress \cite{Chen:2022ymc, Alaee:2023oot, Javadinezhad:2023mtp, Javadinezhad:2022ldc, Fuentealba:2023syb} (see \cite{Mao:2023evc} for an intuitive physical picture) in resolving this issue often relies on a strategic redefinition of the angular momentum charge--a procedure that, in the terminology developed in this paper, is precisely a change-of-slicing in the SPS. {Our formalism suggests that the angular momentum problem may be related to specific slicing choices rather than a fundamental physical ambiguity. It would be interesting to employ our geometric tools}--specifically the SPS connection and the slicing-independent integrability criterion--to rigorously demonstrate how a preferred slicing might naturally emerge to yield a supertranslation-invariant and physically robust definition of angular momentum.
    
     \item {\textbf{Boundary Liouville Theorem and SPS Ergodicity.} In classical Hamiltonian mechanics, Liouville's theorem guarantees the conservation of phase space volume for \emph{closed} systems. However, by explicitly isolating genuine flux (torsion) from gauge artifacts (fake flux), our formalism accommodates \emph{open} boundary parts of gravitational systems where this conservation is dynamically violated. When genuine physical radiation crosses a boundary--such as falling into a black hole or escaping to null infinity--the boundary SPS volume form evolves, reflecting the dissipative nature of the spacetime. This geometric non-conservation inherently breaks ergodicity within the boundary SPS, providing a purely phase-space-geometric arrow of time. It is a fascinating open question to connect this generalized boundary Liouville theorem directly to macroscopic entropy production, thereby basing the second law of thermodynamics in gravitational systems (see \cite{Shajiee:2025cxl} for a novel derivation within CPSF) entirely within the geometric data (torsion and volume forms) of the SPS.}
    
    \item \textbf{Revisiting Black Hole Thermodynamics.} The first law of black hole mechanics is fundamentally an integrability condition relating the variations of conserved charges, naturally formulating itself as a 1-form relation on the SPS. This geometric perspective makes it highly compelling to investigate the precise role of slicing changes within the first law. Physically, it is strongly anticipated that a change-of-slicing in the SPS corresponds to a transition between different thermodynamic ensembles. Furthermore, in \cite{Adami:2021kvx}, the standard framework of black hole thermodynamics was extended from strictly stationary configurations to dynamical regimes, yielding an interpretation of the gravitational equations as an open local thermodynamic system. Revisiting this extended formalism, equipped with the rigorous geometric tools developed in this paper, promises to clarify the underlying thermodynamic structure and provide a robust, coordinate-independent formulation for dynamical black holes.
    
    \item {\textbf{Edge Modes and Fluctuating Boundaries.} The traditional CPSF framework, as utilized in this paper, encounters the issue of non-integrable surface charges. 
    Recently, various approaches—including the introduction of edge modes, the use of dynamical frames, or allowing the boundary itself to fluctuate—have been developed to extend the 
    CPSF such that (among other things) it yields integrable charges \cite{Adami:2024gdx, Ciambelli:2021nmv, Freidel:2021dxw, Carrozza:2022xut, Liu:2026bao}.  
    A notable consequence of these developments is that the inclusion of boundary fluctuations or additional boundary or corner degrees of freedom can render the charge variation integrable, 
    with the resulting integrable part being governed by the Noether charge.} Viewed through the lens of the Frobenius theorem and the geometric SPS framework developed here, it is natural to
    investigate the precise mechanism by which genuine non-integrability is ``absorbed'' into these boundary fluctuations. Specifically, understanding how the SPS connection, torsion, and curvature transform in the presence of such dynamical boundaries remains a compelling direction for future research.

    \item \textbf{Geometric Structure of the BT Bracket.} The Barnich-Troessaert (BT) bracket is the standard tool for handling the algebra of non-integrable charges. However, our work demonstrates that the splitting of the charge variation into an integrable part and a flux term is inherently slicing-dependent. It is therefore crucial to investigate how this freedom is reflected in the overall structure of asymptotic and boundary symmetry algebras. Specifically, future studies should analyze how various components of the BT bracket--including central extensions and cocycles--transform under a change-of-slicing. By utilizing the SPS connection to systematically disentangle quantities that are genuinely covariant from those that are mere artifacts of the slicing choice, one could potentially formulate a rigorously ``slicing covariant''  BT-bracket.

   \item \textbf{Other Examples.} While this paper exclusively analyzed pure Einstein gravity across various spacetime dimensions, a natural next step is to test the robustness of our geometric formulation by incorporating matter fields and extending it to alternative gravitational theories. Furthermore, whereas our primary illustrative example focused on finite-distance null boundaries, it is crucial to apply this machinery to the most widely studied asymptotic regions: the asymptotic timelike boundaries of AdS spacetimes and the null infinity of asymptotically flat spacetimes. In the latter context, we expect that the well-known Bondi news tensor will naturally emerge as the physical source of the SPS torsion.

\end{itemize}

\section*{Acknowledgements}
We are especially grateful to Hamed Adami for his long-standing partnership on related subjects, his valuable contributions during the early stages of this project, and for many insightful discussions. M.G. acknowledges financial support from the Iran National Science Foundation (INSF) under project No. 4034934. The research of V.T. and M.H.V. is supported by the INSF under project No. 4040771. M.M.Sh.J. is supported in part by the INSF Research Chair grant No. 4045163.

\appendix

\section{Useful identities and calculus relations}\label{app:identities}
In this appendix, we summarize the differential geometric identities used throughout the paper. We distinguish between operations defined on the spacetime manifold $\mathcal{M}$ and those defined on the field space manifold $\Gamma$.

\subsection{Spacetime calculus}
We consider the spacetime manifold with coordinates $x^\mu$. The calculus is defined by the exterior derivative $\mathrm{d}$, the interior product (contraction) $i_\xi$ with a vector field $\xi = \xi^\mu \partial_\mu$, and the Lie derivative $\mathcal{L}_\xi$.

\begin{itemize}
    \item \textbf{Cartan Magic Formula:}
    \begin{equation}
        \mathcal{L}_\xi = i_\xi \mathrm{d} + \mathrm{d} i_\xi \,.
    \end{equation}
    
    \item \textbf{Commutator Relations:}
    \begin{subequations}
    \begin{align}
        [\mathcal{L}_\xi, \mathcal{L}_\zeta] &= \mathcal{L}_{[\xi, \zeta]} \,, \\
        [\mathcal{L}_\xi, i_\zeta] &= i_{[\xi, \zeta]} \,, \\
        [\mathcal{L}_\xi, \mathrm{d}] &= 0 \,. 
    \end{align}
    \end{subequations}
    
    \item \textbf{Action on Forms:}
    For a generic $p$-form $\omega$ and $q$-form $\eta$, the Lie derivative satisfies the Leibniz rule:
    \begin{equation}
        \mathcal{L}_\xi (\omega \wedge \eta) = (\mathcal{L}_\xi \omega) \wedge \eta + \omega \wedge (\mathcal{L}_\xi \eta) \,.
    \end{equation}
\end{itemize}

\subsection{Field space calculus}
We consider the infinite-dimensional field space manifold where the ``coordinates'' are the fundamental fields $\phi^I$. The calculus is defined by the variational exterior derivative $\delta$, the variational interior product $\mathrm{I}_{\hat{\xi}}$ with a field space vector $\hat{\xi} = \hat{\xi}^I \fspartial_I$, and the variational Lie derivative $\mathrm{L}_{\hat{\xi}}$.

\begin{itemize}
\item \textbf{Basic Definition, ``Fundamental Link''} (between field space and spacetime).
\begin{equation}
    \hat{\xi}^I := \mathcal{L}_{\xi[\Phi]} \phi^I
\end{equation}
    \item \textbf{Field Space Cartan Formula:}
    \begin{equation}
        \mathrm{L}_{\hat{\xi}} = \mathrm{I}_{\hat{\xi}} \delta + \delta \mathrm{I}_{\hat{\xi}} \,.
    \end{equation}
    
    \item \textbf{Field Space Commutator Relations:}
    Analogous to spacetime, we have:
    \begin{subequations}
    \begin{align}
        [\mathrm{L}_{\hat{\xi}}, \mathrm{L}_{\hat{\zeta}}] &= \mathrm{L}_{\hll \hat{\xi}, \hat{\zeta} \hrr} \,, \\
        [\mathrm{L}_{\hat{\xi}}, \mathrm{I}_{\hat{\zeta}}] &= \mathrm{I}_{\hll \hat{\xi}, \hat{\zeta} \hrr} \,, \\
        [\mathrm{L}_{\hat{\xi}}, \delta] &= 0 \,. 
    \end{align}
    \end{subequations}
    Here $\hll \hat{\xi}, \hat{\zeta} \hrr$ denotes the Lie bracket on the field space.

    \item \textbf{Basis Action:}
    The action on the fundamental coordinates $\phi^I$ and basis 1-forms $\delta\phi^I$ is given by:
    \begin{equation}
        \mathrm{L}_{\hat{\xi}} \phi^I = \mathrm{I}_{\hat{\xi}} \delta \phi^I = \hat{\xi}^I \,, \qquad \mathrm{L}_{\hat{\xi}} (\delta \phi^I) = \delta \hat{\xi}^I \,.
    \end{equation}
\end{itemize}

\subsection{Mixed spacetime-field Space identities}
These identities connect the two manifolds via the fundamental link $\hat{\xi}^I = \mathcal{L}_{\xi[\Phi]} \phi^I$. We assume generic field-dependent parameters $\xi[\Phi]$ and $\zeta[\Phi]$.

\begin{itemize}
    \item \textbf{Commutation of $\delta$ and $\mathcal{L}_\xi$:}
    The variation $\delta$ acts on the field dependence of the spacetime Lie derivative. For any tensor $\phi^I$:
    \begin{equation} \label{eq:delta_Lie_comm}
        \delta \left( \mathcal{L}_{\xi} \phi^I \right) = \mathcal{L}_{\xi} \delta \phi^I + \mathcal{L}_{\delta \xi} \phi^I \,.
    \end{equation}
    
    \item \textbf{Contraction Identity:}
    A crucial identity used in establishing covariance involves the commutation of the field space contraction $\mathrm{I}_{\hat{\eta}}$ with the spacetime Lie derivative $\mathcal{L}_\xi$ when acting on the basis form $\delta\Phi^I$:
    \begin{equation} \label{eq:mixed_contraction}
        \mathrm{I}_{\hat{\eta}} \left( \mathcal{L}_{\xi} \delta \phi^I \right) = \mathcal{L}_{\xi} \left( \mathrm{I}_{\hat{\eta}} \delta \phi^I \right) \,.
    \end{equation}
\end{itemize}

\section{Proof of the adjusted-bracket relation}\label{adj-bracket-proof}

In this appendix, we provide a detailed proof of the relation \eqref{Lie-brackets}, which connects the Lie bracket on field space to the adjusted Lie bracket on spacetime. 

Consider two field space vector fields $\hat{\eta}$ and $\hat{\zeta}$. Based on the fundamental link between spacetime and the field space \eqref{fundmental-rel}, their components are defined by the action of the spacetime transformations $\eta$ and $\zeta$ on the fields:
\begin{equation}
    \hat{\eta}^I[\Phi] = \delta_{\eta[\Phi]} \phi^I \,, \qquad \hat{\zeta}^I[\Phi] = \delta_{\zeta[\Phi]} \phi^I \,.
\end{equation}
Here, $\delta_{\eta}$ represents the spacetime transformation (e.g., Lie derivative $\mathcal{L}_\eta$ for diffeomorphisms or gauge transformation) associated with the parameter $\eta$.
The Lie bracket of these two vectors in field space is defined as \eqref{Lie-deriv-FS}:
\begin{equation} \label{eq:field_bracket_def_gen}
    \hll \hat{\eta}, \hat{\zeta} \hrr^I := \hat{\eta}^J \fspartial_J \hat{\zeta}^I - \hat{\zeta}^J \fspartial_J \hat{\eta}^I \,.
\end{equation}
The operator $\hat{\eta}^J \fspartial_J$ acts as the field space Lie derivative $\mathrm{L}_{\hat{\eta}}$ on scalar functionals. We compute the first term by applying this operator to $\hat{\zeta}^I$. Since $\hat{\zeta}^I$ depends on the fields $\Phi$ both explicitly through the transformation law and implicitly through the parameter $\zeta[\Phi]$, we apply the chain rule:
\begin{equation}
\label{eq:variation_chain_gen}
    \begin{split}
        \hat{\eta}^J \fspartial_J \hat{\zeta}^I &= \mathrm{L}_{\hat{\eta}} \left( \delta_{\zeta[\Phi]} \phi^I \right) \\
        &= \delta_{\zeta} \left( \mathrm{L}_{\hat{\eta}} \phi^I \right) + \delta_{\left( \mathrm{L}_{\hat{\eta}} \zeta \right)} \phi^I \,.
    \end{split}
\end{equation}
In the first term on the right-hand side, the variation acts on the argument of the transformation (the field $\phi^I$), yielding $\mathrm{L}_{\hat{\eta}} \phi^I = \hat{\eta}^I$. In the second term, the variation acts on the parameter $\zeta$. By the definition of the adjusted bracket, we treat the parameters as scalars on the field space, so the variation of the parameter is $\mathrm{L}_{\hat{\eta}} \zeta$.
Substituting these relations into \eqref{eq:variation_chain_gen}, we find:
\begin{equation}
    \hat{\eta}^J \fspartial_J \hat{\zeta}^I = \delta_{\zeta} \big( \hat{\eta}^I \big) + \delta_{\big( \mathrm{L}_{\hat{\eta}} \zeta \big)} \phi^I \,.
\end{equation}
Due to the symmetry between $\eta$ and $\zeta$, the second term of the bracket \eqref{eq:field_bracket_def_gen} is:
\begin{equation}
    \hat{\zeta}^J \fspartial_J \hat{\eta}^I = \delta_{\eta} \big( \hat{\zeta}^I \big) + \delta_{\big( \mathrm{L}_{\hat{\zeta}} \eta \big)} \phi^I \,.
\end{equation}
Subtracting these two results gives the components of the field space bracket:
\begin{equation}
    \hll \hat{\eta}, \hat{\zeta} \hrr^I = \big( \delta_{\zeta} \hat{\eta}^I - \delta_{\eta} \hat{\zeta}^I \big) + \delta_{\big( \mathrm{L}_{\hat{\eta}} \zeta \big)} \phi^I - \delta_{\big( \mathrm{L}_{\hat{\zeta}} \eta \big)} \phi^I \,.
\end{equation}
Recall that $\hat{\eta}^I = \delta_\eta \phi^I$ and $\hat{\zeta}^I = \delta_\zeta \phi^I$. The term in the first parenthesis corresponds to the commutator of the spacetime transformations acting on the field:
\begin{equation}
    \delta_{\zeta} (\delta_\eta \phi^I) - \delta_{\eta} (\delta_\zeta \phi^I) = - [\delta_\eta, \delta_\zeta] \phi^I \,.
\end{equation}
For covariant fields, the commutator of transformations satisfies $[\delta_\eta, \delta_\zeta] \phi^I = \delta_{[\eta, \zeta]} \phi^I$, where $[\eta, \zeta]$ is the standard spacetime Lie bracket of the parameters. Thus, we have:
\begin{equation}
    \begin{split}
        \hll \hat{\eta}, \hat{\zeta} \hrr^I &= -\delta_{[\eta, \zeta]} \phi^I + \delta_{\mathrm{L}_{\hat{\eta}} \zeta} \phi^I - \delta_{\mathrm{L}_{\hat{\zeta}} \eta} \phi^I \\
        &= -\delta_{\left( [\eta, \zeta] - \mathrm{L}_{\hat{\eta}} \zeta + \mathrm{L}_{\hat{\zeta}} \eta \right)} \phi^I \,.
    \end{split}
\end{equation}
The term inside the subscript is precisely the definition of the adjusted bracket $\llbracket \eta, \zeta \rrbracket$. Using the fundamental link $\delta_{\xi} \phi^I = \hat{\xi}^I$, we conclude:
\begin{equation}
    \hll \hat{\eta}, \hat{\zeta} \hrr^I = -\delta_{\llbracket \eta, \zeta \rrbracket} \phi^I = -\widehat{\llbracket \eta, \zeta \rrbracket}^I \,.
\end{equation}
This completes the proof for generic tensor fields.
\section{Covariance in the field space}\label{appen:covariance}
\paragraph{Identity I.} 
We will frequently use the following identity in the equations below
\begin{equation}
   \inbox{ \delta\,\mathcal{L}_{\xi}\Phi=\mathcal{L}_{\xi}\delta\Phi + \mathcal{L}_{\delta\xi}\Phi\, .}
\end{equation}
in which $\delta \xi:=\frac{\fspartial \xi}{\fspartial \Phi} \delta \Phi$.

\textit{Proof.}
\begin{align}
    \delta (\delta_{\xi} \Phi)&=\delta_{\xi+\delta \xi}( \Phi+\delta \Phi)-\delta_{\xi}\Phi\nonumber\\
    &=\delta_{\xi}(\Phi+\delta \Phi)
     +\delta_{\delta {\xi}}(\Phi+\delta \Phi)
     -\delta_{\xi} \Phi\\
     &=\delta_{\xi}\delta \Phi+\delta_{\delta{\xi}} \Phi,\nonumber
\end{align}
here $\delta_{\delta \xi} \delta \Phi=0$ is used. By replacing $\delta_\xi \Phi$ with $\mathcal{L}_\xi \Phi$, we obtain the desired result.
\paragraph{Identity II.}
Another useful identity is as follows
\begin{equation}
   \inbox{ \mathrm{I}_{\hat\xi_2}\mathcal{L}_{\xi_1} \delta\Phi=\mathcal{L}_{\xi_1} \mathrm{I}_{\hat\xi_2}\delta\Phi\, .}
\end{equation}
\textit{Proof.}
\begin{equation}\begin{split}
     \mathrm{I}_{\hat\xi_2}\mathcal{L}_{\xi_1} \delta\Phi&=\mathrm{I}_{\hat\xi_2}\qty(\mathrm{L}_{\hat\xi_1} \delta\Phi - \mathcal{L}_{{\delta\xi_1}}\Phi)=\mathrm{I}_{\hat\xi_2}\mathrm{L}_{\hat\xi_1} \delta\Phi - \mathcal{L}_{\mathrm{I}_{\hat\xi_2}\delta\xi_1}\Phi=\mathrm{I}_{\hat\xi_2}(\mathrm{I}_{\hat\xi_1}\delta+\delta\mathrm{I}_{\hat\xi_1}) \delta\Phi - \mathcal{L}_{(\mathrm{I}_{\hat\xi_2}\delta+\delta\mathrm{I}_{\hat\xi_2})\xi_1}\Phi\\
     &=\mathrm{I}_{\hat\xi_2}\delta\,\mathrm{I}_{\hat\xi_1} \delta\Phi - \mathcal{L}_{\mathrm{L}_{\hat\xi_2}\xi_1}\Phi=\qty(\mathrm{I}_{\hat\xi_2}\delta+\delta\mathrm{I}_{\hat\xi_2})\,\qty(\mathrm{I}_{\hat\xi_1}\delta+\delta\mathrm{I}_{\hat\xi_1} )\Phi - \mathcal{L}_{\mathrm{L}_{\hat\xi_2}\xi_1}\Phi=\mathrm{L}_{\hat\xi_2}\,\mathrm{L}_{\hat\xi_1}\Phi - \mathcal{L}_{\mathrm{L}_{\hat\xi_2}\xi_1}\Phi\\
     &=\mathcal{L}_{\xi_2}\,\mathrm{L}_{\hat\xi_1}\Phi + \mathcal{L}_{[\xi_1,\xi_2]+\mathrm{L}_{\hat\xi_2}\xi_1}\Phi- \mathcal{L}_{\mathrm{L}_{\hat\xi_2}\xi_1}\Phi=\mathcal{L}_{\xi_2} \mathrm{L}_{\hat\xi_1} \Phi+\mathcal{L}_{[\xi_1,\xi_2]}\Phi=\mathcal{L}_{\xi_1} \mathrm{L}_{\hat\xi_2} \Phi=\mathcal{L}_{\xi_1} \mathrm{I}_{\hat\xi_2}\delta\Phi\, .
\end{split}\end{equation}
\paragraph{Transformation of $\delta\Phi$.} 
$\delta\Phi$ transforms in the field space as follows
\begin{equation}
    \inbox{\mathrm{L}_{\hat \xi}(\delta\Phi) = \qty(\mathcal{L}_{\xi} + \mathrm{I}_{\widehat{\delta\xi}})\delta\Phi\, .}
\end{equation}
\textit{Proof:}
\begin{equation}
    \begin{split}
   \mathrm{L}_{\hat \xi}(\delta\Phi)&= (\mathrm{I}_{\hat\xi} \delta + \delta \mathrm{I}_{\hat\xi})\delta\Phi = \delta \mathrm{I}_{\hat\xi}\delta\Phi = \delta(\mathrm{I}_{\hat\xi} \delta + \delta \mathrm{I}_{\hat\xi})\Phi=\delta\;\mathrm{L}_{\hat \xi}\Phi=\delta\;\mathcal{L}_{\xi}\Phi =\mathcal{L}_{\xi}\delta\Phi + \mathcal{L}_{\delta\xi}\Phi \\
    &=\mathcal{L}_{\xi}\delta\Phi + \mathrm{L}_{\widehat{\delta\xi}}\Phi=\mathcal{L}_{\xi}\delta\Phi + \qty(\delta \mathrm{I}_{\widehat{\delta\xi}}+\mathrm{I}_{\widehat{\delta\xi}}\delta)\Phi=\qty(\mathcal{L}_{\xi} + \mathrm{I}_{\widehat{\delta\xi}})\delta\Phi\, .
    \end{split}
\end{equation}
\paragraph{Transformation of $\mathrm{L}_{\hat \xi_1}\Phi$.}
$\mathrm{L}_{\hat \xi_1}\Phi$ transforms as follows
\begin{equation}
    \inbox{\mathrm{L}_{\hat \xi_2}(\mathrm{L}_{\hat \xi_1}\Phi) = \mathcal{L}_{\xi_2} \mathrm{L}_{\hat\xi_1}\Phi+ \mathrm{L}_{\widehat{[\xi_1,\xi_2]+\mathrm{L}_{\hat\xi_2}\xi_1}}\Phi\, .}
\end{equation}
\textit{Proof:}
\begin{equation}
    \begin{split}
   \mathrm{L}_{\hat \xi_2}(\mathrm{L}_{\hat \xi_1}\Phi)&= \mathrm{L}_{\hat \xi_2}(\mathcal{L}_{\xi_1}\Phi)=(\mathrm{I}_{\hat\xi_2} \delta + \delta \mathrm{I}_{\hat\xi_2})\mathcal{L}_{\xi_1}\Phi= \mathrm{I}_{\hat\xi_2} \delta \mathcal{L}_{\xi_1}\Phi= \mathrm{I}_{\hat\xi_2}(\mathcal{L}_{\xi_1} \delta\Phi + \mathcal{L}_{\delta\xi_1}\Phi ) \\
   &=\mathcal{L}_{\xi_1}  \mathrm{I}_{\hat\xi_2}\delta\Phi+ \mathcal{L}_{ \mathrm{I}_{\hat\xi_2}\delta\xi_1}\Phi=\mathcal{L}_{\xi_1}(  \mathrm{I}_{\hat\xi_2}\delta+\delta  \mathrm{I}_{\hat\xi_2})\Phi+ \mathcal{L}_{(\mathrm{I}_{\hat\xi_2}\delta+\delta \mathrm{I}_{\hat\xi_2})\xi_1}\Phi =\mathcal{L}_{\xi_1} \mathrm{L}_{\hat\xi_2}\Phi+ \mathcal{L}_{\mathrm{L}_{\hat\xi_2}\xi_1}\Phi \\
   &=\mathcal{L}_{\xi_1} \mathcal{L}_{\xi_2}\Phi+ \mathcal{L}_{\mathrm{L}_{\hat\xi_2}\xi_1}\Phi=(\mathcal{L}_{\xi_2} \mathcal{L}_{\xi_1}+\mathcal{L}_{[\xi_1,\xi_2]})\Phi+ \mathcal{L}_{\mathrm{L}_{\hat\xi_2}\xi_1}\Phi=\mathcal{L}_{\xi_2} \mathcal{L}_{\xi_1}\Phi+ \mathcal{L}_{[\xi_1,\xi_2]+\mathrm{L}_{\hat\xi_2}\xi_1}\Phi\\
   &=\mathcal{L}_{\xi_2} \mathrm{L}_{\hat\xi_1}\Phi+ \mathrm{L}_{\widehat{[\xi_1,\xi_2]+\mathrm{L}_{\hat\xi_2}\xi_1}}\Phi\, .
    \end{split}
\end{equation}
\paragraph{Transformation of $\mathrm{L}_{\hat \xi_2}\Phi\,\mathrm{L}_{\hat \xi_1}\Phi$.}
$\mathrm{L}_{\hat \xi_2}\Phi\,\mathrm{L}_{\hat \xi_1}\Phi$ transforms as follows
\begin{equation}
    \inbox{\mathrm{L}_{\hat \xi_3}(\mathrm{L}_{\hat \xi_2}\Phi\,\mathrm{L}_{\hat \xi_1}\Phi) = \mathcal{L}_{\xi_3}\qty( \mathrm{L}_{\hat\xi_2}\Phi\,\mathrm{L}_{\hat \xi_1}\Phi)+ \mathrm{L}_{\widehat{[\xi_2,\xi_3]+\mathrm{L}_{\hat\xi_3}\xi_2}}\Phi\,\mathrm{L}_{\hat \xi_1}\Phi+\mathrm{L}_{\hat \xi_2}\Phi\, \mathrm{L}_{\widehat{[\xi_1,\xi_3]+\mathrm{L}_{\hat\xi_3}\xi_1}}\Phi\, .}
\end{equation}
\textit{Proof:}
\begin{equation}
    \begin{split}
   \mathrm{L}_{\hat \xi_3}(\mathrm{L}_{\hat \xi_2}\Phi\,\mathrm{L}_{\hat \xi_1}\Phi)&=
   \mathrm{L}_{\hat \xi_3}(\mathrm{L}_{\hat \xi_2}\Phi)\,\mathrm{L}_{\hat \xi_1}\Phi+\mathrm{L}_{\hat \xi_2}\Phi\,\mathrm{L}_{\hat \xi_3}(\mathrm{L}_{\hat \xi_1}\Phi)\\
   &=\qty(\mathcal{L}_{\xi_3} \mathrm{L}_{\hat\xi_2}\Phi+ \mathrm{L}_{\widehat{[\xi_2,\xi_3]+\mathrm{L}_{\hat\xi_3}\xi_2}}\Phi)\,\mathrm{L}_{\hat \xi_1}\Phi+\mathrm{L}_{\hat \xi_2}\Phi\,\qty(\mathcal{L}_{\xi_3} \mathrm{L}_{\hat\xi_1}\Phi+ \mathrm{L}_{\widehat{[\xi_1,\xi_3]+\mathrm{L}_{\hat\xi_3}\xi_1}}\Phi)\\
   &=\mathcal{L}_{\xi_3}\qty( \mathrm{L}_{\hat\xi_2}\Phi\,\mathrm{L}_{\hat \xi_1}\Phi)+ \mathrm{L}_{\widehat{[\xi_2,\xi_3]+\mathrm{L}_{\hat\xi_3}\xi_2}}\Phi\,\mathrm{L}_{\hat \xi_1}\Phi+\mathrm{L}_{\hat \xi_2}\Phi\, \mathrm{L}_{\widehat{[\xi_1,\xi_3]+\mathrm{L}_{\hat\xi_3}\xi_1}}\Phi\, .
    \end{split}
\end{equation}
\paragraph{Transformation of $\mathrm{L}_{\hat \xi_1}\Phi\,\delta\Phi$.} 
 $\mathrm{L}_{\hat \xi_1}\Phi\,\delta\Phi$ transforms as follows
 \begin{equation}
     \inbox{\mathrm{L}_{\hat \xi_2}(\mathrm{L}_{\hat \xi_1}\Phi\,\delta\Phi) = \mathcal{L}_{\xi_2} \qty(\mathrm{L}_{\hat{\xi_1}}\Phi\,\delta\Phi)+ \mathrm{L}_{\widehat{[\xi_1,\xi_2]+\mathrm{L}_{\hat\xi_2}\xi_1}}\Phi\,\delta\Phi +\mathrm{I}_{\widehat{\delta\xi_2}}\,\qty( \mathrm{L}_{\hat \xi_1}\Phi\,\delta\Phi)\, .}
 \end{equation}
 \textit{Proof:}
\begin{equation}
    \begin{split}
   \mathrm{L}_{\hat \xi_2}(\mathrm{L}_{\hat \xi_1}\Phi\,\delta\Phi)&=\mathrm{L}_{\hat \xi_2}(\mathrm{L}_{\hat \xi_1}\Phi)\,\delta\Phi+\mathrm{L}_{\hat \xi_1}\Phi\,\mathrm{L}_{\hat \xi_2}(\delta\Phi)\\
   &=\qty(\mathcal{L}_{\xi_2} \mathrm{L}_{\hat\xi_1}\Phi+ \mathrm{L}_{\widehat{[\xi_1,\xi_2]+\mathrm{L}_{\hat\xi_2}\xi_1}}\Phi)\,\delta\Phi \,+ \,\mathrm{L}_{\hat{\xi_1}}\Phi\,\qty(\mathcal{L}_{\xi_2} + \mathrm{I}_{\widehat{\delta\xi_2}})\delta\Phi\\
   &=\mathcal{L}_{\xi_2} \qty(\mathrm{L}_{\hat{\xi_1}}\Phi\,\delta\Phi)+ \mathrm{L}_{\widehat{[\xi_1,\xi_2]+\mathrm{L}_{\hat\xi_2}\xi_1}}\Phi\,\delta\Phi + \mathrm{L}_{\hat \xi_1}\Phi\, \mathrm{I}_{\widehat{\delta\xi_2}}\,\delta\Phi\\
   &=\mathcal{L}_{\xi_2} \qty(\mathrm{L}_{\hat{\xi_1}}\Phi\,\delta\Phi)+ \mathrm{L}_{\widehat{[\xi_1,\xi_2]+\mathrm{L}_{\hat\xi_2}\xi_1}}\Phi\,\delta\Phi +\mathrm{I}_{\widehat{\delta\xi_2}}\,\qty( \mathrm{L}_{\hat \xi_1}\Phi\,\delta\Phi)\,  .
    \end{split}
\end{equation}

\paragraph{Commutator of field space Lie derivatives.}
The commutator of the Lie field space Lie derivatives is given as follows
\begin{equation}
    \inbox{[\mathrm{L}_{\hat \xi_2}, \mathrm{L}_{\hat \xi_1}] = \mathrm{L}_{\hll \hat{\xi}_2, \hat{\xi}_1 \hrr} \, .}
\end{equation}
\textit{Proof:}
\begin{equation}
    \begin{split}
   \qty(\mathrm{L}_{\hat \xi_2}\mathrm{L}_{\hat \xi_1}-\mathrm{L}_{\hat \xi_1}\mathrm{L}_{\hat \xi_2})\Phi & = \qty(\mathcal{L}_{\xi_2} \mathcal{L}_{\xi_1}-\mathcal{L}_{\xi_1} \mathcal{L}_{\xi_2})\Phi+ \qty(\mathcal{L}_{[\xi_1,\xi_2]+\mathrm{L}_{\hat\xi_2}\xi_1}-\mathcal{L}_{[\xi_2,\xi_1]+\mathrm{L}_{\hat\xi_1}\xi_2})\Phi\\
   &=\qty(\mathcal{L}_{[\xi_2,\xi_1]})\Phi+ \qty(\mathcal{L}_{[\xi_1,\xi_2]+\mathrm{L}_{\hat\xi_2}\xi_1}-\mathcal{L}_{[\xi_2,\xi_1]+\mathrm{L}_{\hat\xi_1}\xi_2})\Phi\\
   &=\mathcal{L}_{\qty([\xi_1,\xi_2]-\mathrm{L}_{\hat\xi_1}\xi_2+\mathrm{L}_{\hat\xi_2}\xi_1)}\Phi \\
   &=\mathrm{L}_{\widehat{\qty([\xi_1,\xi_2]-\mathrm{L}_{\hat\xi_1}\xi_2+\mathrm{L}_{\hat\xi_2}\xi_1)}}\Phi \\
   & = \mathrm{L}_{\hll \hat{\xi}_2, \hat{\xi}_1 \hrr}\, \Phi\, .
    \end{split}
\end{equation}
One may view this as a consistency check of previous identities. In the above derivation we have used
\begin{equation}
    \begin{split}
   \mathrm{L}_{\hat \xi_2}(\mathrm{L}_{\hat \xi_1}\Phi)&= \mathcal{L}_{\xi_2} \mathcal{L}_{\xi_1}\Phi+ \mathcal{L}_{[\xi_1,\xi_2]+\mathrm{L}_{\hat\xi_2}\xi_1}\Phi\, ,\\
   \mathrm{L}_{\hat \xi_1}(\mathrm{L}_{\hat \xi_2}\Phi)&= \mathcal{L}_{\xi_1} \mathcal{L}_{\xi_2}\Phi+ \mathcal{L}_{[\xi_2,\xi_1]+\mathrm{L}_{\hat\xi_1}\xi_2}\Phi\, .
    \end{split}
\end{equation}

\section{Central extension vanishing decomposition}\label{appen: CEVD}
Eq. \eqref{charge-variation-def} suggests a  specific decomposition into $Q^{\text{\tiny{I}}}(\xi)$ and $F(\xi;\Phi)$, i.e. a specific choice of the $A$-freedom:
\begin{equation}\label{K-free-decomposition}
    Q^{\text{\tiny{I}}}(\xi)=\mathrm{I}_{\hat{\xi}}\Theta\, , \qquad F(\xi;\Phi)= - \mathrm{L}_{\hat\xi} \Theta\, ,
\end{equation}
The reason we did not discuss this simple and suggestive decomposition in the main text is that $\Theta$ is generically a codimension-1 (boundary) integral and therefore, $Q^{\text{\tiny{I}}}(\xi), F(\xi;\Phi)$ are given by codimension-1 integrals, while their sum, $\slashed\delta Q_\xi$ is codimension-2 corner integral. The requirement that $Q^{\text{\tiny{I}}}(\xi)$ is a codimension-2 corner integral is a physically well-motivated and important one. So, this decomposition, while interesting, is not physically motivated. 
With the above decomposition, and \textit{assuming that $\Theta$ is a covariant 1-form as defined in \eqref{covariance-fund-field}}, one can compute the adjusted bracket of two charges starting from the definition in \eqref{MOSPB}, 
\begin{equation}
\begin{split}
\left\{Q^{\text{\tiny{I}}}(\xi_1) , Q^{\text{\tiny{I}}}(\xi_2) \right\}_{\text{\tiny{BT}}}
=& \, \mathrm{I}_{\hat{\xi}_1} \mathrm{I}_{\hat{\xi}_2} \Omega - \mathrm{I}_{\hat{\xi}_2} F({\xi}_1) + \mathrm{{I}}_{\hat{\xi}_1} F(\xi_2)\\
=&\mathrm{I}_{\hat{\xi}_1}\mathrm{I}_{\hat{\xi}_2}\delta\Theta  - (\mathrm{I}_{\hat{\xi}_1} \mathrm{L}_{\hat{\xi}_2}- \mathrm{I}_{\hat{\xi}_2} \mathrm{L}_{\hat{\xi}_1})\Theta\\
=&\mathrm{I}_{\hat{\xi}_1}\mathrm{I}_{\hat{\xi}_2}\delta\Theta  - (\mathrm{I}_{\hll\hat{\xi}_1,\hat{\xi}_2\hrr}+\mathrm{L}_{\hat{\xi}_2}\mathrm{I}_{\hat{\xi}_1} - \mathrm{I}_{\hat{\xi}_2} \mathrm{L}_{\hat{\xi}_1})\Theta\\
=& \qty(\mathrm{I}_{\hat{\xi}_1}\mathrm{I}_{\hat{\xi}_2}\delta-\mathrm{I}_{\hll\hat{\xi}_1,\hat{\xi}_2\hrr}-\mathrm{L}_{\hat{\xi}_2}\mathrm{I}_{\hat{\xi}_1} + \mathrm{I}_{\hat{\xi}_2} \mathrm{L}_{\hat{\xi}_1})\Theta\\
=& \qty(\mathrm{I}_{\hat{\xi}_1}\mathrm{I}_{\hat{\xi}_2}\delta-\mathrm{I}_{\hll\hat{\xi}_1,\hat{\xi}_2\hrr}-\mathrm{I}_{\hat{\xi}_2}\delta\mathrm{I}_{\hat{\xi}_1}-\delta\mathrm{I}_{\hat{\xi}_2}\mathrm{I}_{\hat{\xi}_1} + \mathrm{I}_{\hat{\xi}_2} \mathrm{I}_{\hat{\xi}_1}\delta + \mathrm{I}_{\hat{\xi}_2} \delta\mathrm{I}_{\hat{\xi}_1})\Theta\\
=& \qty(-\mathrm{I}_{\hll\hat{\xi}_1,\hat{\xi}_2\hrr}-\delta\mathrm{I}_{\hat{\xi}_2}\mathrm{I}_{\hat{\xi}_1} )\Theta\\
=& -\mathrm{I}_{\hll\hat{\xi}_1,\hat{\xi}_2\hrr}\Theta=-Q^{\text{\tiny{I}}}\qty({\llbracket}\xi_1,\xi_2{\rrbracket})\, .
\end{split}
\end{equation}  
where in the above we used the fact that for any scalar $\Phi$ in the field space $\mathrm{I}_{\hat{\xi}}\Phi=0$ and that $\mathrm{I}_{\hat{\xi}}\Theta$ is a scalar for any vector on the field space $\hat\xi$. This analysis shows that for the decomposition in \eqref{K-free-decomposition}, the central term vanishes. 

We next show that $K_{\xi_1,\xi_2}$ only gets non-vanishing values through the non-scalar part of the charge $Q_0$, see footnote~\footref{fn:Q0-noncov}. To this end, we first note that the decomposition in \eqref{K-free-decomposition} is invariant under   $A$-freedom or $W$-freedom,  
explicitly,  
\begin{equation}
    \mathrm{I}_{\hat{\xi}}\Theta + A(\xi;\Phi) =  \mathrm{I}_{\hat{\xi}}\tilde\Theta:= \tilde Q^{\text{\tiny{I}}}(\xi)\, , \qquad  - \mathrm{L}_{\hat\xi} \Theta - \delta  A(\xi;\Phi) = - \mathrm{L}_{\hat\xi} \tilde\Theta\, ,
\end{equation}
where 
\begin{equation}
    \tilde\Theta = \Theta + \delta W, \qquad A(\xi;\Phi) =  \mathrm{L}_{\hat\xi} W[\Phi],
\end{equation} 
with $W$ being a scalar on the field space. Therefore,  $\tilde K_{\xi_1,\xi_2}=K_{\xi_1,\xi_2}=0$ and we reach our important statement,
\begin{center}
    \textit{The {two-cocycle} $K_{\xi_1,\xi_2}$ with decomposition \eqref{K-free-decomposition} only receives contribution from the noncovariant, zero or reference -point part of the integrable part of the charge $Q_0$.}
\end{center}
The above dovetails with the discussions and analysis in \cite{Adami:2024gdx, Golshani:2024fry}. 

Despite the elegant and central-term–free property discussed above, its codimension-1 nature is a drawback when defining charges associated with diffeomorphisms.

\section{Surface charge algebra}\label{appen: charge-algebra}

In this appendix, we derive the algebra of integrable surface charges and determine the associated {two-cocycle}. We start from the Noetherian splitting
\begin{equation}
Q^{\text{\tiny I}}(\xi) = Q^{\text{\tiny N}}(\xi) \, ,
\qquad
F(\xi;\delta\Phi)
=
\slashed{\delta}Q(\xi)-\delta Q^{\text{\tiny I}}(\xi)\, ,
\end{equation}
The algebra of integrable charges is defined through the Barnich–Troessaert (BT) bracket \eqref{MOSPB}.
To determine the {two-cocycle} \(K_{\xi_1,\xi_2}\), we compute separately the two contributions entering the BT bracket.

\paragraph{Variation of the integrable charge.}

Using the definition of \(Q^{\text{\tiny I}}\) we obtain
\begin{equation}\label{L_xi-Q^I}
\begin{split}
\mathrm{L}_{\hat{\xi}_2}Q^{\text{\tiny I}}(\xi_1)
&=
\mathrm{L}_{\hat{\xi}_2}Q_{\text{\tiny N}}(\xi_1)
\int_{\cal S}\d x_{\mu\nu}\,
\mathcal{L}_{\xi_2}Q^{\mu\nu}_{\text{\tiny N}}(\xi_1)
+
Q_{\text{\tiny N}}\!\left([\xi_1,\xi_2]+\mathrm{L}_{\hat\xi_2}\xi_1\right)
\end{split}
\end{equation}
In deriving this relation, we used
\begin{equation}
\int_{\cal S}\mathcal{L}_{\xi_2}\!\left(
\d x_{\mu\nu} Q^{\mu\nu}_{\text{\tiny N}}(\xi_1)
\right)
=
\int_{\cal S}\d x_{\mu\nu}\,
\mathcal{L}_{\xi_2}Q^{\mu\nu}_{\text{\tiny N}}(\xi_1), \quad \text{with} \quad \mathcal{L}_{\xi_2}Q^{\mu\nu}_{\text{\tiny N}}(\xi_1)
:=
\big(\mathcal{L}_{\xi_2}Q_{\text{\tiny N}}(\xi_1)\big)^{\mu\nu}.
\end{equation}

\paragraph{Contraction with the flux.}

Next, we evaluate the contraction of the flux with the vector field \(\hat\xi_1\),
\begin{equation}\label{I_xi-F}
\begin{split}
\mathrm{I}_{\hat{\xi}_1}F(\xi_2;\delta\Phi)
& =
-Q_{\text{\tiny N}}(\mathrm{I}_{\hat{\xi}_1}\delta\xi_2)
-
2\int_{\mathcal S}\d x_{\mu\nu}\,
\Big(\xi_2^\mu \mathrm{I}_{\hat{\xi}_1}\Theta^\nu\Big)
\\
& =
-Q_{\text{\tiny N}}(\mathrm{L}_{\hat{\xi}_1}\xi_2)
-
2\int_{\mathcal S}\d x_{\mu\nu}\,
\xi_2^\mu\xi_1^\nu L
-
\int_{\cal S}\d x_{\mu\nu}\,
\mathcal L_{\xi_2}Q^{\mu\nu}_{\text{\tiny N}}(\xi_1).
\end{split}
\end{equation}
In obtaining this expression, we used
\begin{equation}
-\int_{\cal S}\d x_{\mu\nu}\,
\mathcal L_{\xi_2}Q^{\mu\nu}_{\text{\tiny N}}(\xi_1)
=
2\int_{\cal S}\d x_{\mu\nu}\,
\xi_2^\mu \partial_\alpha
Q^{\nu\alpha}_{\text{\tiny N}}(\xi_1).
\end{equation}

\paragraph{Surface charge algebra.}

Combining \eqref{L_xi-Q^I} and \eqref{I_xi-F} we obtain
\begin{equation}
\left\{
Q^{\text{\tiny I}}(\xi_1),
Q^{\text{\tiny I}}(\xi_2)
\right\}_{\text{\tiny BT}}
=
Q^{\text{\tiny I}}\!\left(\llbracket\xi_1,\xi_2 \rrbracket\right)
+
K_{\xi_1,\xi_2}.
\end{equation}
The {two-cocycle} is therefore
\begin{equation}
\inbox{
K_{\xi_1,\xi_2}
=
-
2\int_{\mathcal S}\d x_{\mu\nu}\,
\xi_2^\mu\xi_1^\nu L \; .
}
\end{equation}
A similar result has been also obtained in \cite{Freidel:2021cbc, Freidel:2021dxw}.
This term satisfies the cocycle condition
\[
K_{[\xi_1,\xi_2],\xi_3}
-
\mathrm{L}_{\hat\xi_3}K_{\xi_1,\xi_2}
+
\text{cyclic permutations}
=0 .
\]
The last contribution can be written in a manifestly covariant form as
\begin{equation}
\int_{\mathcal S}\d{} x_{\mu\nu}\,
\xi_2^\mu\xi_1^\nu L
=
\int_{\mathcal S}
i_{\xi_1}i_{\xi_2}
\left(\d{}^d x\, L\right).
\end{equation}

\bibliographystyle{fullsort.bst}
\bibliography{reference}


	\end{document}